\documentclass[aps, prfluids, reprint,superscriptaddress, longbibliography, showkeys, onecolumn]{revtex4}
\pdfoutput=1
\usepackage{graphics}
\usepackage[normalem]{ulem}
\usepackage{graphicx}
\usepackage{epstopdf, epsfig}
\usepackage{amssymb}
\usepackage{amsmath}
\usepackage{csquotes}
\usepackage{color}
\usepackage{xcolor}
\usepackage{subcaption}
\usepackage{natbib}
\usepackage{bm}
\newcommand{\iu}{{i\mkern1mu}}

\begin{document}
\title{The hydrodynamics of slender swimmers near deformable interfaces}
\author{\firstname{Sankalp} \surname{Nambiar}}
\affiliation{Nordita, KTH Royal Institute of Technology and Stockholm University, Stockholm 10691, Sweden}
\author{J. S. \surname{Wettlaufer}}
\email{john.wettlaufer@su.se}
\affiliation{Nordita, KTH Royal Institute of Technology and Stockholm University, Stockholm 10691, Sweden}
\affiliation{Yale University, New Haven, CT 06520-8109, USA}

\begin{abstract}
We study the coupled hydrodynamics between a motile slender microswimmer and a deformable interface that separates two Newtonian fluid regions. From the disturbance field generated by the swimming motion, we quantitatively characterize the interface deformation and the manner in which the coupling modifies the microswimmer translation itself. We treat the role of the swimmer type (pushers and pullers),  size and model an interface that can deform due to both surface tension and bending elasticity. Our analysis reveals a strong dependence of the hydrodynamics on the swimmer orientation and position. Given the viscosities of the two fluid media, the interface properties and the swimmer type, a swimmer can either migrate towards or away from the interface depending on its configurations. When the swimmer is oriented parallel to the interface, a pusher-type swimmer is repelled from the interface at short times if it is swimming in the more viscous fluid. At long times however, pushers are always attracted to the interface, and pullers are always repelled from it. On the other hand, swimmers oriented orthogonal to the interface exhibit a migration pattern opposite to the parallel swimmers. In consequence, a host of complex migration trajectories emerge for swimmers arbitrarily oriented to the interface. We find that confining a swimmer between a rigid boundary and a deformable interface results in regimes of attraction towards both surfaces depending on the swimmer location in the channel, irrespective viscosity ratio. The differing migration patterns are most prominent in a region of order the swimmer size from the interface, where the slender swimmer model yields a better approximation to the coupled hydrodynamics.
\end{abstract}

\maketitle


\section{\label{sec_intro}Introduction}
Swimming microorganisms are often found in fluid environments near interfaces that can be either rigid \cite{rothschild1963non, Frymier1995, diluzio2005escherichia, berke_2008, diLeonardo_2011, Ishikawa_2013, Stocker_2014} or compliant \cite{montecucco2001living, plos_pathogen_2008, hosoi_lauga_2008, physio_rev_2010}. The hydrodynamics in such systems involves a coupling between the intrinsic swimming motion of the microswimmer and the boundary \cite{hosoi_lauga_2_2008, spagnolie2012hydrodynamics, dias2013, Yeomans_2013, lowen_volpe_rev_2016, shaik_ardekani_2017, Ishikawa_2019, abdallah2019, morozov_2020, Misbah_2020}, due to which a host of rich and complex dynamical responses emerge \cite{aranson_goldsetin_spatial_2009, gollub_2011, woodhouse_goldstein_2012, Lushi2014, costanzo2014motility, Wioland_2016, yeomans2016, lowen_volpe_rev_2016, saintillan_2018}. For instance, observations near rigid interfaces include aspects of confinement induced microswimmer migration such as upstream swimming and boundary accumulation \cite{berke_2008, Guanglai_2009, rusconi2014bacterial, ezhilan2015distribution, bearon_hazel_2015, stocker_2015_1, ezhilan_saintillan_2015, yeomans2016, Wioland_2016, manabe_omori_ishikawa_2020, vennamneni2020shear}, changes in the confinement pressure and fluctuation forces \cite{yan_brady_2015, ezhilan2015distribution, Lee_Wettlaufer_2017, morozov_2020}, as well as boundary-induced changes in the swimmer trajectory \cite{Stocker_2014, hu2015physical, Wioland_2016, manabe_omori_ishikawa_2020}. When swimming near complaint interfaces however, hydrodynamic effects drive interfacial deformation. The resulting coupled dynamics depends on surface tension and/or elasticity \cite{leal2007advanced, bickel2007, pozrikidis_2007, abdallah2016, rallabandi2018membrane}. Depending on the properties of the deformable interface, the fluid medium and the swimmer model adopted, simulations and continuum theories have shown enhanced pumping of the disturbance flow field \cite{dias2013}, as well as deformation-induced enhancements \cite{Yeomans_2013, abdallah2019} and retardation \cite{Facui_1995, Misbah_2020} in the swimmer translation.

Analytical studies that have investigated the hydrodynamics of microswimmers and passive particles near deformable interfaces, have treated the bodies as singularities (point forces, force-dipoles, quadropoles, etc.) \cite{bickel2007, hosoi_lauga_2_2008, abdallah2016, rallabandi2018membrane, abdallah2019}, two-dimensional swimming sheets \cite{dias2013}, or spherical particles/squirmers \cite{Berdani_leal_1982, shaik_ardekani_2017}. Each of these unique approaches can explore observations in the regime of their applicability. For example, \citet{dias2013} showed that a two-dimensional sheet swimming near a deformable interface can generate a pumping velocity that can be either attractive or repulsive to the interface depending on the viscosity ratio of the two fluid regions. Moreover, they found that an increase in the bending stiffness resulted in enhanced swimmer translation but reduced the fluid pumping. \citet{abdallah2019} considered higher order terms of a multipole analysis for determining the effect of the coupled hydrodynamics on the translational and rotational motion of a microswimmer near a deformable interface. By considering an interface which exhibits resistance to both bending and shear, they established that the swimmer translation was enhanced due to interface bending and suppressed due to shear resistance. On the other hand \citet{shaik_ardekani_2017} modeled the swimmer as a sphere with a prescribed surface slip velocity (the spherical squirmer approach \cite{blake_1971}) and quantitatively characterized both the interface deformation and the swimmer translation for different squirmer configurations. They found that, depending on the initial orientation of the squirmer relative to the interface, the squirmer either swam towards or away from the interface.

Analytical progress in such approaches has been contingent upon a small parameter, such a small capillary number $Ca$ (the ratio of viscous fluid stress to surface tension) \cite{hosoi_lauga_2_2008, shaik_ardekani_2017}, or small sheet deformations in a two-dimensional domain \cite{dias2013}. Alternatively, the small parameter emerges from the separation of scales of the swimmer size relative to its distance from the interface; when it is sufficiently far away from the interface to evoke a multipole analysis \cite{hosoi_lauga_2_2008, abdallah2019}. There are, however, systems where the microswimmers are at distances of order their size $L$ from the interface, that aren't necessarily constrained as in a quasi-two-dimensional setting \cite{montecucco2001living, plos_pathogen_2008}. Moreover, several bacteria, such as $H. pylori$, $B. subtilis$, have a strongly orientable geometry. The orientability can be characterized in terms of an aspect ratio $\kappa$, which is defined as the total swimmer length (head+flagellar bundle combined) to its lateral extent. Typical values of the aspect ratio are of the order $\kappa\sim 10$, implying that $\kappa\gg 1$ \cite{montecucco2001living, berg2008coli, Elgeti_2015, rusconi2014bacterial}. One would therefore like to relax some of the constraints and develop a more general framework to model microscopic swimmers near deformable interfaces. These could include aspects of the finite swimmer size and geometry, characterizing the hydrodynamics for swimmer distances of $O(L)$ from the interface, and going beyond the limit where the ratio of the viscous fluid stress to the deformation driving stress is small.

Here, we consider the motion of a slender microswimmer moving in the proximity to a deformable interface that separates two Newtonian fluid regions of different viscosities. By modulating the distribution of forces along the swimmer length, we are able to model both pusher- and puller-type swimmers \cite{kasyapkoch2014, nambiar2021enhanced}, hereafter referred to as pushers and pullers, respectively. Importantly, using slender-body theory enables us to resolve the hydrodynamics in regions of order the swimmer size $L$ owing to the weak inverse logarithmic scaling of the disturbance flow field in $\kappa$ \cite{batchelor1970slender}. In particular, this weak scaling also allows us to avoid the smallness of the stress ratios discussed above and thereby access a wider parameter regime of interfacial properties. We characterize the nature of interface deformation and swimmer translation due to the coupled hydrodynamics for different swimmer orientations. We treat an interface that can deform due to both surface tension and bending elasticity, and our study extends the framework developed to model such systems by \citet{bickel2007} and  \citet{abdallah2016}.

The manuscript is organized as follows. In Sec. \ref{sec_govrn}, we formulate the governing equations and the boundary conditions for a slender swimmer disturbing the fluid medium in the vicinity of a deformable interface. Specifically, in Sec. \ref{subsec_fluid_equations} we describe the Stokes equations of the two fluid region along with the relevant velocity and stress boundary conditions at the interface, and in Sec. \ref{subsec_nondim} we describe the non-dimensional system, approximations to the field variables and the boundary conditions that exploit slender swimmer model. These non-dimensional equations and boundary conditions are solved in Appendix \ref{sec_appendix_A}, for swimmers oriented parallel to the interface and Appendix \ref{sec_appendix_B} for swimmers oriented orthogonal to the interface. In Appendix \ref{sec_appendix_C}, we validate the approximations to the boundary condition for swimmers oriented orthogonal to the interface. We discuss the results for the swimmer translation and interface deformation, first for swimmers oriented parallel to the interface in Sec. \ref{subsec_swimmers_parallel_results}, and for swimmers oriented orthogonal to the interface in Sec. \ref{subsec_swimmers_orthogonal_results}. In each case, we describe the results pertaining to the interface deformation and of the effect on the swimming motion due to the hydrodynamic coupling. In Sec. \ref{subsec_swimmers_arbit_results} we generalize our analysis for arbitrary swimmer orientations, focusing only on the swimmer translation and rotation. In Sec. \ref{sec_swimmers_confined_results} we analyze the change in the swimmer translation when it is confined between a deformable interface and a rigid boundary, and oriented parallel to both. Lastly, in Sec. \ref{sec_conclusion} we summarize the results and present concluding remarks on our analysis.

\section{\label{sec_govrn}Slender swimmers near a deformable interface: a coupled hydrodynamics framework}
\begin{figure}
\includegraphics[width=0.85\columnwidth]{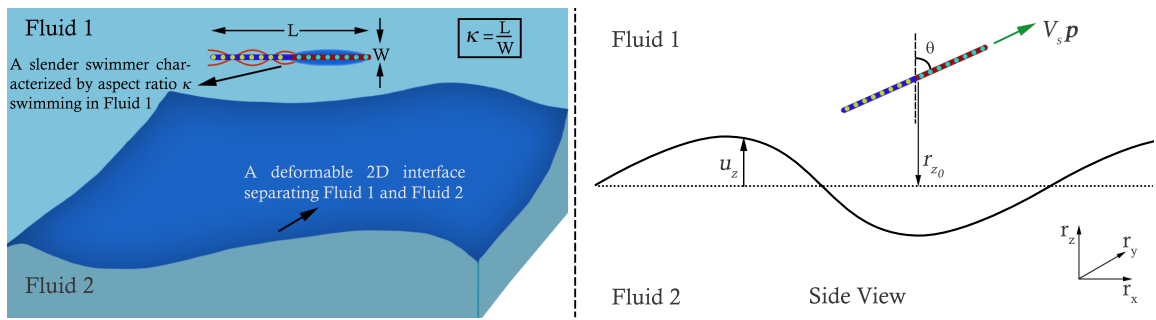}
\caption{A schematic representation of a slender swimmer translating near a deformable interface with speed $V_s$ along its director vector $\bm{p}$. The swimmer is characterized by an aspect ratio $\kappa$ defined as the ratio of its total length $L$ to its lateral extent $W$. The circles along the axial length of the fore-aft symmetric swimmer represent a line distribution of stokeslets characterizing the head and tail for pushers. The disturbance flow field generated by the swimming motion deforms the interface, and $u_z$ is the interface deformation (solid gray line) relative to its initially flat undeformed (dotted line) state $z_0$.}
\label{fig1}
\end{figure}

In this section, we derive the coupled set of differential equations that characterize the disturbance flow field and the interface deformation due to a force- and torque-free microswimmer translating in the vicinity of a deformable interface. As described in Sec. \ref{sec_intro} we model a slender microswimmer moving near a deformable interface at distances of order its own body length away from it. The interface separates two density-matched Newtonian fluid regions that have different viscosities, and is modeled such that it can deform due to surface tension and bending elasticity as is the case for typical lipid bilayer membranes or vesicles and cells \cite{faucon1989bending, Libchaber_1997, rosen2005ultralow, freund2014numerical}. A schematic is given in figure \ref{fig1}.

\subsection{\label{subsec_fluid_equations}Equations for the fluid velocity and the boundary condition}
The equations governing the two fluid regions are the Stokes equations and the continuity equations which characterize the disturbance flow field $\bm{v}$ and the pressure field $P$, and are:
\begin{subequations}
\label{eq:stokes_continuity}
	\begin{eqnarray}
		-\nabla P_\alpha + \eta_\alpha \nabla^2 \bm{v}_\alpha &=& \int_{-\frac{L}{2}}^{\frac{L}{2}}\bm{f}_\alpha \bm{\delta}(\bm{x}-\bm{x}_s - V_s \bm{p} t-s\bm{p})\mathrm{d}s,\;\; \text{and} \label{eq:stokes_continuity_a}\\
		\bm{\nabla}\cdot \bm{v}_\alpha &=& 0. \label{eq:stokes_continuity_b}
	\end{eqnarray}
\end{subequations}
Here, the subscript $\alpha\in [1, 2]$ represents the two fluid regions, $\eta$ is the fluid viscosity, $\bm{\delta}(\bm{x})$ is the Dirac-delta function \cite{lighthill1958}, $L$ is the total swimmer length, $V_s$ is the swimming speed, and $s$ refers to the distance along the slender swimmer axial coordinate, whose center-of-mass is at position $\bm{x}_s$ and is oriented along the direction $\bm{p}$. The orientation vector $\bm{p}$ is characterized by a polar angle $\theta$ relative to the vertical and azimuthal angle $\phi$ in the plane of the undeformed interface. The kernel in the integral on the right-hand side of Eq. \eqref{eq:stokes_continuity_a}, $\bm{f}_\alpha$, represents the forcing due to the slender swimmer and characterizes a line distribution of Stokeslets along the axial coordinate of the swimmer. For a fore- and aft-symmetric swimmer, which disturbs in the surrounding fluid medium as it moves, one may write $\bm{f}_\alpha$ as \cite{kasyapkoch2014, nambiar2021enhanced}:
\begin{equation}
\bm{f}_\alpha = 
    \begin{cases}
      D\, \eta_\alpha V_s \bm{p}\, \text{sgn}(s)/(\ln\kappa) \!\!& ; \alpha = 1 \\
      0 \!\! & ; \alpha = 2,\\
    \end{cases} \\ \\   
\label{eq:fAlpha}
\end{equation}
where $\kappa$ refers to the slender swimmer aspect ratio, and is defined as the ratio of the total swimmer length to its lateral extent, as shown in Fig. \ref{fig1}. The above form of $\bm{f}_\alpha$ describes two swimming mechanisms, namely, pushers (rear propelled) and pullers (fore propelled). In the far-field both swimming mechanisms exhibit a disturbance flow field akin to a point-force dipole with opposing dipole strength \cite{Lauga_2009, subramanian2011fluid}. This opposing character is encoded in the parameter $D = -1\,(+1)$ in Eq. \eqref{eq:fAlpha} which describes the specific nature of the force-dipole swimming mechanism, that is extensile for pushers (contractile for pullers). Typically, pusher-type bacteria such as $E. coli$, $B. subtilis$ are fairly slender with $\kappa \sim 10$ when the swimmer length is measured based on the cell body and the flagellar bundle length combined \cite{brennen_annrev77, patteson2016particle, kasyapkoch2014, rusconi2014bacterial}.

Equation \eqref{eq:stokes_continuity} must be solved subject to appropriate boundary conditions at the interface. We consider an impenetrable no-slip interface of infinitesimal thickness within the continuum framework \cite{leal2007advanced}. This translates to the following velocity boundary conditions: $\bm{v}_1\cdot \bm{n} = \bm{v}_2\cdot \bm{n}$ and $\bm{v}_1\cdot (\bm{I} - \bm{n n}) = \bm{v}_2\cdot (\bm{I} - \bm{n n})$, where $\bm{n} = \bm{\nabla} F/\vert\bm{\nabla} F\vert$ is the unit normal to the interface $F = z-z_0-u_z$ \cite{leal2007advanced}, and $\bm{I}$ is the identity tensor. The $z$-component of the interface deformation relative to its undeformed planar state is $u_z$ (see Fig. \ref{fig1}). As described in Sec. \ref{sec_intro}, the interface can deform when there is a flow in the fluid medium surrounding it, since it has a finite surface tension and can to bend elastically. Therefore, the normal component of the stress undergoes a jump across the interface: $\sigma_1^{zz}\vert_{u_z^+} - \sigma_2^{zz}\vert_{u_z^-} = \gamma \bm{\nabla}\cdot\bm{n} + \delta F_{\text{bend}}$, where $\bm{\sigma}_\alpha = -P \bm{I} + \eta_\alpha (\bm{\nabla}\cdot\bm{v} + \bm{\nabla}\cdot\bm{v}^t)$ is the stress tensor, $\gamma$ the isotropic surface tension and $\delta F_{\text{bend}}$ represents the stress jump due to bending. Here, we consider linear elastic bending, and use the Helfrich model \cite{Helfrich1973} to relate the interface deformation to the bending stress. This model has been used extensively in the literature for analyzing swimming sheets, Stokeslets and other higher-order singularities  near deformable interfaces  \cite[see][and also references therein]{dias2013, abdallah2016, rallabandi2018membrane, abdallah2018, abdallah2019}. For simplicity we consider the tangential component of the stress jump across the interface to be continuous, implying that $\sigma_1^{xz}\vert_{u_z^+} = \sigma_2^{xz}\vert_{u_z^-}$ and $\sigma_1^{yz}\vert_{u_z^+} = \sigma_2^{yz}\vert_{u_z^-}$. However, we note that the present analytical framework allows incorporation of a shear resistance on the interface \cite{abdallah2019}. Now, in addition to these fluid equations and boundary conditions, we also have the kinematic condition at the interface, which provides a direct relation between the interface deformation and the disturbance flow field measured at the interface \cite{Berdani_leal_1982, leal2007advanced, shaik_ardekani_2017}, and is:
\begin{equation}
\frac{\partial F}{\partial t} + \bm{v}\cdot\bm{\nabla} F = 0.
\label{eq:Interface}
\end{equation}

With the interface deformation and the disturbance field characterized, we express the swimmer translation velocity and rate of rotation due to the hydrodynamic interaction with the interface as follows. From viscous slender-body theory \cite{batchelor1970slender, kim2005microhydrodynamics} the velocity $\bm{V}$ for a slender swimmer is:
\begin{equation}
\label{eq:swim_vel_1}
\bm{V}(s\bm{p}) - \bm{v}(s\bm{p})\vert_{\text{swim location}} = \frac{\bm{f}_1}{4\pi\eta_1}\cdot (\bm{I} + \bm{pp}) \ln\kappa,
\end{equation}
where the second term on the left-hand side of Eq. \eqref{eq:swim_vel_1} is the disturbance flow field at the location of the swimmer. For a rigid slender swimmer, $\bm{V} = \bm{V}^T + \bm{\omega}\wedge\bm{p}s$, with $\bm{V}^T$ being the translational velocity and $\bm{\omega}$ the angular velocity of the swimmer emanating from the coupled hydrodynamics. The force-free ($\int_{-\frac{1}{2}}^{\frac{1}{2}}\bm{f}_1 \mathrm{d}s = 0$) and torque-free ($\int_{-\frac{1}{2}}^{\frac{1}{2}}s\bm{p}\wedge\bm{f}_1 \mathrm{d}s = 0$) swimming constraints translate to the following expressions for the swimmer translational velocity $\bm{V}^T$ and rotation rate $\dot{\bm{p}} = \bm{\omega}\wedge\bm{p}$ \cite{nambiar2021enhanced},
\begin{subequations}
\label{eq:trans_rot}
\begin{eqnarray}
\bm{V}^T &=& \int_{-\frac{1}{2}}^{\frac{1}{2}} \bm{v}(s\bm{p}) \mathrm{d}s, \;\; \text{and}\\
\dot{\bm{p}} &=& 12  \int_{-\frac{1}{2}}^{\frac{1}{2}}(\bm{I} - \bm{pp})\cdot \bm{v}(s\bm{p}) s \mathrm{d}s. \label{eq:trans_rot_pdot}
\end{eqnarray}
\end{subequations}
Therefore, in addition to the swimmers intrinsic motion, a slender swimmer translates along the length averaged disturbance field at its center and rotates due to the first moment of the disturbance field along its axial coordinate.

\subsection{\label{subsec_nondim}Non-dimensionalization}

We non-dimensionalize the equations and the boundary conditions using $L$ as a length scale, $V_s$ as a velocity scale and $\eta_1 V_s/L$ as a scale for the fluid stress. It is convenient to solve the problem in a reference frame moving with the swimmer centre: $\bm{r} = \bm{x}-\bm{x}_s - (V_s t_c/L) \bm{p}\, t$; $\bm{\nabla} \equiv \bm{\nabla}_{\bm{r}}$ and $\partial(\cdot)/\partial t = \partial (\cdot)/ \partial t - (V_s t_c/L) \bm{p}\cdot\nabla_{\bm{r}}(\cdot)$, where we have allowed the time scale $t_c$ to remain arbitrary for the moment. This yields the following non-dimensional form of the Stokes equations and the continuity equations, Eq. \eqref{eq:stokes_continuity}, in the two fluid regions:
\begin{subequations}
\label{eq:stokes_continuity_nd}
	\begin{eqnarray}
	-\nabla P_1 + \nabla^2_r \bm{v}_1 &=& \frac{D\, \bm{p}}{\ln\kappa}\int_{-\frac{1}{2}}^{\frac{1}{2}} \text{sgn}(s) \bm{\delta}\left(\bm{r}-s\bm{p}\right)\mathrm{d}s, \label{eq:stokes_continuity_nd_a} \\
	\bm{\nabla}_{\bm{r}}\cdot \bm{v}_1 &=& 0, \label{eq:stokes_continuity_nd_b}\\
	-\nabla_{\bm{r}} P_2 + \lambda\nabla^2_r \bm{v}_2 &=& 0, \; \text{and} \label{eq:stokes_continuity_nd_c}\\		
	\bm{\nabla}_{\bm{r}}\cdot \bm{v}_2 &=& 0, \label{eq:stokes_continuity_nd_d}				
	\end{eqnarray}
\end{subequations}
where Eqs. \eqref{eq:stokes_continuity_nd_a} and \eqref{eq:stokes_continuity_nd_c} correspond to Eq. \eqref{eq:stokes_continuity_a}, and Eqs. \eqref{eq:stokes_continuity_nd_b} and \eqref{eq:stokes_continuity_nd_d} to Eq. \eqref{eq:stokes_continuity_b} for $\alpha \in [1, 2]$. Note that, for brevity, we retain the original notation of the field variables in the non-dimensional Eqs. \eqref{eq:stokes_continuity_nd_a}-\eqref{eq:stokes_continuity_nd_d}. Now, the leading order flow field generated by a slender-body is $O(\ln \kappa)^{-1}$  \cite{batchelor1970slender, kim2005microhydrodynamics, Lauga_2009, kasyapkoch2014}, where we emphasize that the aspect ratio $\kappa \gg 1$. As the disturbance field $\bm{v}$ in the Stokes equations \eqref{eq:stokes_continuity_nd_a} is linear in the forcing, it will also scale as $O(\ln \kappa)^{-1}$, and is therefore asymptotically small. Hence, we can consider small interface deformations without having to invoke the smallness of the ratio of viscous fluid stresses to the deformation driving stresses on the interface (surface tension and bending). In other words, owing to the weak flow field of a slender-body, the slender swimmer model enables simplifications of the boundary conditions.

For small deformations, we can linearize the expressions for the unit normal and the curvature at the interface as: $\boldsymbol{n} \approx \mathbf{1}_{z}$ and $\bm{\nabla}_{\bm{r}} \cdot \bm{n} \approx-\Delta_{r_{\|}} u_{z}$, where $\mathbf{1}_{z}$ is the $r_{z}$-component of the unit normal \cite{bickel2007, abdallah2016}.  Within the linear response framework the height function, $u_{z}$, depends only upon the $r_{x}$ and $r_{y}$ coordinates, and an analytical expression for the stress jump due to bending using the Helfrich elastic energy function dimensionally is: $\delta F_{\text {bend }}=\kappa_{\beta} \Delta_{\|}^{2} u_{z}$, where $\boldsymbol{\Delta}_{\|}$ represents the Laplacian in the $r_{x}-r_{y}$ plane and $\kappa_{\beta}$ the bending modulus of the interface \cite{abdallah2016}. Thus, the non-dimensional form of the boundary conditions specified above Eq. \eqref{eq:Interface} are:
\begin{subequations}
\label{eq:bc_nd}
	\begin{eqnarray}
		v_{1z}\vert_{r_{z_0}^+} &=& v_{2z}\vert_{r_{z_0}^-}, \label{eq:bc_nd_a} \\
		\bm{v}_1\cdot\left.(\bm{I}-\bm{1}_z\bm{1}_z)\right\vert_{r_{z_0}^+} &=& \bm{v}_2\cdot\left.(\bm{I}-\bm{1}_z\bm{1}_z)\right\vert_{r_{z_0}^-}, \label{eq:bc_nd_b} \\
		\left.\left(\frac{\partial v_{1z}}{\partial r_x} + \frac{\partial v_{1x}}{\partial r_z}\right)\right\vert_{r_{z_0}^+} &=& \lambda\left.\left(\frac{\partial v_{2z}}{\partial r_x} + \frac{\partial v_{2x}}{\partial r_z}\right)\right\vert_{r_{z_0}^-}, \label{eq:bc_nd_c} \\	
		\left.\left(\frac{\partial v_{1z}}{\partial r_y} + \frac{\partial v_{1y}}{\partial r_z}\right)\right\vert_{r_{z_0}^+} &=& \lambda\left.\left(\frac{\partial v_{2z}}{\partial r_y} + \frac{\partial v_{2y}}{\partial r_z}\right)\right\vert_{r_{z_0}^-}, \; \text{and} \label{eq:bc_nd_d}\\	
		-\left(P_1\vert_{r_{z_0}^+} - P_2\vert_{r_{z_0}^-}\right) + 2\left(\left.\frac{\partial v_{1z}}{\partial r_z}\right\vert_{r_{z_0}^+}-\lambda \left.\frac{\partial v_{2z}}{\partial r_z}\right\vert_{r_{z_0}^-}\right) &=& -\frac{\gamma}{\eta_1 V_s}\Delta_{r_{\Vert}}u_z + \frac{\kappa_\beta L^2}{\eta_1 V_s}\Delta^2_{r_{\Vert}}u_z,	\label{eq:bc_nd_e}	
	\end{eqnarray}
\end{subequations}
where $\lambda = \eta_2/\eta_1$ is the viscosity ratio. We are studying an interfacial deformation process and to non-dimensionalize the kinematic condition Eq. \eqref{eq:Interface}, we can choose between the scale for the elastic bending and surface tension, corresponding to $t_{c1} = \eta_1 L^3/\kappa_\beta$ and $t_{c2} = \eta_1 L/\gamma$, respectively. For ultra-soft interfaces, the surface tension can be $\gamma \sim O(10)^{-5} N / m$ or smaller \cite{faucon1989bending, Henk_2001, rosen2005ultralow}, and for example the bending rigidity of vesicles is $\kappa_{\beta} \sim O(10)^{-19} J$ \cite{faucon1989bending, bending_1990, bending_1993, Libchaber_1997}. In a fluid medium with a viscosity approximately that of water, $\eta_{1} \sim O(10)^{-3} \mathrm{Ns} / \mathrm{m}^{2}$, and a swimmer of size $L \sim O(10)^{-5} m$, this implies that $t_{c 1} \sim O(1)-O(10)^{2}$ and $t_{c 2} \sim O(10)^{-3}-O(1)$. Independent of the time scale chosen, the relative importance of the surface tension and bending stress on the right-hand side of Eq. \eqref{eq:bc_nd_e} can then be characterized by the ratio of the two scales, namely $\gamma L^{2} / \kappa_{\beta}$. Depending on the stiffness of the interface, for typical values of the time scales, $\gamma L^{2} / \kappa_{\beta}$ can vary from $O(1)$ to $\gamma L^{2} / \kappa_{\beta} \gg 1$. Therefore, we choose $t_{c 1}$ as a time scale noting that choosing $t_{c 2}$  is equally plausible. In the literature, $\gamma L^{2} / \kappa_{\beta}$ is also treated as an elasto-capillary length \cite{Style_2017, Snoijer_review_2020}, to characterize the length scales over which surface tension or elasticity dominate. The dimensionless kinematic boundary condition Eq. \eqref{eq:Interface} becomes
\begin{equation}
\frac{\partial u_z}{\partial t} + \left(\frac{\eta_1 L^2 V_s}{\kappa_\beta}\right)\bm{v}\cdot\bm{\nabla}_{\bm{r}_{\Vert}} u_z = \left(\frac{\eta_1 L^2 V_s}{\kappa_\beta}\right)\Big[\left. v_z\right\vert_{u_z} + p_z \Big],
\label{eq:Interface_2}
\end{equation}
where $p_z = \cos\theta$, and the right-hand side of Eq. \eqref{eq:Interface_2} applies at the location of the deformed interface $u_z$.

In principle, there is a third time scale in the problem if we account for the intrinsic reorientation time scale of a microswimmer. For instance, $t_{c3} = \tau$ or $t_{c3} = D_r^{-1}$, where $\tau$ is the mean run duration of a run-and-tumble particle (RTP) \cite{berg2008coli, koch2011collective, Elgeti_2015, ezhilan2015distribution, Lauga_annRev_2016, berg2018random} and $D_r$ the rotary diffusivity coefficient of an active Brownian particle (ABP) \cite{koch2011collective, rusconi2014bacterial, takatori_2014_PRL, berg2018random, vennamneni2020shear}. For simplicity, we neglect this third scale and treat only straight-swimmers, that is, the class of swimmers that do not tend to intrinsically reorient. In Sec. \ref{sec_conclusion}, we briefly discuss the relevance of the swimmer reorientation time and how it compares to the response scale of the interface deformation.

For small deformations, one can use the method of domain perturbation \cite{leal2007advanced, abdallah2018} to represent this term at the location of the planar undeformed interface $z_0$ instead. Namely, $v_z\vert_{u_z}$, can be perturbed about $r_{z_0}$ as: $v_z\vert_{u_z} \approx v_z\vert_{r_{z_0}} + u_z (\partial v_z/\partial r_z) \vert_{r_{z_0}} + O(u_z)^2+\ldots$ , and therefore considering terms up to $O(u_z)$ in the velocity, one may rewrite Eq. \eqref{eq:Interface_2} as:
\begin{equation}
\frac{\partial u_z}{\partial t} + \left(\frac{\eta_1 L^2 V_s}{\kappa_\beta}\right)\bigg[\bm{v}\cdot\bm{\nabla}_{\bm{r}_{\Vert}} u_z - u_z\frac{\partial v_z}{\partial r_z}\bigg] =\left(\frac{\eta_1 L^2 V_s}{\kappa_\beta}\right) \Big[\left. v_z\right\vert_{r_{z_0}} + p_z \Big].
\label{eq:Interface_3}
\end{equation}
Note that the second and third nonlinear terms on the left-hand side of Eq. \eqref{eq:Interface_3} involve products of the field variable ($u_z$, $\bm{v}$) and their associated gradients. These terms are usually neglected if the forcing is sufficiently far away from the interface so that terms involving their products are asymptotically small in the far-field \cite{bickel2007, abdallah2016}. Given that we consider swimmers at $O(1)$ distances from the interface, these terms can not be neglected a-priori.
However, as noted below Eq. \ref{eq:stokes_continuity_nd}, the disturbance field $\bm{v}$ is $O(\ln \kappa)^{-1}$. Therefore, if we express the interface deformation as a perturbation series in $(\ln \kappa)^{-1}$, $u_{z} = u_{z 1} \times(\ln \kappa)^{-1}+u_{z 2} \times(\ln \kappa)^{-2}+\ldots$, the second and third terms on the left-hand side of Eq. \eqref{eq:Interface_3} will be $O(\ln \kappa)^{-2}$, and hence asymptotically smaller than the disturbance field $\boldsymbol{v}$. A similar argument, owing to the asymptotically weak flow field of a slender swimmer, has recently been used to treat the tracer diffusivity and the velocity variance in a suspension of interacting slender swimmers in bulk \cite{nambiar2021enhanced}. Thus, to a leading order in $(\ln\kappa)^{-1}$, one can simplify Eq. \eqref{eq:Interface_3} as:
\begin{equation}
\frac{\partial u_z}{\partial t} \approx \left(\frac{\eta_1 L^2 V_s}{\kappa_\beta}\right)  \Big[\left. v_z\right\vert_{r_{z_0}} + p_z \Big],
\label{eq:Interface_4}
\end{equation}
for a slender-body near a deformable interface. We can solve the above system of linear equations \eqref{eq:stokes_continuity_nd_a}-\eqref{eq:stokes_continuity_nd_d} and the boundary conditions \eqref{eq:bc_nd_a}-\eqref{eq:bc_nd_e} in Fourier space \cite{lighthill1958, bickel2007, abdallah2016}, and we define the two-dimensional Fourier transform of any variable $A(r_x, r_y)$ as: $\hat{A}(\bm{k}) = \int_{-\infty}^{\infty}\int_{-\infty}^{\infty}\mathrm{d}x\mathrm{d}y A(r_x, r_y) \exp(-2\pi \iu \bm{k}\cdot\bm{r}_{\Vert})$. Determining the Fourier transformed disturbance field $\hat{v}_{z}$ is straightforward, as shown in Appendix \ref{sec_appendix_A}, \ref{sec_appendix_B} for swimmers oriented parallel and perpendicular to the interface, respectively. However, here we show the Fourier transformed kinematic boundary condition Eq. \eqref{eq:Interface_4}:
\begin{equation}
\frac{\text{d} \hat{u}_z}{\text{d} t} = \left(\frac{\eta_1 L^2 V_s}{\kappa_\beta}\right)\left[  \hat{v}_z\vert_{r_{z_0}} + p_z \frac{\delta(k)}{\pi k} \right],
\label{eq:Interface_6}
\end{equation}
which is solved in the next section for different swimmer configurations relative to the interface. For swimmers translating parallel to the interface $p_z = 0$, in which case the right-hand side of Eq. \eqref{eq:Interface_6} is only the Fourier transformed  disturbance velocity at the undeformed interface. This approximation holds as long as the ratio of the viscous to bending stress: $\eta_1 V_s L^2/\kappa_\beta$ remains $O(1)$. However, if $\eta_1 V_s L^2/\kappa_\beta\gg 1$, the nonlinear boundary condition given by Eq. \eqref{eq:Interface_3} would have to be used. In Appendix \ref{sec_appendix_C}, we validate the above approximation for the interface deformation by comparing $u_z$ derived from solving Eq. \eqref{eq:Interface_3} with that from Eq. \eqref{eq:Interface_4}, for the specific case of swimmers oriented orthogonal to the interface over a range of $\eta_1 V_s L^2/\kappa_\beta$. In the following sections, we fix the aspect ratio $\kappa = 10$ and the ratio of the viscous to bending stresses $\eta_1 V_s L^2/\kappa_\beta =1$.

\section{\label{sec_deformation} Swimmers near a single interface}
In general, simultaneously solving Eq. \eqref{eq:trans_rot} and Eq. \eqref{eq:Interface_6} results in an inherently unsteady problem, with the coupling involving a dependence on both the distance of the swimmer to the interface as well as its orientation. Therefore, in what follows, we first discuss the results for a swimmer oriented parallel to the interface, then for a swimmer orthogonal to the interface, and finally for a swimmer with an arbitrary orientation. In Sec. \ref{sec_swimmers_confined_results} we consider a swimmer confined between a rigid boundary and a deformable interface.

\subsection{\label{subsec_swimmers_parallel_results} Microswimmers swimming parallel to the interface}

To understand the coupled hydrodynamics, it is useful to refer to the equation for the interface deformation and for the swimmer motion for swimmers oriented parallel to the interface. For the former, we use the expression for $\hat{v}_{1z}\vert_{r_{z_0}}$ from Eq. \eqref{eq:vhat_z} in Appendix \ref{subsec_appendix_A_solution} and obtain the following equation for the time evolution of $\hat{u}_z$:
\begin{eqnarray}
\frac{\text{d} \hat{u}_z}{\text{d} t} &+& \frac{\pi k}{(1+\lambda)} \left(4\pi^2 k ^2 + \Gamma\right) \hat{u}_z  =\left(\frac{\eta_1 L^2 V_s}{\kappa_\beta}\right)\left[  \frac{D}{\pi k (1+\lambda) \ln\kappa}\sin^2\left(\frac{\pi}{2} k p_l\right) r_{z_0} \exp(2\pi k r_{z_0}) \right],
\label{eq:Interface_7}
\end{eqnarray}
where $\Gamma \equiv \gamma L^2/\kappa_\beta$ characterizes the relative importance of surface tension to bending stress. Note that the interface deformation is linked to the swimmers instantaneous configuration via $r_{z_0}$ and $\bm{p}$, both of which can evolve in time. Here, $\dot{\bm{p}} = 0$, as is also true for a generic force-dipole swimmer near a planar or weakly deforming boundary \cite{two_sphere_2011, spagnolie2012hydrodynamics, abdallah2019}. However, the same is not true for $\bm{V}^T$, since the $z$-component of the translation is responsible for the coupling with the interface deformation in Eq. \eqref{eq:Interface_7} via $r_{z_0}$. We only focus on $V_z^T$, which is,
\begin{eqnarray}
V_z^T \equiv \frac{\mathrm{d} r_{z_0}}{\mathrm{d} t} &=& -\int \mathrm{d}\bm{k}\frac{\sin(\pi k p_l)}{\pi k p_l}\left[\frac{2 D r_{z_0}^2}{\ln\kappa}\left(\frac{1-\lambda}{1+\lambda}\right)\sin^2\left(\frac{\pi}{2}k p_l\right)\exp(4\pi k r_{z_0}) \right. \nonumber\\ && \left. \qquad + \left(\frac{\kappa_\beta}{\eta_1 V_s L^2}\right)\frac{\pi k}{(1+\lambda)}\left(4\pi^2 k^2 + \Gamma\right)(1-2\pi k r_{z_0})\hat{u}_z \exp(2\pi k r_{z_0})\right],
\label{eq:VzT_parallel}
\end{eqnarray}
where we have used the definition of the inverse Fourier transform: $A(r_x, r_y) = \int \mathrm{d} \bm{k} \hat{A}$ $\exp(2\pi \iu \bm{k}\cdot\bm{r}_{\Vert})$; $\bm{k}= (k_x, k_y)$ and $\bm{r}_{\Vert} = (r_x, r_y)$. Thus, characterizing the coupled hydrodynamics due to a swimmer translating parallel to the interface reduces to simultaneously solving Eqs. \eqref{eq:Interface_7} and \eqref{eq:VzT_parallel} to obtain $\hat{u}_z$ and $r_{z_0}$, and then the disturbance flow field given by Eq. \eqref{eq:vhat_t} and Eq. \eqref{eq:vhat_z} in Appendix \ref{subsec_appendix_A_solution}. 

\begin{figure}
    \centering
    \begin{subfigure}[t]{0.495\textwidth}
        \centering
        \includegraphics[width=\textwidth]{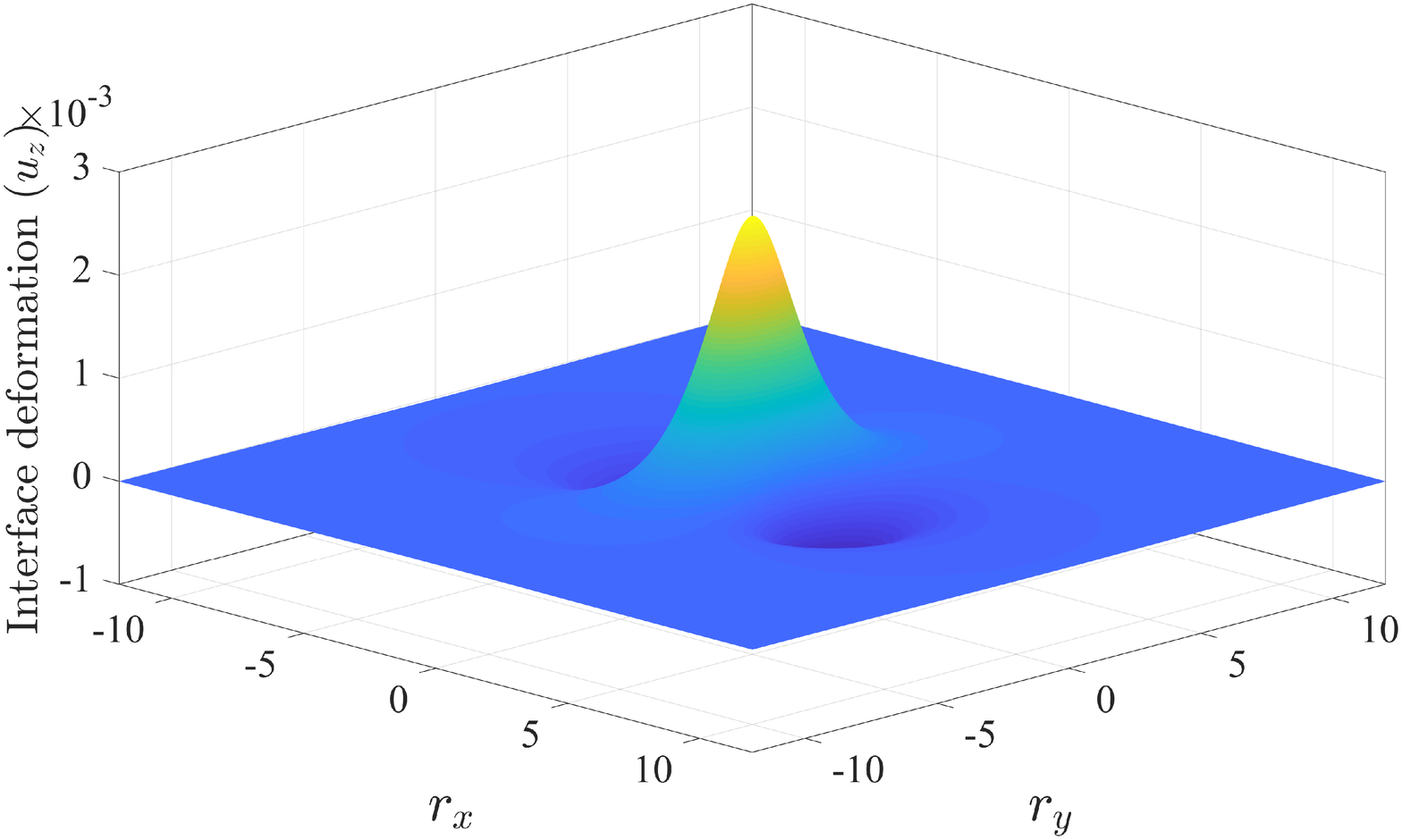}
        \caption{Three-dimensional contour of the interface deformation}
    \end{subfigure}%
    ~
    \begin{subfigure}[t]{0.495\textwidth}
        \centering
        \includegraphics[width=\textwidth]{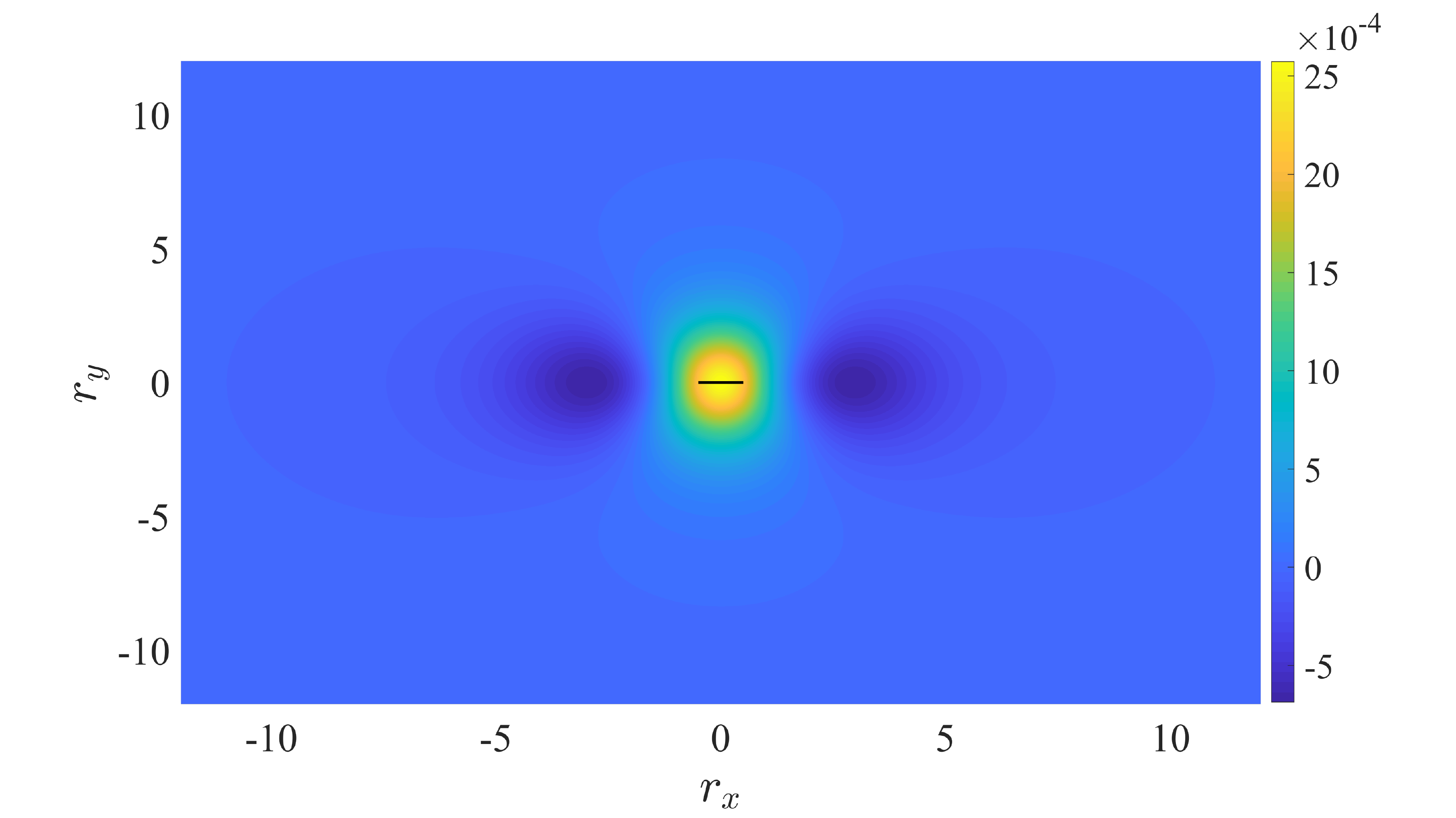}
        \caption{Two-dimensional contour of the interface deformation}
    \end{subfigure}   
    \caption{The non-dimensional interface deformation plotted for a swimmer of unit length oriented along the $r_x-$direction (a) viewed at an angle of 30$^\circ$ from the undeformed interface and (b) top view. The black line in (b) spanning from $r_x\in[-1/2, 1/2]$ signifies the slender swimmer. The colorbar given on the right-hand side of (b) applies to both figures. The plot is for $\Gamma = 0.1$, $\lambda = 0.5$ at $t\approx 2$, $\kappa = 10$ and $\eta_1 V_s L^2/\kappa_\beta =1$.}
\label{fig:int_def_swim_para}
\end{figure}

We now interpret the nature of the deformation influenced by the disturbance field generated from the swimmer. In Fig. \ref{fig:int_def_swim_para}a we plot the three-dimensional contour of the interface deformation due to a pusher in fluid region 1 for the parameters $\lambda = 0.5$ and $\Gamma = 0.1$. As expected, for pushers (or pullers) oriented parallel to the interface, the deformation is not radially symmetric \cite{shaik_ardekani_2017}. The qualitative nature of the deformation remains the same for all values of $\lambda$ (not shown), with the magnitude of the deformation decreasing as $\lambda$ increases, as expected from Eq. \eqref{eq:Interface_7}, noting that for $\lambda\gg 1$, $u_z\propto 1/\lambda$.

\begin{figure}
    \centering
    \begin{subfigure}[t]{0.495\textwidth}
        \centering
        \includegraphics[width=\textwidth]{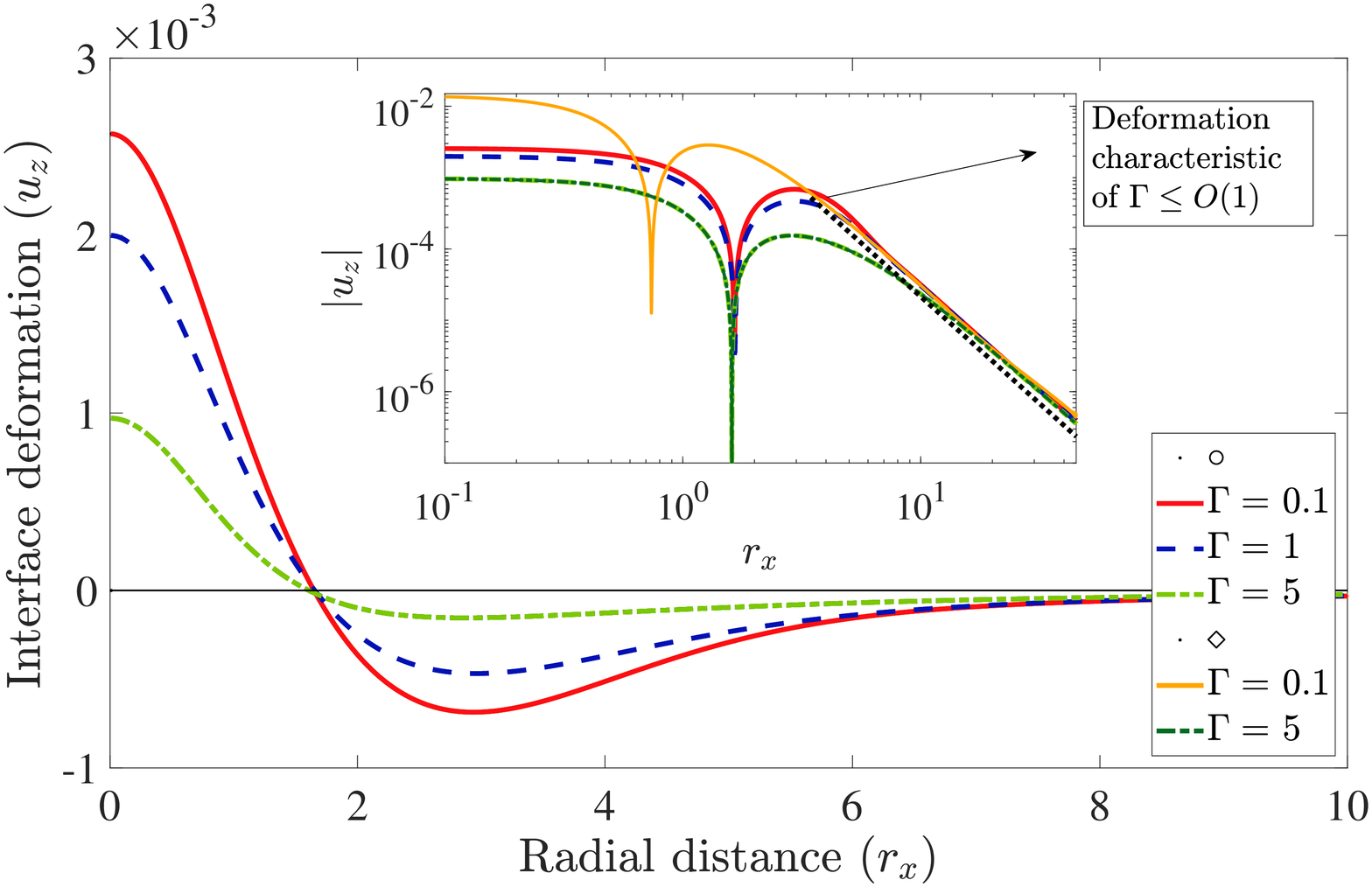}
        \caption{Along swimmer orientation}
    \end{subfigure}%
    ~
    \begin{subfigure}[t]{0.495\textwidth}
        \centering
        \includegraphics[width=\textwidth]{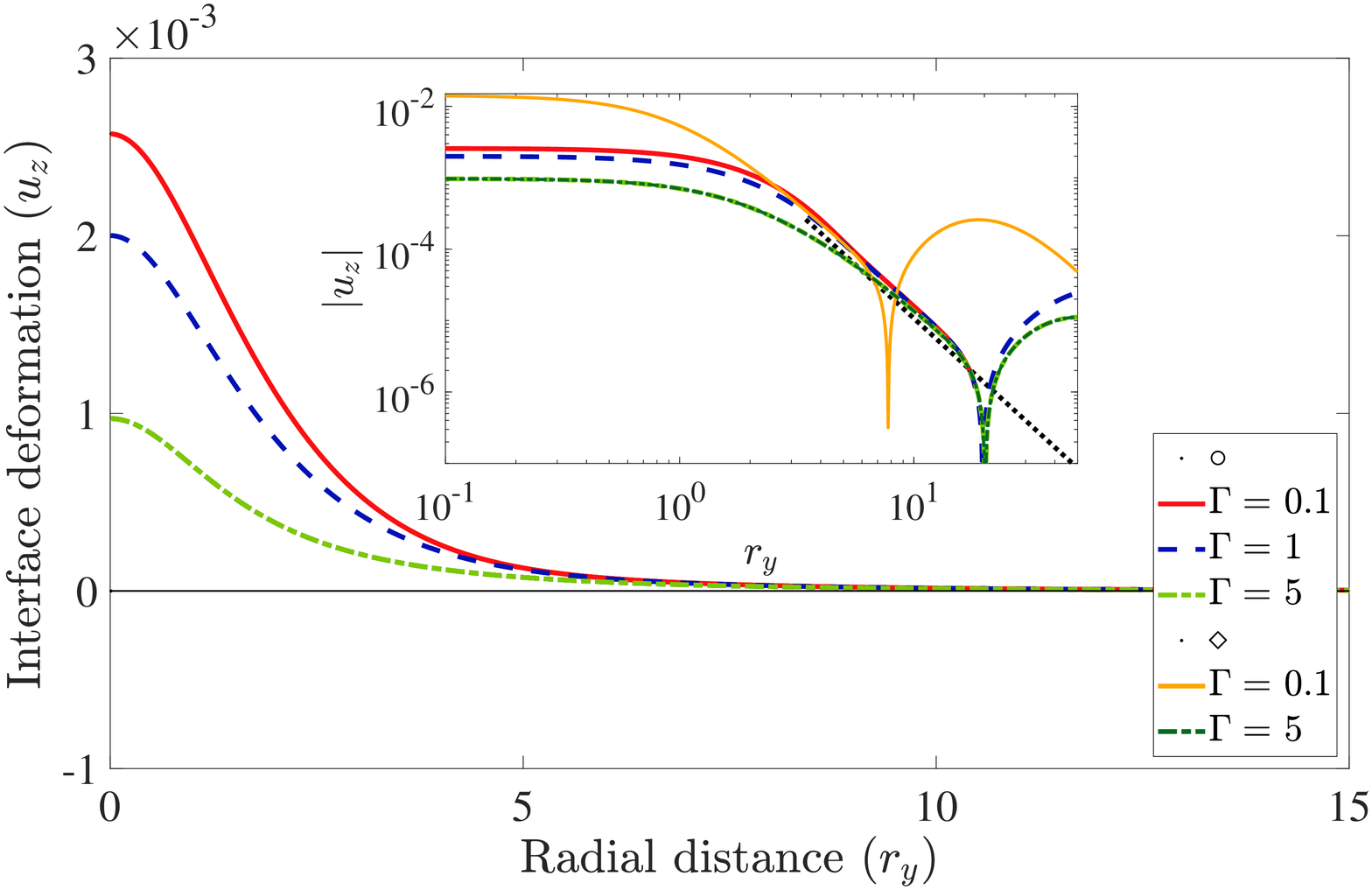}
        \caption{Orthogonal to swimmer orientation}
    \end{subfigure}   
    \caption{The interface deformation, $u_z$, plotted for a swimmer oriented along the $r_x-$coordinate for three values of $\Gamma =$ 0.1, 1 and 5 (a) along $r_x$ (b) along $r_y$. In both the figures $\lambda = 0.5$, $r_{z_0}(0) = 1$, $t\approx 2$, $\kappa = 10$ and $\eta_1 V_s L^2/\kappa_\beta =1$. Legends under $\circ$ refer to $u_z$ obtained from solving the pair of equations \eqref{eq:Interface_7}, \eqref{eq:VzT_parallel}, and those under $\diamond$ from solving the approximate equations \eqref{eq:Interface_8}, \eqref{eq:VzT_parallel2}. The (black) dotted line in the insets is the far-field approximation from Eq. \eqref{eq:u_z_approx_1}.}
\label{fig:int_def_swim_para_rx_y}
\end{figure}

To characterize the interface response with $\Gamma$, the deformation is plotted as a function of the radial distance parallel and orthogonal to the swimmer axis in Fig. \ref{fig:int_def_swim_para_rx_y} for $\lambda = 0.5$ and $\Gamma =$ $0.1, 1, 5$. It is clear that increasing the importance of surface tension, decreases the magnitude of the deformation. In the far-field,  however, at any finite time, the deformation is independent of $\Gamma$ and scales as $O(r_\Vert)^{-3}$, as shown in the insets of Fig. \ref{fig:int_def_swim_para_rx_y}. In fact, for any small but finite time, one can obtain an analytical expression for the interface deformation assuming $r_{z_0}$ to be nearly constant; this constraint of a constant $r_{z_0}$ is reasonable up to an $O(1)$ change in time, as will be seen below while interpreting the swimmer translation. The resulting expression for $u_z$ is:
\begin{equation}
\left. u_z\right\vert_{r_\Vert\gg 1} \approx \left(\frac{\eta_1 L^2 V_s}{\kappa_\beta}\right) \frac{D}{(1+\lambda)\ln\kappa}\frac{r_{z_0}t}{16 \pi}\;\frac{r_\Vert^2-3r_x^2}{r_\Vert^5},
\label{eq:u_z_approx_1}
\end{equation}
highlighting independence from $\Gamma$, and the $O(r_\Vert)^{-3}$ scaling in the far-field. The approximate form of $u_z$ given by Eq. \eqref{eq:u_z_approx_1} is shown in the insets of Fig. \ref{fig:int_def_swim_para_rx_y}, and it agrees well with the numerically determined values.

Now, the interface deformation is also a function of the distance of a swimmer from the interface: At distances of order the swimmer size, one needs to numerically solve Eqs. \eqref{eq:Interface_7} and \eqref{eq:VzT_parallel} simultaneously, whereas simplifications can be made when the swimmer is farther away. We rescale the wavevector in the interface deformation, Eq. \eqref{eq:Interface_7}, as $\bar{k} \equiv k \bar{r}_{z_{0}}$ and time as $\bar{t}=t / \bar{r}_{z_{0}}$, which is valid in the regime $\bar{r}_{z_{0}} \equiv\left|r_{z_{0}}(0)\right| \gg 1$. Seeking simplifications appropriate for $\bar{k}$, $\bar{t}\sim O(1)$ \cite{hinch_1991} yields the following reduced equation for the interface deformation:
\begin{eqnarray}
\frac{\text{d} \hat{u}_z}{\text{d} \bar{t}} + \frac{\pi \bar{k}}{(1+\lambda)} \Gamma \hat{u}_z   = \left(\frac{\eta_1 L^2 V_s}{\kappa_\beta}\right)\left[  \frac{\pi D r_{z_0} }{4  (1+\lambda) \ln\kappa} \bar{k} p_l^2 \exp\left(2\pi \bar{k} \frac{r_{z_0}}{\bar{r}_{z_0}}\right) \right].
\label{eq:Interface_8}
\end{eqnarray}
The simplified expression for the vertical component of the swimmer translation is
\begin{equation}
V_z^T = -\frac{3 D}{256\pi\ln\kappa}\left(\frac{1-\lambda}{1+\lambda}\right)\frac{1}{r_{z_0}^2} + \left(\frac{\kappa_\beta}{\eta_1 V_s L^2}\right)\int\mathrm{d}\bm{k}\frac{\pi k \Gamma}{(1+\lambda)}\left(1-2\pi \bar{k}\frac{r_{z_0}}{\bar{r}_{z_0}}\right)\hat{u}_z \exp\left( 2\pi \bar{k}\frac{r_{z_0}}{\bar{r}_{z_0}}\right).
\label{eq:VzT_parallel2}
\end{equation}
Eq. (\ref{eq:Interface_8}) highlights two important qualitative changes in the interface deformation for swimmers that are far from the interface. First, the interface deformation is controlled by surface tension alone, as the term proportional to bending is $O(\bar{r}_{z_0})^{-2}$ smaller than $\Gamma$ on the left-hand side of Eq. \eqref{eq:Interface_7}, and hence is absent from the leading order approximation. Second, the above rescalings of $k$ and $t$ recover the original dependence of the rescaled variables on $r_{z_0}$, as in Eq. \eqref{eq:Interface_7}. This implies that, for $r_{z_0}\gg 1$, the time evolution slows down in proportion to $r_{z_0}$. In other words, swimmers at an $O(1)$ distance from the interface influence an $O(1)$ growth of the interface deformation at $O(1)$ times. However, a swimmer that is farther away ($r_{z_0}\gg 1$) will influence an $O(1)$ growth of the interface deformation only when $t\sim O(r_{z_0})$. In the insets of Fig. \ref{fig:int_def_swim_para_rx_y} we plot the interface deformation determined from simultaneously solving Eqs. \eqref{eq:Interface_8} and \eqref{eq:VzT_parallel2}. The agreement of both equations is excellent for $\Gamma = 5$ for $r_\Vert\sim O(1)$, but the deviation of the approximate solution is large for $\Gamma = 0.1$. This is expected as Eq. \eqref{eq:Interface_8} relies on $\Gamma$ being larger than the bending term, and hence, for $\bar{r}_{z_0} = 1$ it relies solely on the largeness of $\Gamma$. At large $r_\Vert$, the agreement is good for the same reason underlying Eqs. \eqref{eq:Interface_8} and \eqref{eq:VzT_parallel2}, but now for $r_\Vert\gg 1$ instead of $r_{z_0}\gg 1$. For large $\bar{r}_{z_0}$ (not shown), the approximations in Eqs. \eqref{eq:Interface_8} and \eqref{eq:VzT_parallel2} agree well even for $\Gamma< O(1)$. We note that the far-field approximation corresponds to modeling the swimmer as a point force-dipole, and the disagreement at small to moderate values of $\Gamma$, particularly for $r_{z_0} \sim O(1)$ is indicative of the dipolar swimmer overestimating the interface deformation.

A result arising from the neglect of the bending term in Eq. \eqref{eq:Interface_8} is a subtle qualitative change in the deformation behavior when bending dominates $(\Gamma \ll 1)$ versus when surface tension dominates $(\Gamma \gg 1)$. In particular, following a zero-crossing at an $O(1)$ distance away from the swimmer, when $\Gamma \gg 1$, the deformation monotonically transitions to an $O\left(r_{\|}\right)^{-3}$ far-field character, whereas for $\Gamma$ up to $O(1)$, the deformation exhibits an intermediate scaling before transitioning to the far-field decay. As shown in the insets of Fig. \ref{fig:int_def_swim_para_rx_y}, there is a characteristic maxima in $\left|u_{z}\right|$ for $\Gamma=0.1$ (see the red curve corresponding to the solution of Eqs. \eqref{eq:Interface_7} and \eqref{eq:VzT_parallel}). These features are absent in the approximate Eqs. \eqref{eq:Interface_8} and \eqref{eq:VzT_parallel2}, and are most evident in Fig. \ref{fig:int_def_swim_para_rx_y}a (see orange and dark green curves in the inset). This qualitative difference emerges since the bending mode, which dominates in this regime, depends on higher order derivatives in the curvature as specified by the stress boundary condition, Eq. \eqref{eq:bc_nd_e}.

\begin{figure}
    \centering
    \begin{subfigure}[t]{0.495\textwidth}
        \centering
        \includegraphics[width=\textwidth]{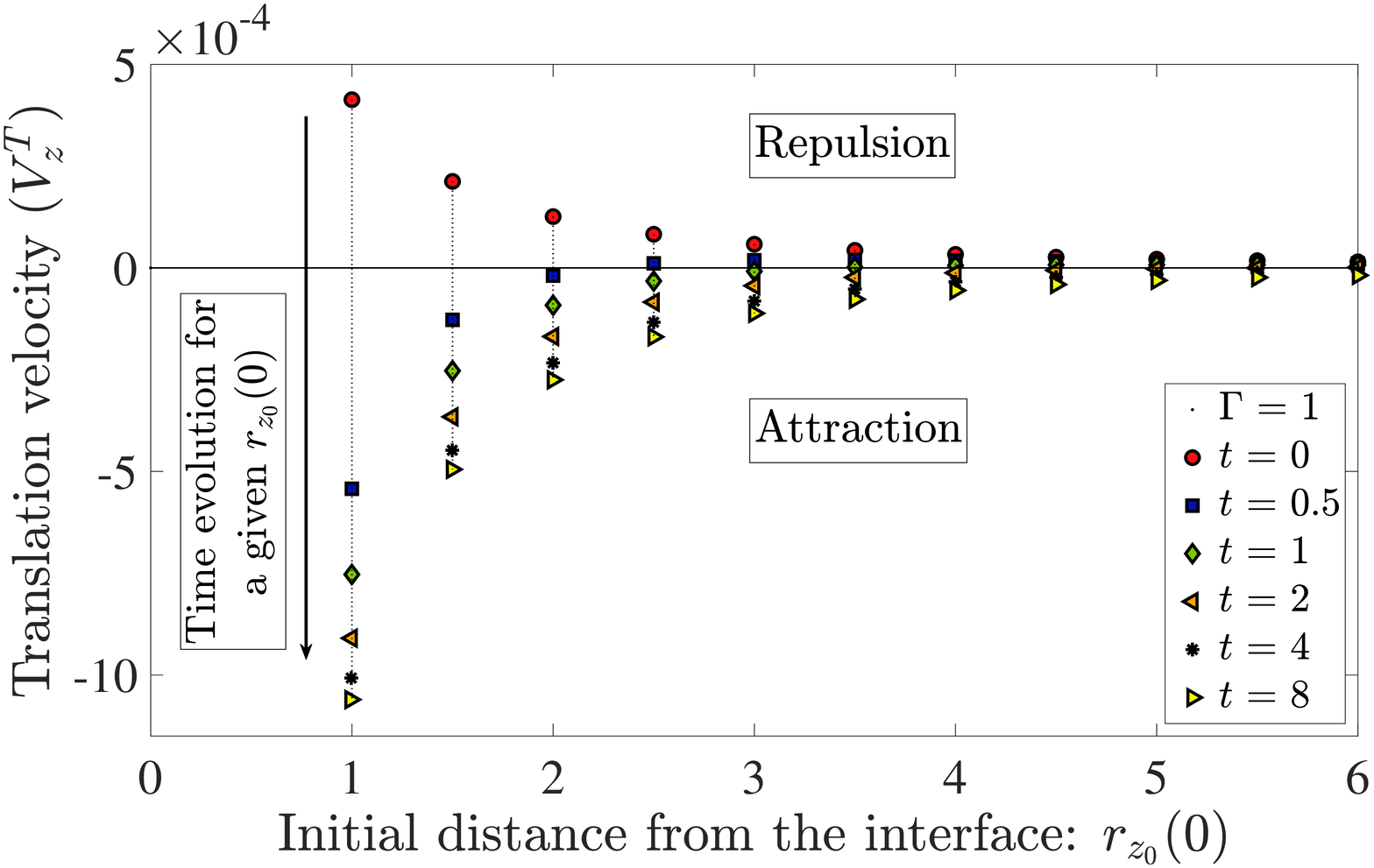}
        \caption{$\lambda = 0.5$}
    \end{subfigure}%
    ~
    \begin{subfigure}[t]{0.495\textwidth}
        \centering
        \includegraphics[width=\textwidth]{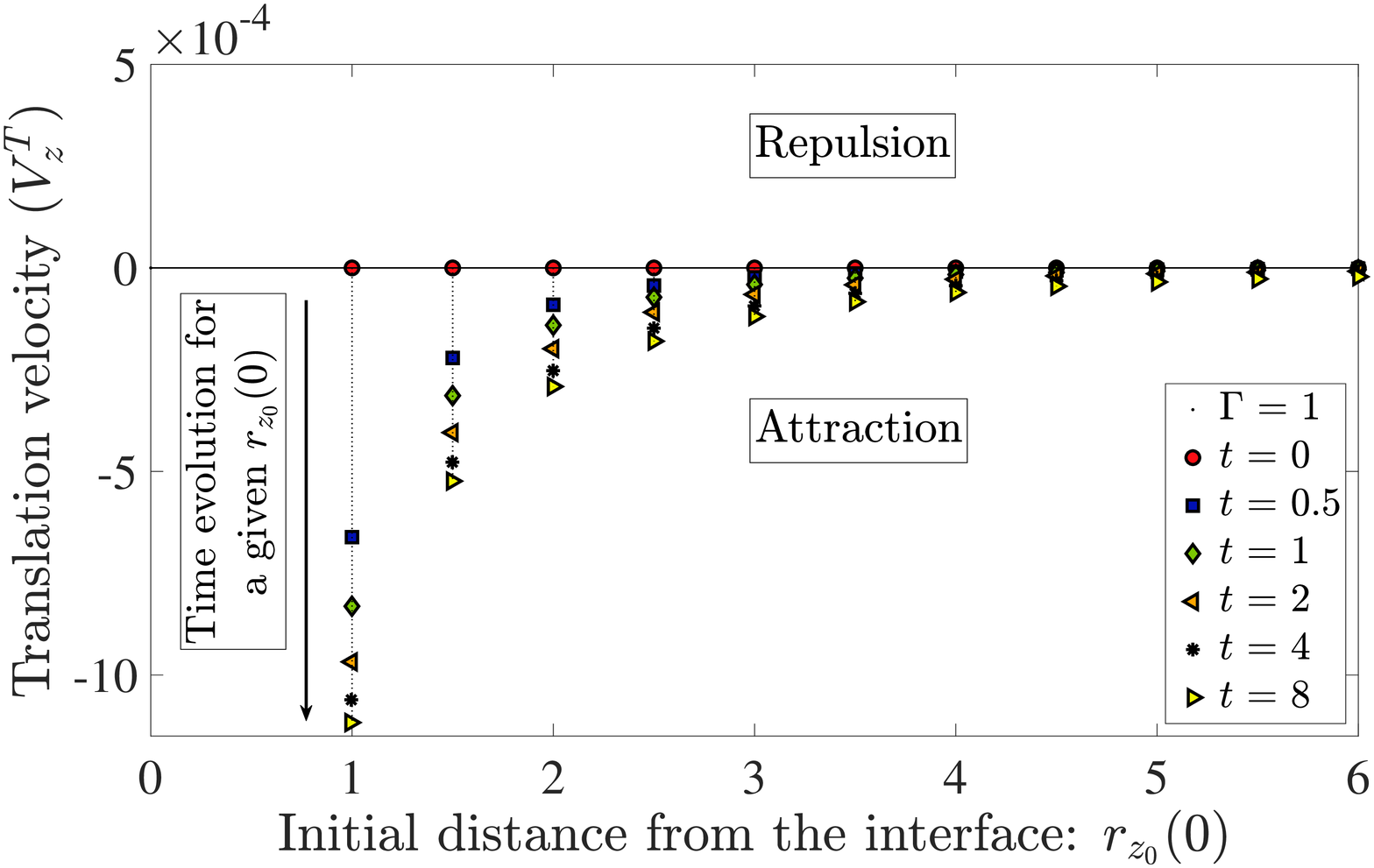}
        \caption{$\lambda = 1$}
    \end{subfigure}
    ~
    \centering
    \begin{subfigure}[t]{0.495\textwidth}
        \centering
        \includegraphics[width=\textwidth]{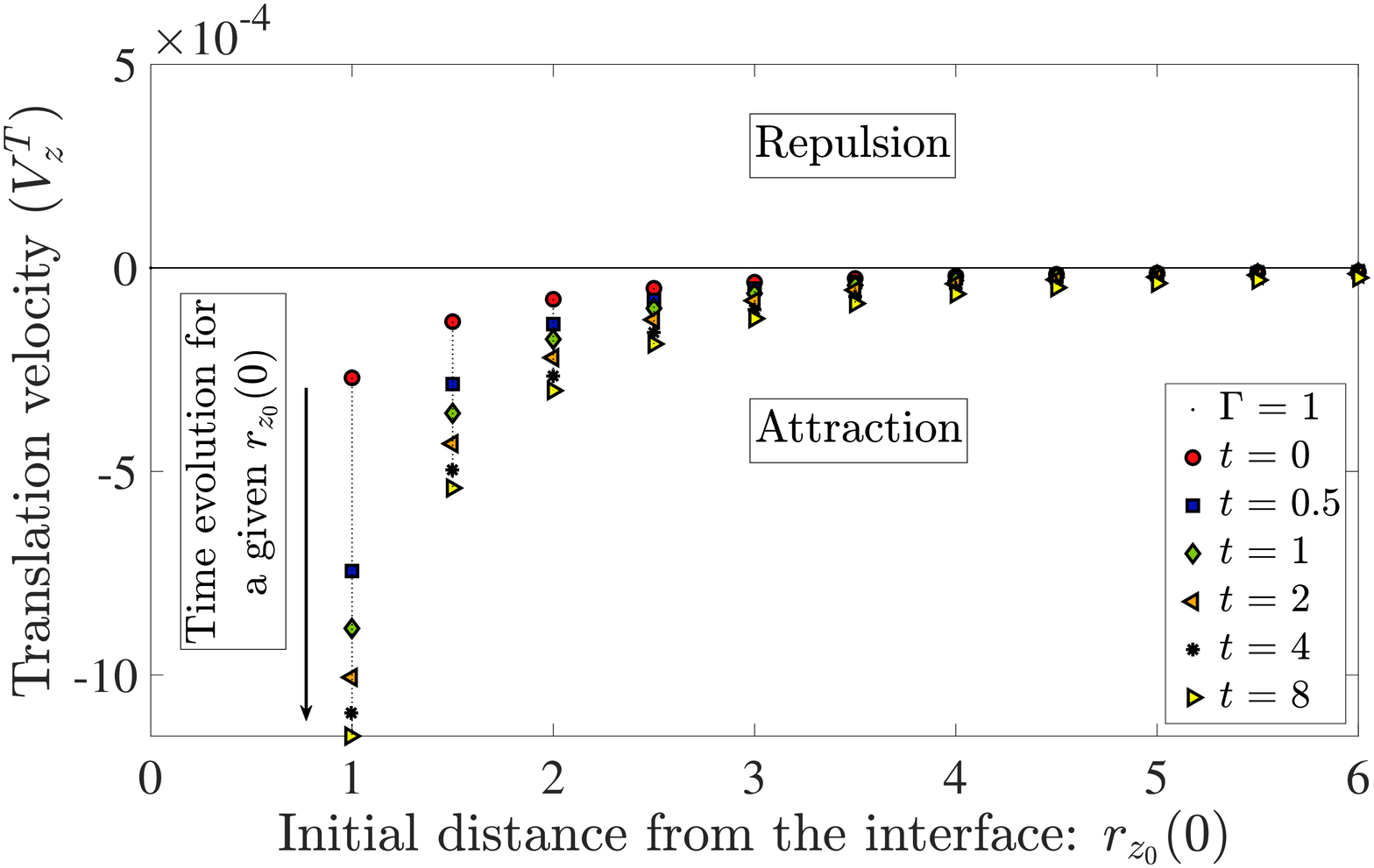}
        \caption{$\lambda = 1.5$}
    \end{subfigure}    
    \caption{The vertical swimmer translation velocity $V_z^T$ of pushers plotted as a function of the initial distance of the swimmer from the interface $r_{z_0}$ at different time instants for the viscosity ratios (a) $\lambda= 0.5$, (b) $\lambda=1$ and (c) $\lambda=1.5$. The plot is to be interpreted as follows: at each $r_{z_0}$, the vertical black dotted lines trace the time evolution of $V_z^T$. In all the plots $\kappa = 10$, $\Gamma = 1$ and $\eta_1 V_s L^2/\kappa_\beta =1$.}
\label{fig:vel_vs_dist}
\end{figure}

In Fig. \ref{fig:vel_vs_dist}, we plot the vertical translation velocity $V_z^T$ of a pusher oriented parallel to the interface for three specific viscosity ratios $\lambda =$ 0.5, 1 and 1.5. We note two interesting features of the vertical migration. First, as shown in Fig. \ref{fig:vel_vs_dist}a, when $\lambda < 1$, $V_z^T$ is positive for short times and negative for long times. That is, a pusher is repelled from the interface at short times, while at long times it is eventually attracted to the interface. Second, the crossover time from repulsion to attraction is shorter for swimmers closer to the interface. For $r_{z_0}\sim O(1)$, the swimmer undergoes a change in its vertical translational velocity at a $t_{\text{transition}}<O(1)$, whereas, when $r_{z_0}\gg 1$, $t_{\text{transition}}\sim O(r_{z_0})$; here, $t_{\text{transition}}$ is the time taken for $V_z^T$ to vanish. For instance, when $r_{z_0} \in [1, 2]$, $t_{\text{transition}}<0.5$ and when $r_{z_0} \geq 4$, $t_{\text{transition}}>1$. This is unlike the case of $\lambda \geq 1$ as shown in Fig. \ref{fig:vel_vs_dist}b and c, where the pusher is monotonically attracted to the interface. Pullers (not shown) behave in the exact opposite manner, that is, for $\lambda<1$ they are attracted to the interface at short times and repelled at long times, with the crossover time for attraction to repulsion having the same character; for $\lambda\geq 1$ pullers are monotonically repelled from the interface.

\begin{figure}
    \centering
    \includegraphics[width=\textwidth]{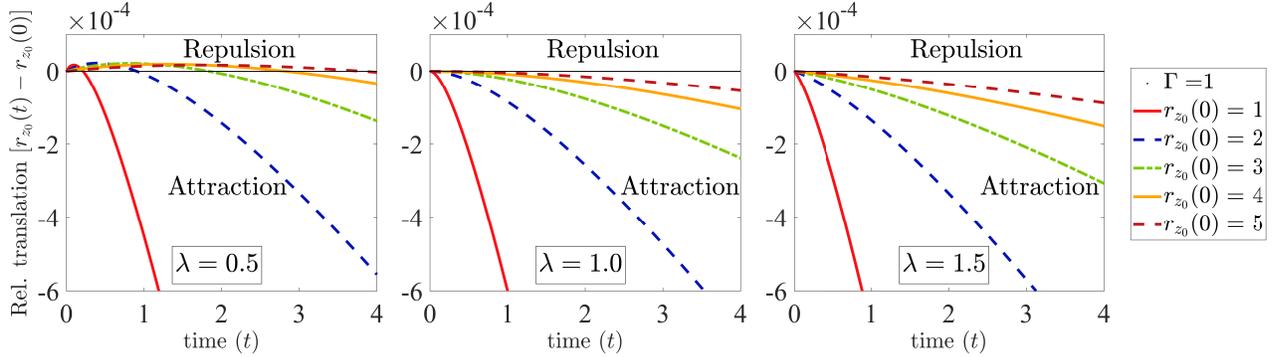}  
    \caption{The relative vertical swimmer trajectory $r_{z_0}(t)-r_{z_0}(0)$ of pushers plotted as a function of  time, for the viscosity ratios $\lambda= 0.5$, 1 and 1.5. In all the plots $\kappa = 10$, $\Gamma = 1$ and $\eta_1 V_s L^2/\kappa_\beta =1$.}
\label{fig:swim_traj}
\end{figure}

In Fig. \ref{fig:swim_traj}, we plot the relative swimmer trajectories of pushers as a function of time for swimmers starting from different initial locations $r_{z_0}(0)$, for $\lambda =$ 0.5, 1 and 1.5 as well. Owing to the repulsion ($V_z^T > 0$) at short times when $\lambda<1$ as highlighted in Fig. \ref{fig:vel_vs_dist}a, the relative swimmer trajectory: $r_{z_0}(t)-r_{z_0}(0)\geq 0$ in Fig. \ref{fig:swim_traj}a. Moreover, as expected from the discussion surrounding Fig. \ref{fig:vel_vs_dist}, the time it takes for the swimmer to return to its initial location depends sensitively upon its initial distance from the interface. For instance, when $r_{z_0}(0) = 1$, the return time is about $t\sim 0.25$, whereas, when $r_{z_0}(0) = 5$, return time is $t\sim 3.7$. Such a non-monotonic vertical swimmer translation for $\lambda<1$ stands in contrast to what has been observed for swimmers near rigid boundaries \cite{berke_2008, Guanglai_2009, spagnolie2012hydrodynamics, Elgeti_2015, ezhilan2015distribution}, where the pushers are, hydrodynamically, only attracted to the interface. Even for model swimmers near deformable interfaces, the swimmer migration has been reported to behave monotonically for a given $\lambda$ \cite{dias2013, shaik_ardekani_2017}. For instance, \citet{dias2013} noted that the average pumping velocity between a swimming sheet and the interface was of negative for $\lambda <1$ and positive for $\lambda >1$. \citet{shaik_ardekani_2017} calculated the vertical translational velocity of a spherical squirmer over a range of $\lambda$, and did not find a qualitative change in the swimmer migration across $\lambda = 1$ (see Figs. 2 and 8 therein).

The first of the above observations regarding the non-monotonic swimmer translational for $\lambda<1$ can be explained from a closer inspection of $V_z^T$ given by Eq. \eqref{eq:VzT_parallel}. The translational velocity has two principal contributions: (1) the instantaneous contribution from Stokes flow that is proportional to $(1-\lambda)/(1+\lambda)$, (2) the time dependent term associated with the interface deformation $\hat{u}_z$. In fact, at $t=0$, $V_z^T$ (red circles in Fig. \ref{fig:vel_vs_dist}) is precisely the instantaneous contribution before the interface deforms. The time dependent term remains negative at all times $t>0$, and its magnitude gradually increases as the interface deformation grows. However, the instantaneous term changes sign depending on whether $\lambda$ is greater than or less than unity, although its magnitude remains nearly constant. For pushers, when $\lambda<1$ ($>1$), the instantaneous term is positive (negative), and it vanishes when $\lambda = 1$. Its magnitude remains nearly constant because it changes only via the hydrodynamic interaction-induced change in the swimmer vertical motion $r_{z_0}(t)$, which does not change appreciably over $O(1)$ times implying that $\Delta r_{z_0}(t)\equiv r_{z_0}(t) - r_{z_0}(0)\ll r_{z_0}(0)$. Hence, among the two fluid regions, if a pusher is present in the more viscous fluid ($\lambda < 1$), it starts off being repelled at short times due to the dominance of the repulsive instantaneous term. At $t = t_{\text{transition}}$, the growing time dependent deformation term exactly matches the nearly constant instantaneous Stokes term, whence $V_z^T$ vanishes, and this is characterized by the peak in the relative swimmer trajectory in Fig. \ref{fig:swim_traj} (left panel). For $t > t_{\text{transition}}$ the swimmer experiences an attraction to the interface owing to the dominance of the time dependent term. This explains the repulsion of pushers from the interface at short times and attraction at long times only when it is in the more viscous fluid.

To explain the second observation of the time dependence of the swimmer translation as a function of its distance from the interface, we note that the time dependence in the problem originates from the kinematic boundary condition of the interface deformation given by Eq. \eqref{eq:Interface_7}. The swimmer translation velocity $V_z^T$(and thus, $r_{z_0}$) obeys the same time dependence as $u_z$ due to the linear relationship between the two field variables in Eq. \eqref{eq:VzT_parallel}, independent of $\lambda$. Therefore, such a delay in the time to return back to its original position when $\lambda<1$ is expected following the arguments explained in the context of the interface deformation. That is, when $r_{z_0}\sim O(1)$, the interface deformation, and hence, $V_z^T$ evolves on $O(1)$ time units. However, when $r_{z_0}\gg 1$, the interface deformation evolves on a time scale of $O(r_{z_0})$. Therefore, as shown in Fig. \ref{fig:swim_traj} for $\lambda =0.5$, the farther the pusher (puller) is from the interface, the longer it will sense a repulsion (an attraction), albeit with reduced intensity.

\begin{figure}
    \centering
    \begin{subfigure}[t]{0.495\textwidth}
        \centering
        \includegraphics[width=\textwidth]{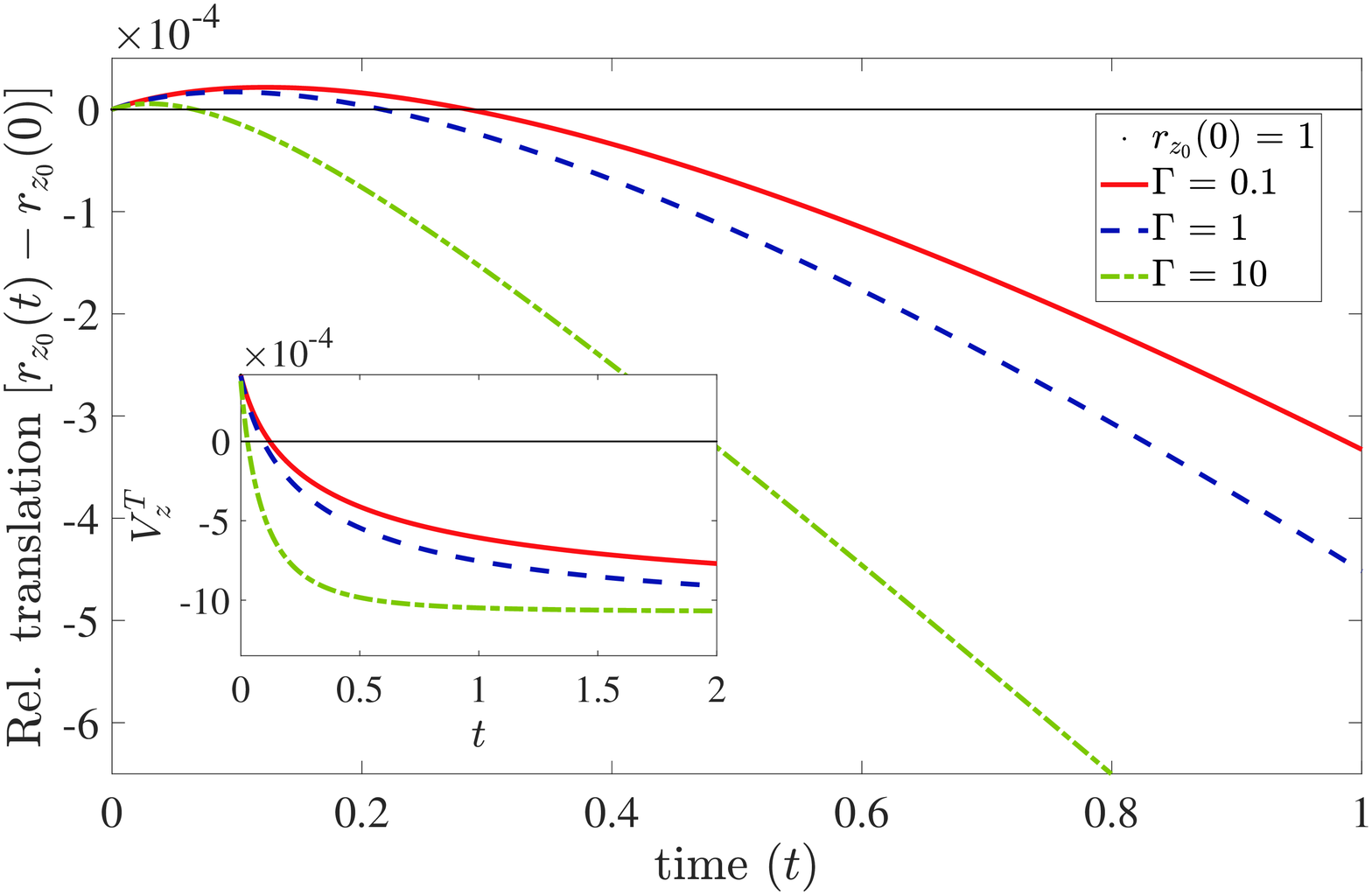}
        \caption{}
    \end{subfigure}%
    ~
    \begin{subfigure}[t]{0.495\textwidth}
        \centering
        \includegraphics[width=\textwidth]{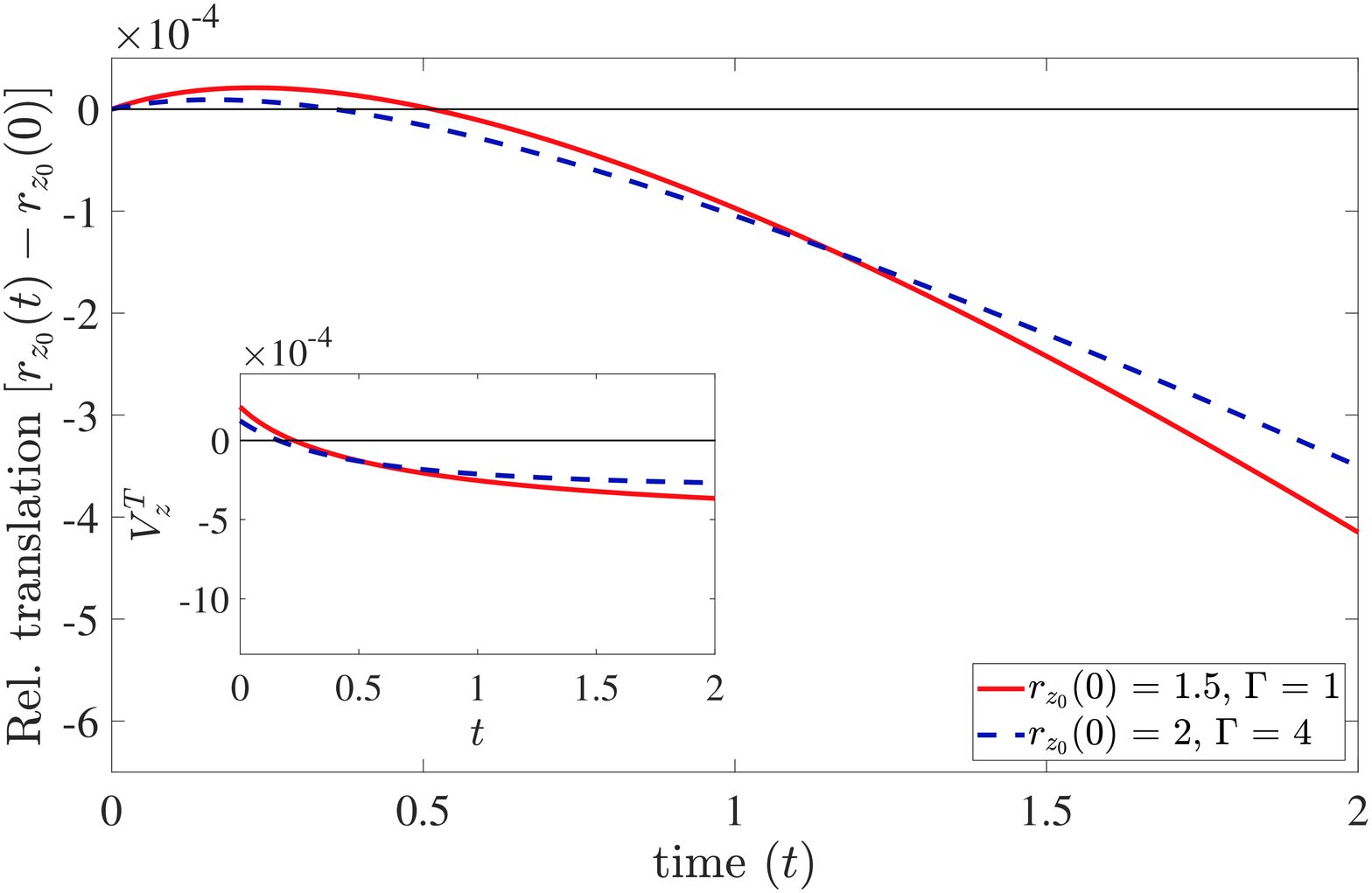}
        \caption{}
    \end{subfigure}  
    \caption{The relative swimmer trajectory $r_{z_0}(t)-r_{z_0}(0)$ of a pusher plotted as a function of time. In (a) the initial swimmer distance to the interface is fixed: $\left\vert r_{z_0}\right\vert = 1$ and the ratio of surface tension to bending stress is varied: $\Gamma =$ 0.1, 1, 10. In (b) both $r_{z_0}(0)$ and $\Gamma$ are varied so as to yield similar swimmer translation, with $r_{z_0}(0) = 1.5, 2$ and $\Gamma = 1, 4$, respectively. In both figures $\lambda = $ 0.5, $\kappa = 10$ and $\eta_1 V_s L^2/\kappa_\beta =1$. In the inset the corresponding swimmer translation velocities $V_z^T$ are plotted for the same parameters.}
\label{fig:swimmer_par_gl2bkb_vary}
\end{figure}

To show the relative importance of surface tension to bending stress on the swimmer motion, in Fig. \ref{fig:swimmer_par_gl2bkb_vary}a we plot the relative swimmer translation as a function of time for different values of $\Gamma (\equiv \gamma L^2/\kappa_\beta)$. Clearly, increasing the surface tension results in a more rapid translation, with a dramatic reduction in the time spent in the repulsive state. For instance, as $\Gamma$ is increased from 0.1 to 10, the return time decreases from $t\approx 0.28$ to $t\approx 0.06$. Moreover, as evident from the inset of Fig. \ref{fig:swimmer_par_gl2bkb_vary}a, the swimmer translation velocity approaches a quasi-steady state when $\Gamma\gg 1$. Note the emphasis on \enquote*{quasi-steady state}, as the problem is inherently an unsteady one, even though $V_z^T$ appears to be almost steady for $\Gamma = 10$ following the initial transient. This is because as the swimmer is moving closer to the interface, it will continue to accelerate, given that the interface undergoes a stronger deformation from swimmers closer to it. In turn, the translation velocity will continue to grow. Nevertheless, this acceleration remains small after sustaining an initial growth in the deformation, as the driving force is still small (see the scales along the ordinate of Fig. \ref{fig:swimmer_par_gl2bkb_vary}a).

This analysis shows that the swimmer translation must be analyzed in terms of both distance to the interface and interface properties. That is, both $r_{z_0}$ and $\Gamma$ have comparable effects on the swimmer translation when these quantities are large. Note, however, that owing to the exponential damping of $u_z$ and $V_z^T$ with respect to $r_{z_0}$ even an $O(1)$ change in the distance of the swimmer to the interface results in a large reduction of the swimmer translational velocity. Therefore, a swimmer near an interface with relatively low surface tension can exhibit similar translational dynamics to that of a swimmer that is slightly farther away from the interface  with higher surface tension. An example of such a trajectory is highlighted in Fig. \ref{fig:swimmer_par_gl2bkb_vary}b, where the initial locations of the swimmers are displaced by 0.5 units of $r_{z_0}$, but $\Gamma$ is more than doubled. For an $O(1)$ change in time, the trajectories in two scenario are similar.

\subsection{\label{subsec_swimmers_orthogonal_results} Microswimmers swimming orthogonal to the interface}

We now consider swimmers oriented orthogonal to the interface ($p_z = \pm 1$ and $p_x=p_y=0$), since the results are easier to interpret relative to swimmers with arbitrary orientation. Here too, the rate of rotation $\dot{\bm{p}}=0$, and therefore characterizing the swimmer motion requires simultaneously solving for the interface deformation and the vertical translation.

The equation governing the interface deformation is obtained using the expression for $\hat{v}_{1z}\vert_{r_{z_0}}$ from Eq. \eqref{eq:vhat_z_ortho} (Appendix \ref{sec_appendix_B}) in Eq. \eqref{eq:Interface_6}, and is:
\begin{eqnarray}
\frac{\partial \hat{u}_z}{\partial t} + \frac{\pi k}{1+\lambda}\left(\Gamma + 4\pi^2 k^2\right) \hat{u}_z &=& \left(\frac{\eta_1 L^2 V_s}{\kappa_\beta}\right) \frac{D}{(1+\lambda)\ln\kappa}\frac{\exp(2\pi k r_{z_0})}{4\pi^2 k^2} \big[2(1-\pi k r_{z_0}) \left(\cosh(\pi k)-1\right) - \pi k \sinh(\pi k)\big]\nonumber\\ && + \left(\frac{\eta_1 L^2 V_s}{\kappa_\beta}\right) p_z \frac{\delta(k)}{\pi k}.
\label{eq:Interface_9}
\end{eqnarray}
The expression for the vertical component of the swimmer translation velocity $V_z^T$ is:
\begin{eqnarray}
V_z^T &=& \left(\frac{2\pi}{1+\lambda}\right)\left(\frac{\kappa_\beta}{\eta_1 L^2 V_s}\right)\int_0^\infty \mathrm{d}k\;  k\left(\Gamma + 4\pi^2 k^2\right)\exp(2\pi k r_{z_0})\left[\pi k \cosh(\pi k) - 2(1-\pi k r_{z_0}) \sinh(\pi k) \right]\hat{u}_z\nonumber\\ && +\frac{D}{4\pi\ln\kappa}\left(\frac{1-\lambda}{1+\lambda}\right)\Bigg [\frac{1}{2 - 40 r_{z_0}^2 + 128 r_{z_0}^4}\bigg( 12 r_{z_0}^2 - \left(5 - 4 r_{z_0} - 100 r_{z_0}^2 + 80 r_{z_0}^3 + 320 r_{z_0}^4 \right. \nonumber\\ && \left. \qquad - 256 r_{z_0}^5\right) \ln[-2 + 4 r_{z_0}] + \left(5 - 8 r_{z_0} - 100 r_{z_0}^2 + 160 r_{z_0}^3 + 320 r_{z_0}^4 - 512 r_{z_0}^5\right)\ln[-1 + 4 r_{z_0}] \nonumber\\ &&\qquad  + \left(5 + 8 r_{z_0} - 100 r_{z_0}^2 - 160 r_{z_0}^3 + 320 r_{z_0}^4 + 512 r_{z_0}^5\right) \ln[1 + 4 r_{z_0}] \nonumber\\ && \qquad+ \left( - 5 - 4 r_{z_0} + 100 r_{z_0}^2 + 80 r_{z_0}^3 - 320 r_{z_0}^4 -  256 r_{z_0}^5\right) \ln[2 + 4 r_{z_0}] \nonumber\\ && \qquad -24 r_{z_0}^2 + 3(1 - 20 r_{z_0}^2 + 64 r_{z_0}^4) \ln\left[\frac{4 - 16 r_{z_0}^2}{1 - 16 r_{z_0}^2}\right] \bigg) + \frac{1}{2(4 r_{z_0}^2-1)}\Bigg ]; \left \vert r_{z_0} \right\vert > \frac{1}{2}.
\label{eq:VzT_perp}
\end{eqnarray}
Here, the interface deformation and the swimmer translation exhibit a radial symmetry in the plane of the interface, which enables a convenient analytical formulation of the instantaneous part of $V_z^T$ as given in Eq. \eqref{eq:VzT_perp}. Such a  radial symmetry is specific to swimmers orthogonal to the interface, and is a consequence of the linearity of the problem. Note that we have not incorporated the self-swimming term in $V_z^T$, which would involve addition of a constant factor $p_z$ to Eq. \ref{eq:VzT_perp}. Therefore, the discussion here pertains to shakers - swimmers that disturb the fluid with a flow field $V_{s} / \ln \kappa$ but do not self-propel \cite{morozov_2017}.

\begin{figure}
    \centering
    \begin{subfigure}[t]{0.495\textwidth}
        \centering
        \includegraphics[width=\textwidth]{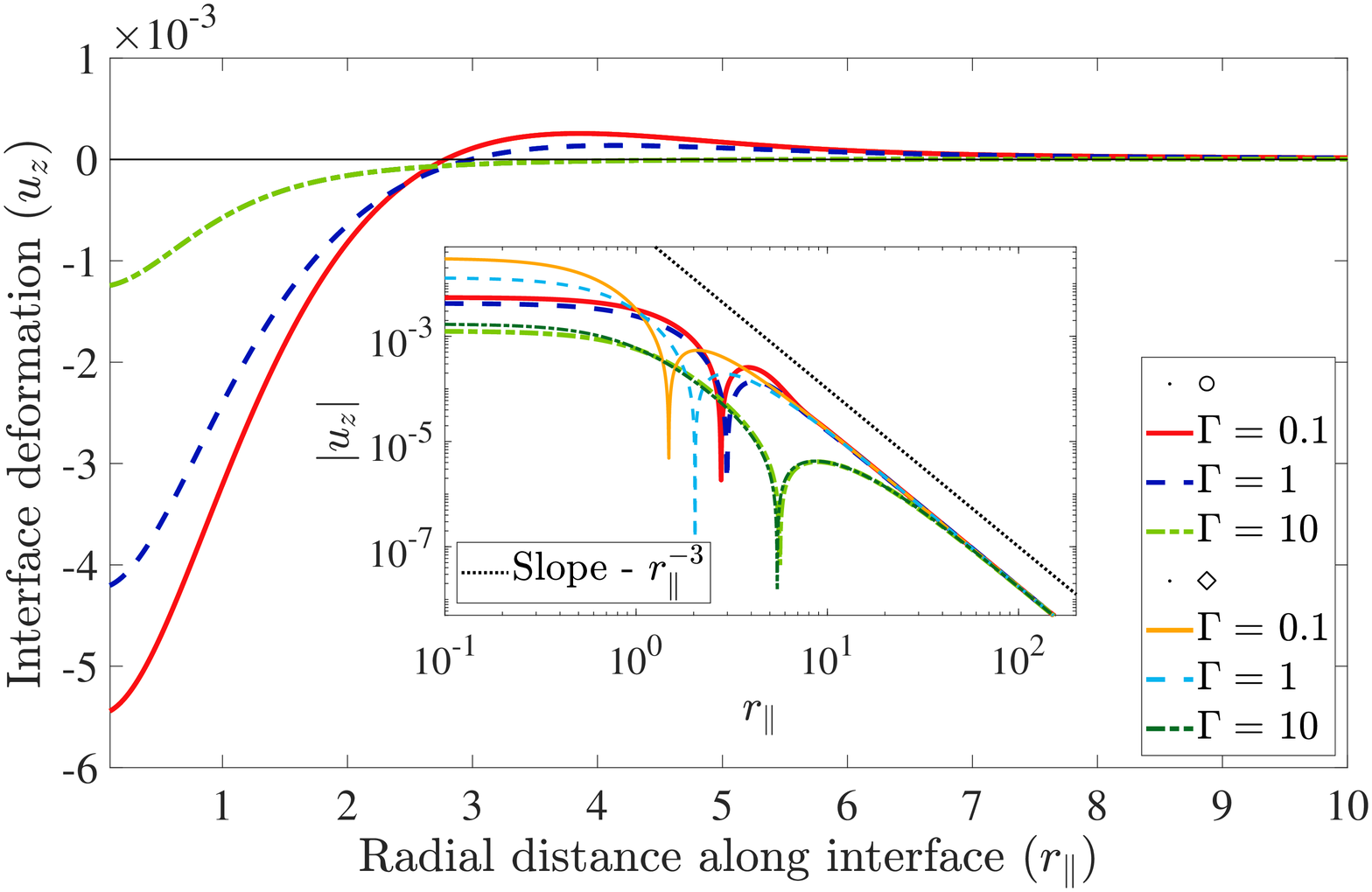}
        \caption{$r_{z_0}(0) = 1$}
    \end{subfigure}%
    ~
    \begin{subfigure}[t]{0.495\textwidth}
        \centering
        \includegraphics[width=\textwidth]{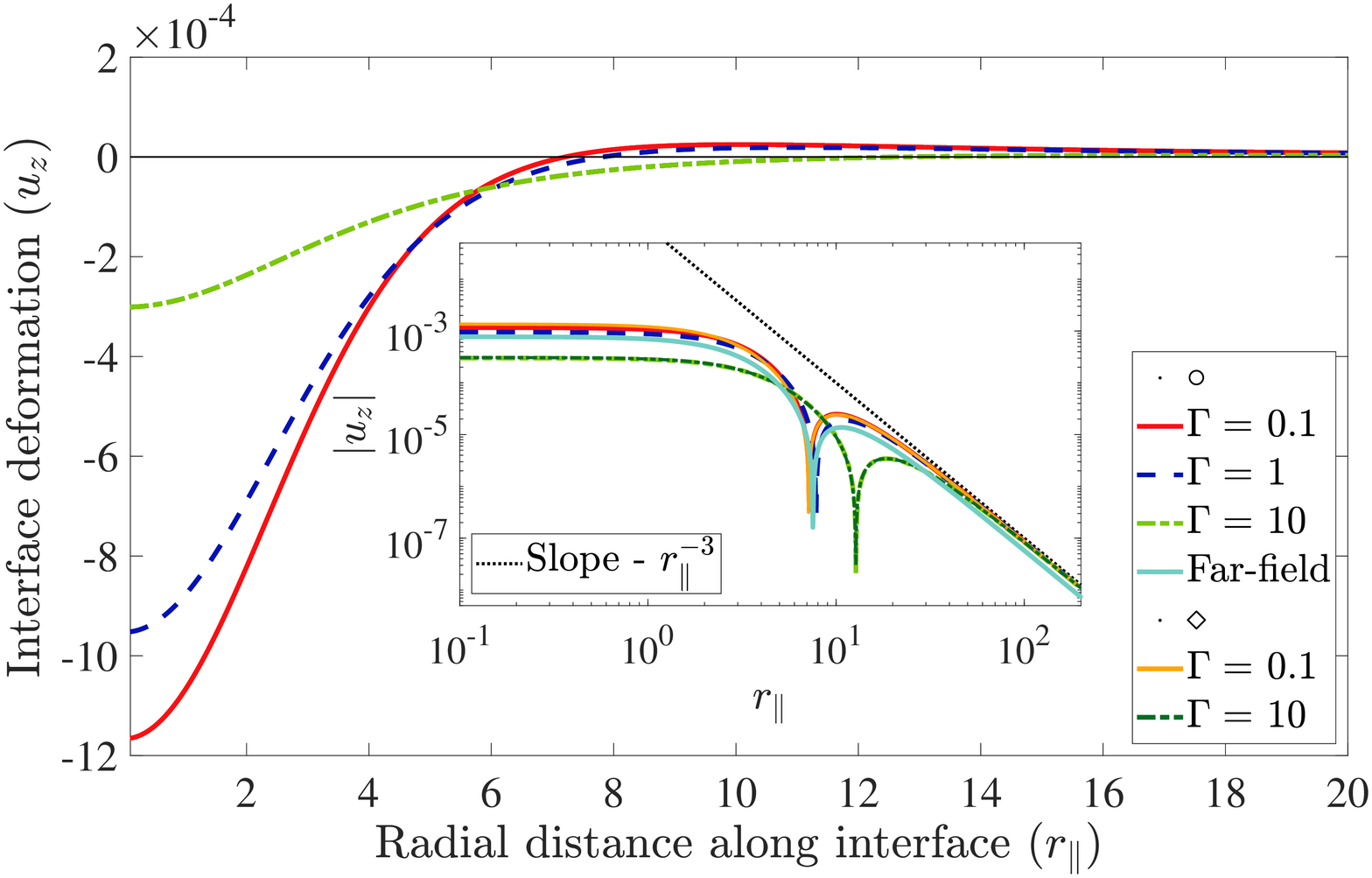}
        \caption{$r_{z_0}(0) = 5$}
    \end{subfigure}
    \caption{The interface deformation, $u_z$, due to a pusher plotted as a function of the radial distance along the interface $r_{\Vert}$ for $\Gamma = $ 0.1, 1 and 10 at time $t=2$ (a) for $r_{z_0}(0) = 1$ and (b) $r_{z_0}(0) = 5$. The dotted lines in the inset represent the far-field $O(r_\Vert)^{-3}$ scaling. Legends under $\circ$ refer to $u_z$ obtained from solving the pair of equations \eqref{eq:Interface_9}, \eqref{eq:VzT_perp}, and those under $\diamond$ from solving \eqref{eq:Interface_10}, \eqref{eq:VzT_perp_2}. The \enquote*{Far-field} in the inset of (b) refers to $u_z$ from Eq. \eqref{eq:u_z_approx_2} for $\Gamma=1$.}
\label{fig:int_def_perp}
\end{figure}

We now analyze the interface deformation created by the swimming motion. In Fig. \ref{fig:int_def_perp}, we plot the interface deformation due to a pusher as a function of the radial distance along the plane of the interface, $r_\Vert$, for $\Gamma = 0.1$, 1 and 10. As defined previously, $r_\Vert = 0$ is the swimmer location on the interface. For large $r_{\|}$ the deformation $u_{z}$ is independent of $\Gamma$, similar to that of swimmers parallel to the interface (see the collapse in the inset of Fig. \ref{fig:int_def_perp} for $r_{\|} \gg 1$). This far-field scaling is $O\left(r_{\|}\right)^{-3}$, and can be identified by solving Eq. \eqref{eq:Interface_9} analytically, applying the approach used to obtain Eq. \eqref{eq:u_z_approx_1} in Sec. \ref{subsec_swimmers_parallel_results}, yielding:
\begin{equation}
\left. u_z\right\vert_{r_\Vert\gg 1} \approx \frac{1}{2\pi \Gamma }\left(\frac{\eta_1 L^2 V_s}{\kappa_\beta}\right) \frac{D}{\ln\kappa}\bigg[ \frac{r_{z_0}^2}{4(r_\Vert^2 + r_{z_0}^2)^{\frac{3}{2}}} +\frac{  r_{z_0}(1+\lambda)\Big ( \Gamma t - 2 r_{z_0}(1+\lambda) \Big) }{\Big ( 4 r_\Vert^2(1+\lambda)^2 + (\Gamma t - 2 r_{z_0}(1+\lambda))^2\Big )^{\frac{3}{2}} } \bigg].
\label{eq:u_z_approx_2}
\end{equation}
The above approximation also holds for $r_{z_0}\gg 1$ at $r_\Vert = O(1)$, and is plotted in the inset of Fig. \ref{fig:int_def_perp}b.

However, the interface deformation for this swimmer configuration exhibits an opposite trend to that of swimmers oriented parallel to the interface; namely, for a pusher the deformation occurs away from the swimmer (into fluid region 2) at short separations, whereas, its positive for $r_{\|} \gg 1$, as also noted previously in \citet{hosoi_lauga_2008} for point force-dipoles and \citet{shaik_ardekani_2017} for spherical squirmers. This is expected as the disturbance field generated by a pusher exits along the ends of its long-side and is thereby pushing onto the interface when it is oriented orthogonal to it. Given that the interface deformation also contributes to the evolution of the vertical translation, this opposing trend will have important consequences for the nature of the migration of swimmers oriented orthogonal to the interface, as will be discussed below.

The radial symmetry in the plane of the interface enables simplifications to both $u_z$ and $V_z^T$ for swimmers positioned far away from the interface ($r_{z_0}\gtrsim O(1)$). Following the discussion in Sec. \ref{subsec_swimmers_parallel_results} for swimmers oriented parallel to the interface, we rescale the wavevector $\bar{k} = k \bar{r}_{z_0}$ and time $\bar{t} = t/\bar{r}_{z_0}$ in Eq. \ref{eq:Interface_9} to find
\begin{equation}
\frac{\partial \hat{u}_z}{\partial \bar{t}} + \frac{\pi \Gamma \bar{k}}{1+\lambda} \hat{u}_z = \left(\frac{\eta_1 L^2 V_s}{\kappa_\beta}\right) \frac{\pi D r_{z_0}}{4(1+\lambda)\ln\kappa} \bar{k}\exp\left(2\pi \bar{k}\frac{r_{z_0}}{\bar{r}_{z_0}}\right),
\label{eq:Interface_10}
\end{equation}
with the corresponding expression for vertical component of the translation velocity becoming
\begin{equation}
V_z^T = -\left(\frac{2\pi^2 \Gamma }{(1+\lambda) \bar{r}_{z_0}^3}\right)\left(\frac{\kappa_\beta}{\eta_1 L^2 V_s}\right)\int_0^\infty \mathrm{d} \bar{k}\;  \bar{k}^2 \exp\left(2\pi \bar{k}\frac{r_{z_0}}{\bar{r}_{z_0}}\right)\hat{u}_z +\frac{3 D}{128\pi\ln\kappa}\left(\frac{1-\lambda}{1+\lambda}\right) \left(\frac{1}{r_{z_0}^2}\right).
\label{eq:VzT_perp_2}
\end{equation}
In the insets of Fig. \ref{fig:int_def_perp}, we plot $\vert u_z\vert$ using both the full equations \eqref{eq:Interface_9} and \eqref{eq:VzT_perp}, and the approximate equations \eqref{eq:Interface_10} and \eqref{eq:VzT_perp_2}. Similar to the discussion in Sec. \ref{subsec_swimmers_parallel_results}, for $r_{z_0}(0)=1$ in Fig. \ref{fig:int_def_perp}a, the agreement of both pair of equations is excellent for $\Gamma = 10$, but the deviation of the approximate solution is large for $\Gamma = 1$ and 0.1. For $r_{z_0} = 5$ however, this holds for $\Gamma< O(1)$, as seen in the inset of Fig. \ref{fig:int_def_perp}b.

\begin{figure}
    \centering
    \begin{subfigure}[t]{0.495\textwidth}
        \centering
        \includegraphics[width=\textwidth]{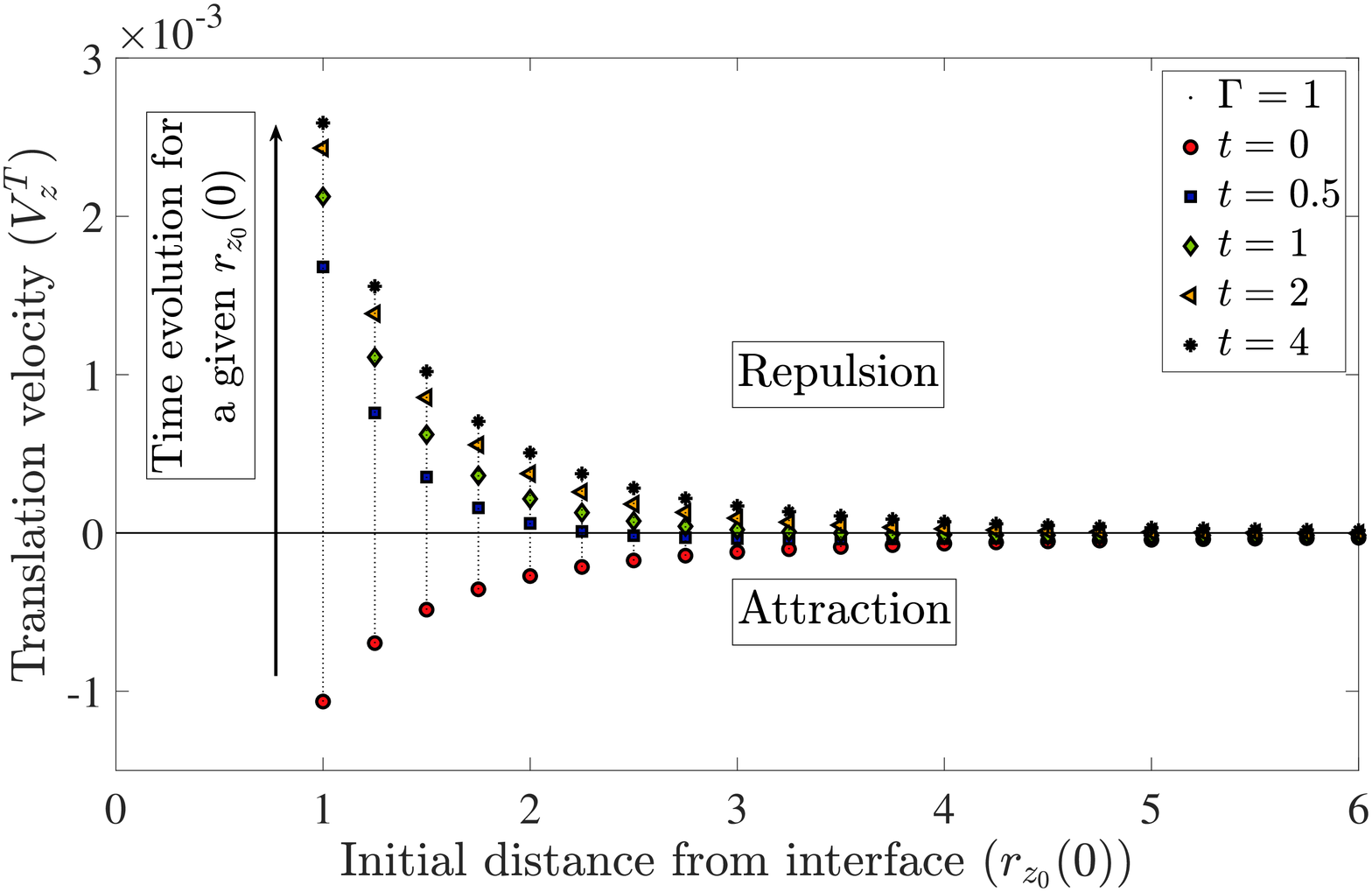}
        \caption{$\lambda = 0.5$}
    \end{subfigure}%
    ~
    \begin{subfigure}[t]{0.495\textwidth}
        \centering
        \includegraphics[width=\textwidth]{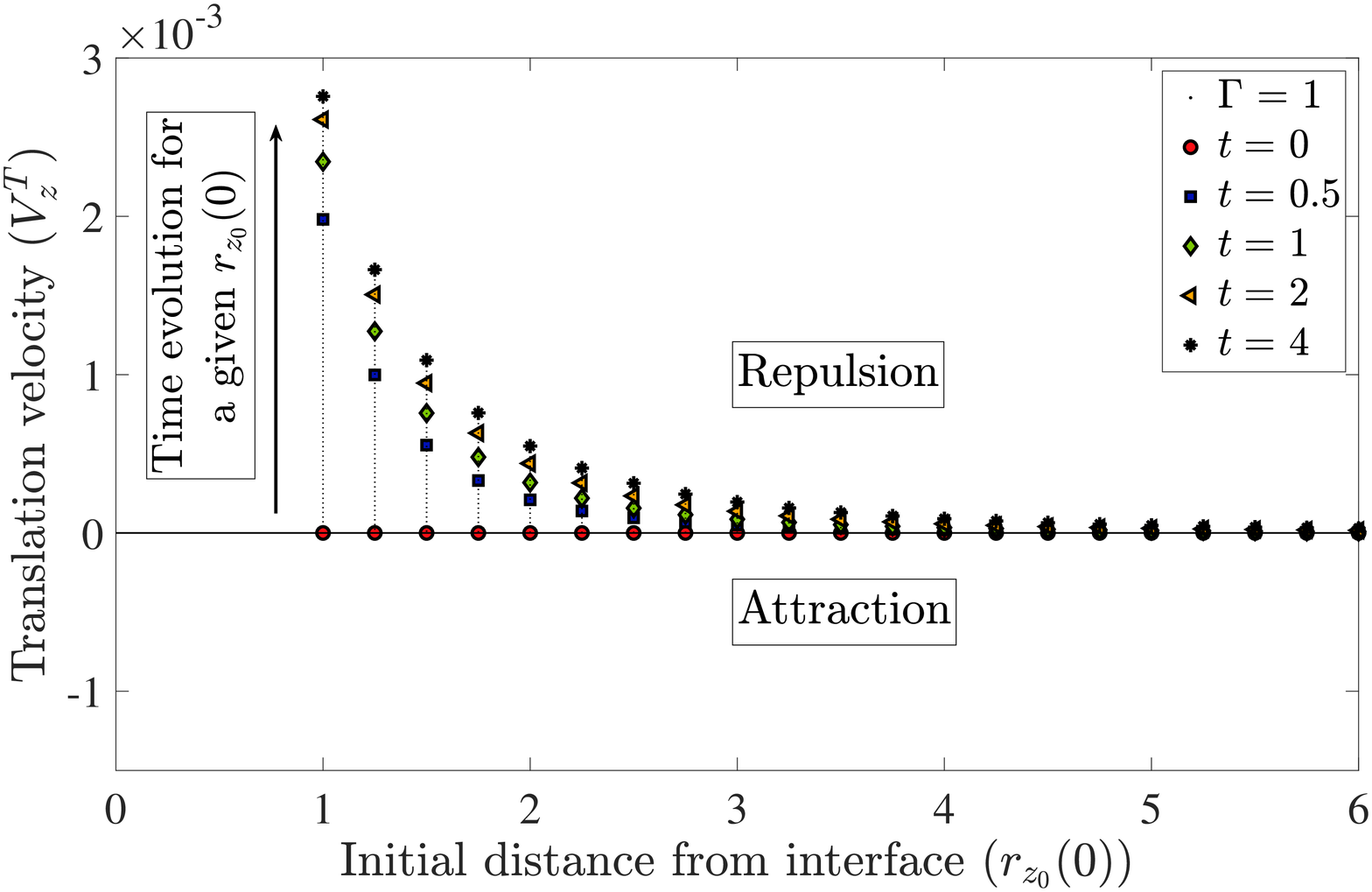}
        \caption{$\lambda = 1$}
    \end{subfigure}
    ~
    \centering
    \begin{subfigure}[t]{0.495\textwidth}
        \centering
        \includegraphics[width=\textwidth]{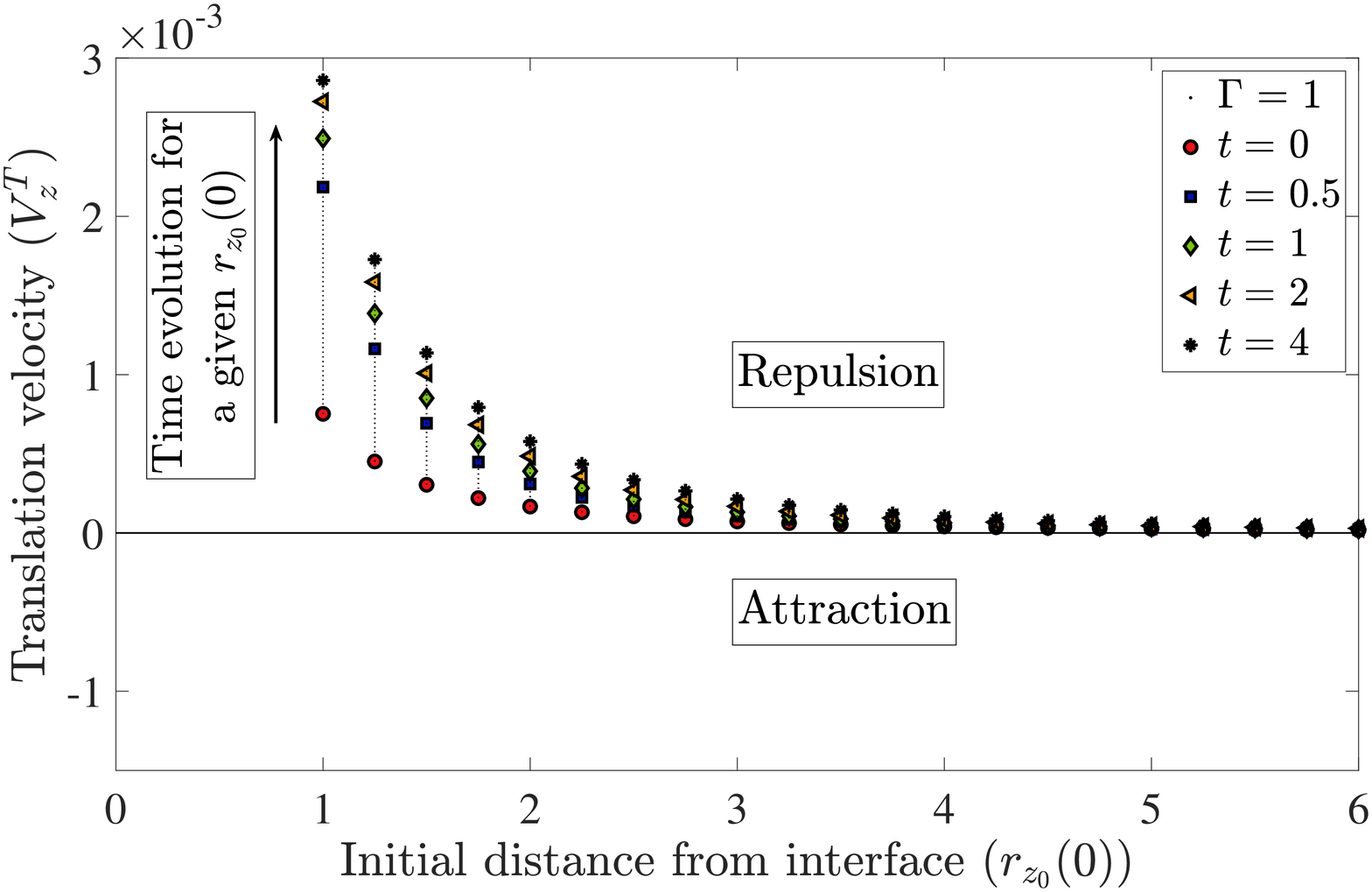}
        \caption{$\lambda = 1.5$}
    \end{subfigure}    
    \caption{The vertical swimmer translation velocity $V_z^T$ of pushers plotted as a function of the initial distance of the swimmer from the interface $r_{z_0}$ at different time instants for the viscosity ratios (a) $\lambda= 0.5$, (b) $\lambda=1$ and (c) $\lambda=1.5$. The plot is to be interpreted as follows: at each $r_{z_0}$, the vertical black dotted lines trace the time evolution of $V_z^T$ as indicated by the black arrow. In all the plots $\kappa = 10$, $\Gamma = 1$ and $\eta_1 V_s L^2/\kappa_\beta =1$.}
\label{fig:vel_vs_dist2}
\end{figure}

\begin{figure}
    \centering
    \includegraphics[width=\textwidth]{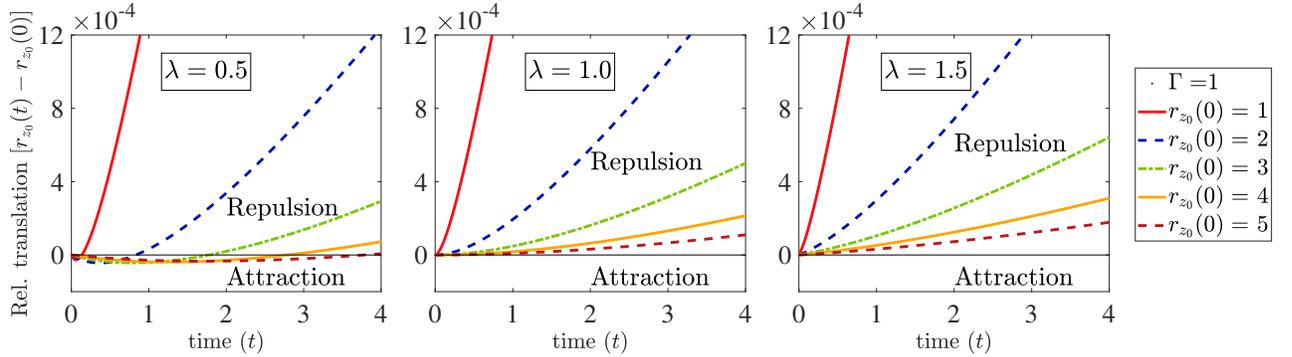} 
    \caption{The relative vertical swimmer trajectory $r_{z_0}(t)-r_{z_0}(0)$ of pushers plotted as a function of  time, for the viscosity ratios $\lambda=$ 0.5, 1 and 1.5. In all the plots $\kappa = 10$, $\Gamma = 1$ and $\eta_1 V_s L^2/\kappa_\beta =1$.}
\label{fig:swim_traj2}
\end{figure}

In Fig. \ref{fig:vel_vs_dist2}, we plot the time trace of the vertical component of the swimmer translation velocity $V_z^T$ of a pusher at different initial swimmer distances from the interface $r_{z_0}(0)$, for $\lambda =$ 0.5, 1, and 1.5. Similar to swimmers parallel to the interface, for $\lambda<1$, both repulsive and attractive regimes are exhibited by $V_{z}^{T}$ as a function of time. Here, the deformation term is repulsive owing to the opposite character of the interface deformation, as discussed below Eq. \eqref{eq:VzT_perp}. The instantaneous term in $V_{z}^{T}$ is attractive (repulsive) when $\lambda<1$ $(\lambda>1)$. Therefore, when $\lambda<1$, for short times, $V_{z}^{T}$ is attractive (negative), and is repulsive (positive) at long times. In turn, the relative swimmer trajectory $r_{z_{0}}(t)-r_{z_{0}}(0)$ shown in Fig. \ref{fig:swim_traj2} is the qualitative mirror image about $r_{z_{0}}(t)-r_{z_{0}}(0)=0$, to that of a swimmer parallel to the interface in Fig. \ref{fig:swim_traj}.

The transition time from attraction to repulsion, $t_{\text{transition}}$, exhibits a character similar to that of swimmers parallel to the interface, and is expected from the similar scaling's of Eq. \eqref{eq:Interface_8} and Eq. \eqref{eq:Interface_10}. Namely, for $r_{z_0}\sim O(1)$, $t_{\text{transition}}\sim O(1)$, whereas for large $r_{z_0}$, $t_{\text{transition}} \sim O(r_{z_0})$. However, the magnitude of $V_z^T$ is larger for swimmers oriented orthogonal to the interface compared to swimmers oriented parallel to the interface. At $r_{z_0}(0) = 1$, for a swimmer parallel to the interface $V_z^T$ changes from $+ 0.5 \times 10^{-3}$ to about $-10^{-3}$ in a time interval $t\in [0, 4]$, as can be seen in Fig. \ref{fig:vel_vs_dist}a (see the dotted trace of the first vertical line). In contrast, for swimmers oriented orthogonal to the interface, $V_z^T$ transitions from about $- 10^{-3}$ to about $+2.5 \times 10^{-3}$ in the same time interval, as shown in Fig. \ref{fig:vel_vs_dist2}a. Since the velocity at $t=0$, is that prescribed by the instantaneous component, and that at large times is dominated by the deformation, such character shows that both terms of $V_{z}^{T}$ are larger for swimmers oriented orthogonal to the interface. Therefore, in a given time interval, the relative swimmer vertical translation of an orthogonally oriented swimmer as shown in Fig. \ref{fig:swim_traj2} is larger than that of a swimmer parallel to the interface (see Fig. \ref{fig:swim_traj}). In other words, the coupled hydrodynamics more strongly repels a pusher oriented orthogonal to the interface than it attracts a pusher oriented parallel to the interface.

The relative importance of surface tension to bending stress on the swimmer motion (not shown), is similar to that of swimmers parallel to the interface, as discussed in Sec. \ref{subsec_swimmers_parallel_results} and as is evident in Eqs. \eqref{eq:Interface_9} and \eqref{eq:VzT_perp}. Increasing the surface tension results in a more rapid translation, and for $\lambda<1$ this implies a reduction in the time spent being attracted to the interface. Thus, this monotonic response in $\Gamma$ implies that swimmers farther away from an interface with larger $\Gamma$ can exhibit migration similar to those closer to an interface with lower $\Gamma$.  This is qualitatively similar in character to Fig. \ref{fig:swimmer_par_gl2bkb_vary}b, but again of an opposite trend.

\subsection{\label{subsec_swimmers_arbit_results} Microswimmers swimming arbitrarily oriented to the interface}

We now generalize the coupled hydrodynamics to arbitrary swimmer orientations. As explained in Appendix \ref{sec_appendix_B}, the derivation procedure for any non-parallel swimmer orientation remains the same, and the changes to the fluid equations due to a non-trivial in-plane orientation component are discussed in Appendix \ref{sec_appendix_B2}. Unlike the previous cases, here the vertical component of the rotation rate is non-zero; $ \dot{p}_{z} \neq 0$. For brevity we omit the large equations for $\hat{u}_{z}$, $V_{z}^{T}$ and $\dot{p}_{z}$, the girth of latter two associated with the contributions from all three orientation components. In the following, we only focus on the role of the coupled hydrodynamics on the swimmer vertical translation and rotation. We study how the change in the swimmer orientation affects the nature of the swimmer migration, and the final swimmer orientation given an arbitrary initial orientation. As in Sec. \ref{subsec_swimmers_orthogonal_results}, here too we do not account for the self-swimming term in $V_{z}^{T}$, which is instantaneously shifted by constant factor $p_{z}$.

\begin{figure}
    \centering
    \begin{subfigure}[t]{0.495\textwidth}
        \centering
        \includegraphics[width=\textwidth]{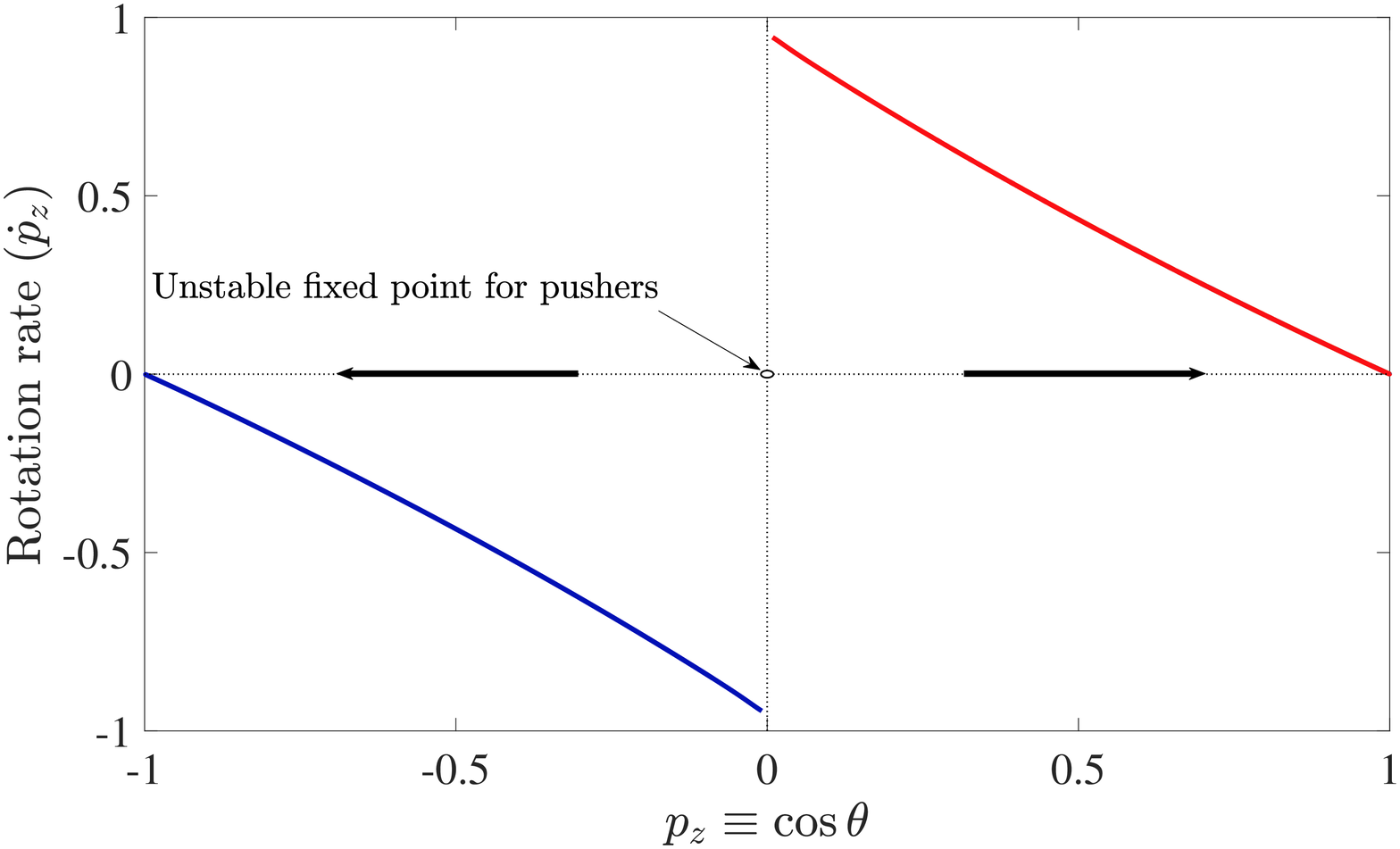}
        \caption{}
    \end{subfigure}%
    ~
    \begin{subfigure}[t]{0.495\textwidth}
        \centering
        \includegraphics[width=\textwidth]{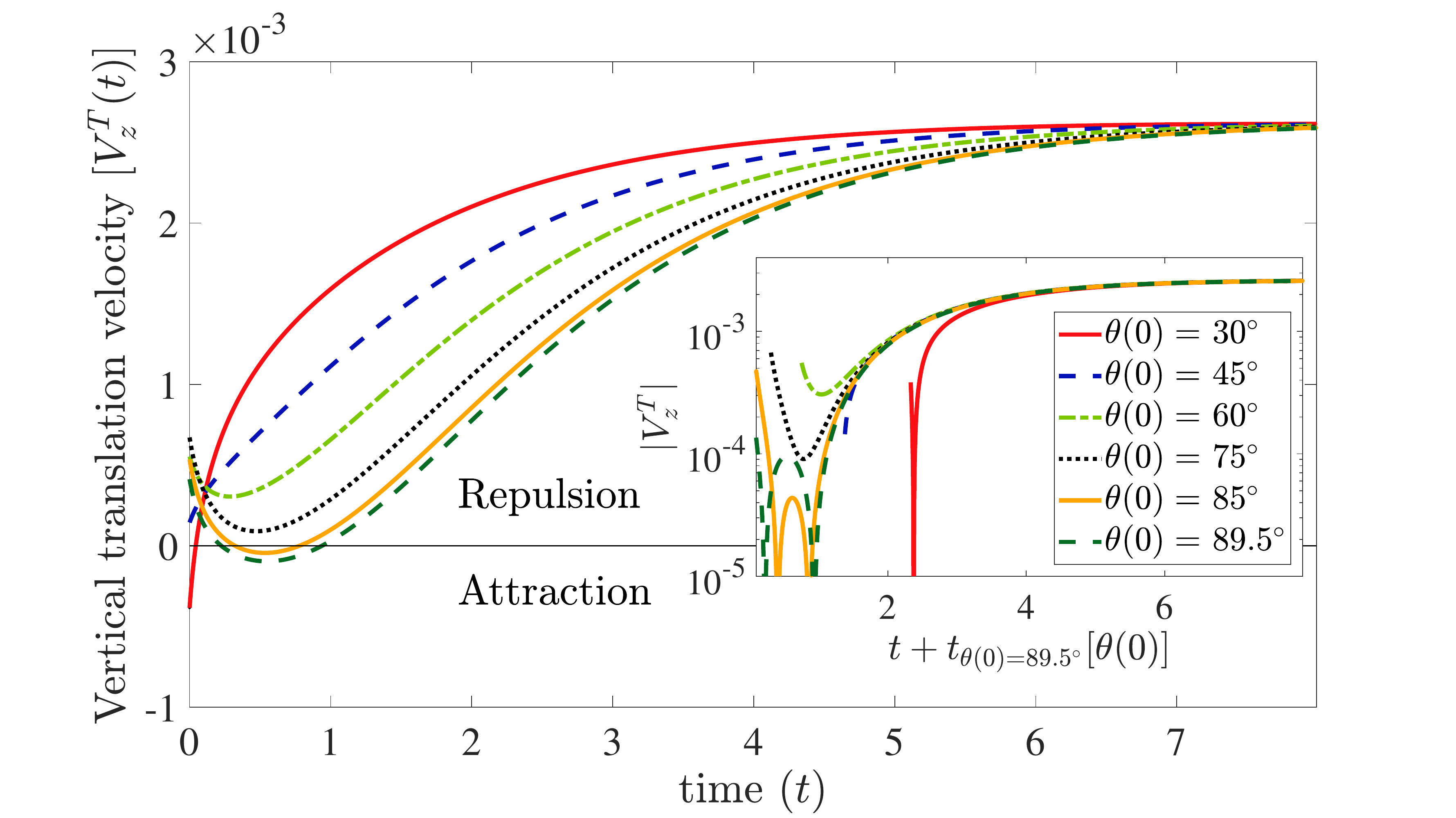}
        \caption{}
    \end{subfigure}    
    \caption{(a) The rotation rate $\dot{p}_z$ plotted as a function of $p_z (\equiv \cos\theta)$. (b) The vertical component of the translation velocity $V_z^T(t)$ of pushers plotted as a function of  time, for a range of initial swimmer orientation relative to the $r_z$ axis $\theta(0) =$ $30^\circ$, $45^\circ$, $60^\circ$, $75^\circ$, $85^\circ$ and $89.5^\circ$. In the inset, $V_z^T$ is shifted by the time it takes the $\theta(0)=89.5^\circ$ curve to attain the corresponding $\theta(0)$. In all the plots $\lambda=$ 0.5, $\kappa = 10$, $\Gamma = 1$ and $\eta_1 V_s L^2/\kappa_\beta =1$.}
\label{fig:arbitrary_rotation}
\end{figure}

As in the case of swimmer translation, the rotation rate has two parts; one from the time dependent interface deformation and the other from the instantaneous Stokes flow field. However, because the time dependent interface deformation term in $\dot{p}_z$ is smaller than the instantaneous term, the rotation rate is dominated by the latter. This quasi-steady behavior allows us to construct a phase portrait in orientation space, as shown in Fig. \ref{fig:arbitrary_rotation}a \cite{strogatz2015nonlinear}.  
For pushers, only trajectories in the first and third quadrants are admissible and hence any initial orientation $p_z\neq 0$ approaches $p_z= \pm 1$ depending on whether its directed away from ($p_z\to 1$ or $\theta\to 0$) or towards the interface ($p_z\to -1$ or $\theta\to 180^\circ$). Now, from the analysis of swimmers parallel to the interface in Sec. \ref{subsec_swimmers_parallel_results}, we know that $\dot{p}_z = 0$, for $p_z=0$. Therefore, $p_z = 0$ is an unstable fixed point for pusher, and hence a stable fixed point for pullers. Similar rotational preferences have been reported by \citet{shaik_ardekani_2017} for spherical squirmers near a deformable interface.

In Fig. \ref{fig:arbitrary_rotation}b the time dependence of the vertical component of the swimmer translation velocity $V_z^T$ is shown for pushers starting from different initial orientations $\theta(0)$. When a swimmer is nearly aligned with the interface ($\theta(0) = 89.5^\circ$), $V_z^T$ decreases until $t\sim O(1)$, owing to the dominant contribution from the parallel configuration, as anticipated from Sec. \ref{subsec_swimmers_parallel_results}. On the other hand, $V_z^T$ grows at large times, owing to the dominant contribution of the orthogonal configuration as $\theta(t)\to 0$. Therefore, the translation velocity $V_z^T$ of swimmers nearly aligned with the interface goes through a minimum, which is negative for nearly parallel swimmers (see curves of $\theta(0) =$ $85^\circ$ and $89.5^\circ$ ) and positive for those that start at a smaller initial orientation, say $\theta(0) \lesssim 75^\circ$ and $\gtrsim 45^\circ$. At $\theta(0)\lesssim 45^\circ$, $V_z^T$ only increases in time with a $\theta(0)$ dependent $t=0$ intercept. Therefore, the early time trajectories are qualitatively and quantitatively dependent upon $\theta(0)$ but all of the curves collapse at late times.

\begin{figure}
    \centering
    \begin{subfigure}[t]{0.495\textwidth}
        \centering
        \includegraphics[width=\textwidth]{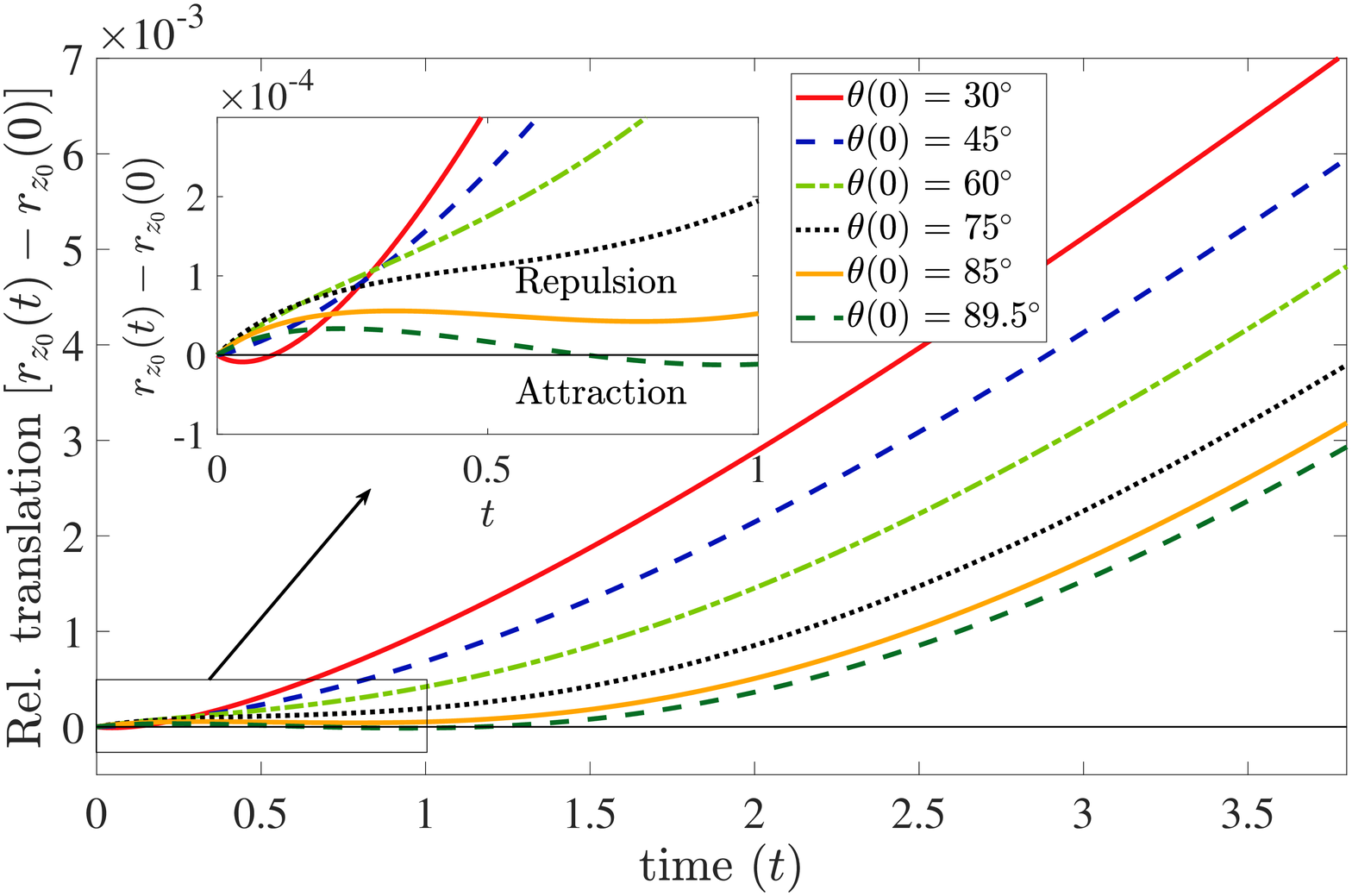}
        \caption{}
    \end{subfigure}%
    ~
    \begin{subfigure}[t]{0.495\textwidth}
        \centering
        \includegraphics[width=\textwidth]{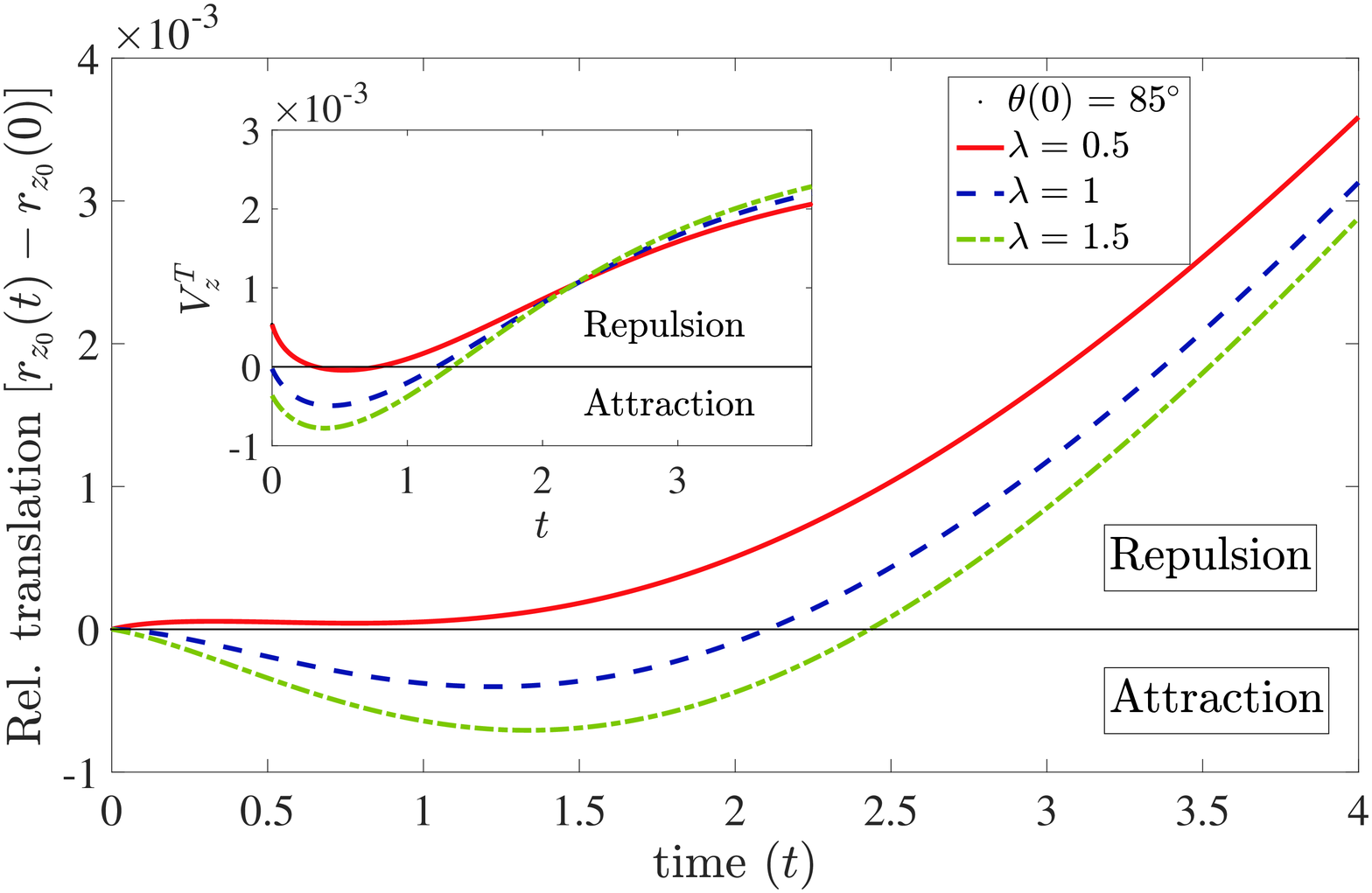}
        \caption{}
    \end{subfigure}    
    \caption{(a) The relative vertical swimmer trajectory $r_{z_0}(t)-r_{z_0}(0)$ of pushers plotted as a function of time for a range of initial swimmer orientation relative to the $r_z$ axis $\theta(0) =$ $30^\circ$, $45^\circ$, $60^\circ$, $75^\circ$, $85^\circ$ and $89.5^\circ$. The viscosity ratio is $\lambda = 0.5$, and the inset is a zoomed version of the abscissa for $t\in [0, 1]$. (b) The relative vertical swimmer trajectory $r_{z_0}(t)-r_{z_0}(0)$ of pushers plotted as a function of  time, for the viscosity ratios $\lambda=$ 0.5, 1 and 1.5. In the inset the corresponding vertical swimmer translation velocity $V_z^T$ of pushers plotted as a function of time. In all the plots $\kappa = 10$, $\Gamma = 1$ and $\eta_1 V_s L^2/\kappa_\beta =1$.}
\label{fig:arbitrary_VzT_rz0}
\end{figure}

Owing to the sensitive dependence of $V_z^T(t)$ on $\theta(0)$ at short times, the swimmer migration has a rich set of trajectories up to times of $O(1)$. In Fig. \ref{fig:arbitrary_VzT_rz0}a, we plot the relative vertical swimmer trajectories $r_{z_0}(t)-r_{z_0}(0)$ for the same values of $\theta(0)$.  The inset shows short time oscillations for $\theta(0) =$ $85^\circ$ and $89.5^\circ$, the latter of which briefly becomes weakly attractive, whereas the curvature gradually changes as $\theta(0)$ decreases. At long times the slopes asymptote for all $\theta(0)$ and the overall displacement increases as $\theta(0)$ decreases.

In Fig. \ref{fig:arbitrary_VzT_rz0}b, we plot the relative vertical swimmer trajectory $r_{z_0}(t)-r_{z_0}(0)$, for three values of the viscosity ratio $\lambda =$ 0.5, 1 and 1.5 for a pusher that starts off initially almost parallel to the interface ($\theta = 85^\circ$). Although the three trajectories are qualitatively similar, here a swimmer exhibits regimes of attraction and repulsion for $\lambda\geq 1$. The underlying principle for such a response can be explained as follows. For swimmers that start off nearly aligned with the interface, at short times $V_z^T$ is positive for $\lambda <1$, zero for $\lambda=1$ (no contribution from instantaneous component), and negative for $\lambda > 1$. As time evolves the swimmer rotates to an orthogonal configuration and $V_z^T$ increases, independent of $\lambda$. In turn, $V_z^T (t=0)$ of a pusher nearly aligned to the interface positive, zero or negative if $\lambda <1$, $\lambda=1$ or $\lambda\geq 1$ respectively. Therefore, for $t< O(1)$, a swimmer will always have an attractive component for $\lambda\geq 1$, and $r_{z_0}(t)-r_{z_0}(0)<0$.  

\begin{figure}
    \centering
    \includegraphics[width=0.7\textwidth]{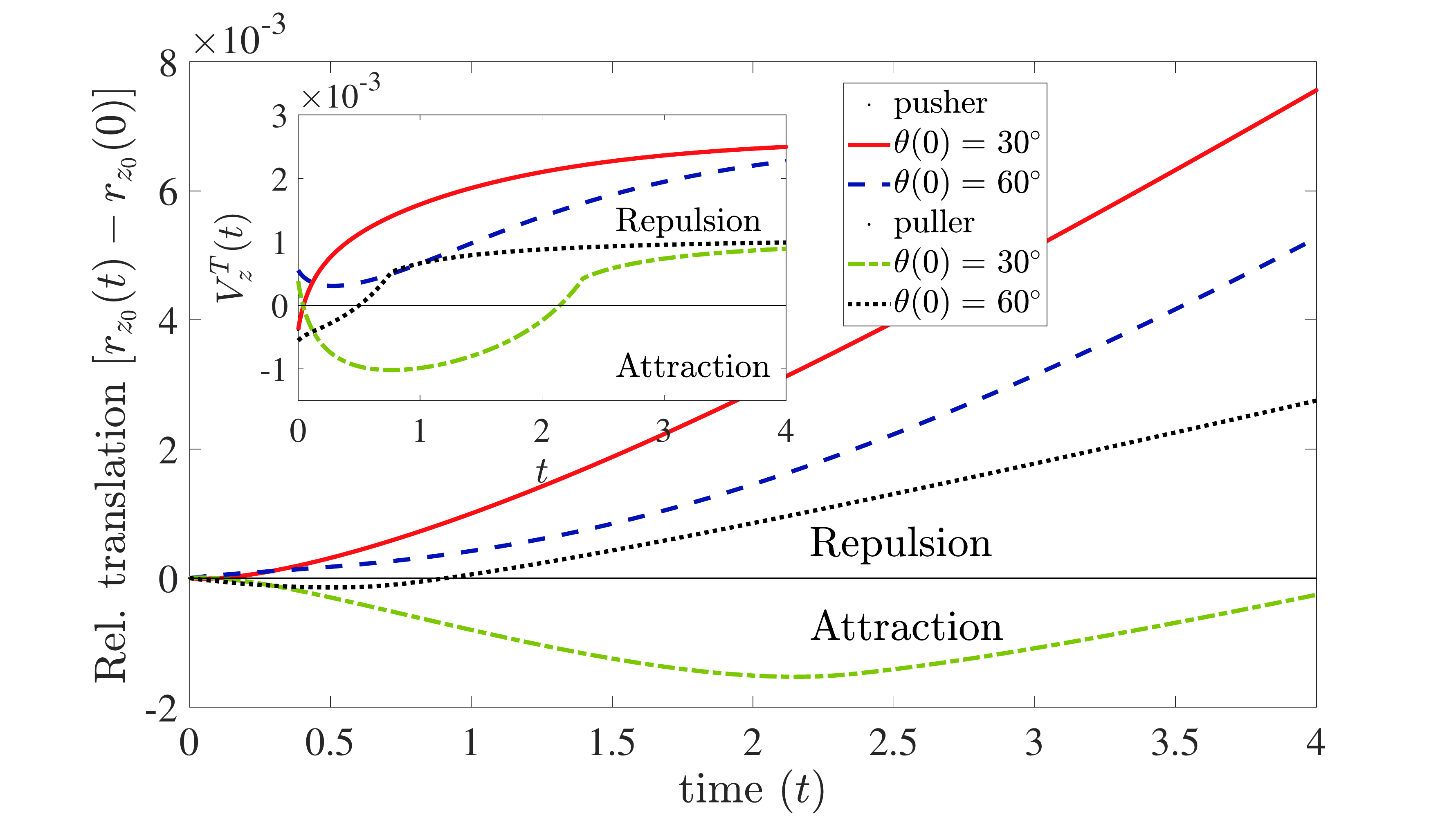} 
    \caption{The relative vertical swimmer trajectory $r_{z_0}(t)-r_{z_0}(0)$ of pushers and pullers plotted as a function of  time, initial orientations $\theta(0) = 30^\circ$ and $60^\circ$. In the insets, the corresponding $V_z^T(t)$ plotted as a function of time. In all the plots $\lambda = 0.5$, $\kappa = 10$, $\Gamma = 1$ and $\eta_1 V_s L^2/\kappa_\beta =1$.}
\label{fig:swim_push_pull}
\end{figure}

We now understand that hydrodynamic interactions with a deformable interface lead to pushers reorienting into an orthogonal configuration, pullers reorient into a parallel configuration. Therefore, unlike their response to a rigid boundary, both swimmer types seek a configuration that drives them away from the interface. As shown in Fig. \ref{fig:swim_push_pull}, pushers and pullers that start at the same location will eventually segregate spatially, independent of their initial orientation but to a degree that increases as $\theta(0)$ decreases. Note that $\bm{p}=0$ is a stable configuration for pushers only the absence of an imposed flow, which may be present in many microswimmer settings, as we discuss in Sec. \ref{sec_conclusion}.

\section{\label{sec_swimmers_confined_results} Microswimmers confined between a rigid boundary and an interface - Swimming parallel to boundaries }

\begin{figure}
\includegraphics[width=0.5\textwidth]{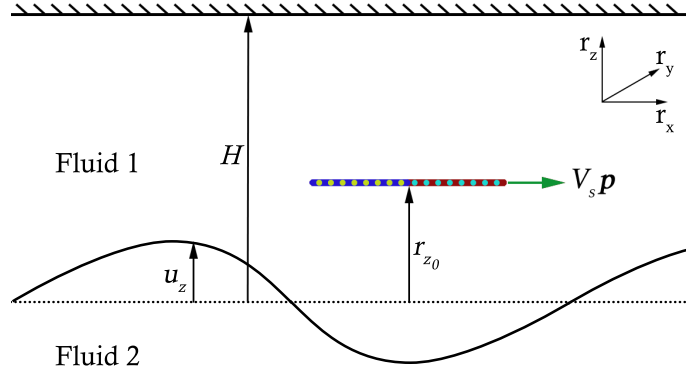}
\caption{A sectional schematic representation of a slender swimmer translating with speed $V_s$ along its director vector $\bm{p}$ in a channel with a deformable interface below and a rigid boundary on top. The distance between the plane of the undeformed interface and the rigid boundary is $H$. The circles along the axial length of the fore-aft symmetric swimmer represent a line distribution of stokeslets characterizing the head and tail for pushers. The disturbance flow field generated by the swimming motion deforms the interface, and $u_z$ is the interface deformation (solid gray line) relative to its initially flat undeformed (dotted line) state $z_0$.}
\label{fig1b}
\end{figure}

Here, we generalize our approach to treat the coupled hydrodynamics of a swimmer confined between a rigid boundary and an underlying deformable interface (Fig. \ref{fig1b}). We note that in this configuration the rotation rate $\dot{\bm{p}}=0$, similar to a swimmer oriented parallel to a single deformable interface (see Sec. \ref{subsec_swimmers_parallel_results}), or to a single rigid boundary \cite{berke_2008}. Therefore, the hydrodynamics is characterized by the interface deformation $u_z$ and the vertical component of the translation velocity $V_z^T$. It is convenient to choose a reference frame moving with the swimmer but shifted to the plane of the undeformed interface, and hence the definition of $r_{z_0}$ used here is the opposite of that used for a swimmer near a single interface in Sec. \ref{sec_deformation} (see the direction of arrows in Figs. \ref{fig1} and \ref{fig1b}).

The equation for the Fourier transformed interface deformation $\hat{u}_z$ is
\begin{eqnarray}
\frac{\text{d} \hat{u}_z}{\text{d} t} &+& \frac{\pi k }{\mathcal{E}} \left(4\pi^2 k ^2 + \Gamma\right) \big[1+\lambda - (1-\lambda)\exp(-8\pi k H) - 2 (\lambda + 4\pi k H + 8\pi^2 k^2 H^2 \lambda) \exp(-4\pi k H) \big]\hat{u}_z \nonumber\\ &&  =\left(\frac{\eta_1 L^2 V_s}{\kappa_\beta}\right)\left(\frac{D}{\pi k \ln\kappa}\right)\sin^2\left(\frac{\pi}{2} k p_l\right) \frac{1}{\mathcal{E}} \bigg[  (1+\lambda)r_{z_0} \exp(-2\pi k r_{z_0}) - (1-\lambda)r_{z_0} \exp(-2\pi k(4H-r_{z_0}))\nonumber\\ && \quad +\; (1+\lambda)(-r_{z_0} - 4\pi k H r_{z_0} + 4\pi k H^2)\exp(-2\pi k (2H-r_{z_0})\nonumber\\ && \quad +\; (1-\lambda)(r_{z_0} - 4\pi k H r_{z_0} + 4\pi k H^2)\exp(-2\pi k (2H+r_{z_0}) \bigg],
\label{eq:Interface_1rigid}
\end{eqnarray}
and the expression for $V_z^T$ is
\begin{eqnarray}
V_z^T \equiv \frac{\mathrm{d} r_{z_0}}{\mathrm{d} t} &=& -\frac{2 D}{\ln\kappa}\int \mathrm{d}\bm{k}\frac{\sin(\pi k p_l)}{\pi k p_l}\frac{1}{\mathcal{E}}\sin^2\left(\frac{\pi}{2}k p_l\right)\bigg[(1-\lambda^2)r_{z_0}^2\big[\exp(-4\pi k r_{z_0}) + \exp(-4\pi k (2H-r_{z_0}))\big] \nonumber \\&& \qquad + (H-r_{z_0})^2 \big[ (1+\lambda)^2\exp(-4\pi k (H - r_{z_0})) + (1-\lambda)^2 \exp(-4\pi k (H + r_{z_0}))\big] \nonumber \\ \quad &&\qquad + 2H(1-\lambda^2)(H-2 r_{z_0})\exp(-4\pi k H)  \bigg]  \nonumber\\ &&  + \left(\frac{\kappa_\beta}{\eta_1 V_s L^2}\right) \int \mathrm{d}\bm{k}\frac{\sin(\pi k p_l)}{\pi k p_l} \pi k \left(4\pi^2 k^2 + \Gamma\right)\frac{\hat{u}_z}{\mathcal{E}} \bigg[-(1+\lambda)(1+2\pi k r_{z_0})\exp(-2\pi k r_{z_0})\nonumber\\ &&\qquad + (1-\lambda)(1-2\pi k r_{z_0})\exp(-2\pi k(4 H - r_{z_0}))\nonumber\\ &&\qquad -(1-\lambda)\big[(1-4\pi k H)(1+2\pi k r_{z_0}) + 8\pi^2 k^2 H^2\big]\exp(-2\pi k (2H+ r_{z_0}))\nonumber\\ && \qquad + (1+\lambda)\big[(1+4\pi k H)(1-2\pi k r_{z_0}) + 8\pi^2 k^2 H^2\big]\exp(-2\pi k (2H- r_{z_0})) \bigg],
\label{eq:VzT_parallel_1rigid}
\end{eqnarray}
where, $\mathcal{E}$ in both Eqs. \eqref{eq:Interface_1rigid} and \eqref{eq:VzT_parallel_1rigid} is
\begin{eqnarray}
 \mathcal{E} = (1+\lambda)^2 + (1-\lambda)^2\exp(-8\pi k H)  - 2(1-\lambda)^2(1+8\pi^2 k^2 H^2)\exp(-4\pi k H).
\label{eq:VzT_Denom}
\end{eqnarray}
We note that the additional $H$-dependent terms in Eq. \eqref{eq:Interface_1rigid} for the interface deformation are at least exponentially smaller by a factor $\exp(-2\pi k (2H-r_{z_0})$. Unless $H\to r_{z_0}$ (extreme confinement), the contributions from these terms remains small, apart from there being an $H$-dependent amplitude, as the terms with the smaller exponents are proportional to $1-\lambda$. Hence, varying $\lambda$ about unity does not yield a qualitatively different deformation than that of a swimmer parallel a single interface. Therefore, in the following we focus only on the change in the swimmer translation effected by the coupled hydrodynamics between the swimmer and the boundaries.

\begin{figure}
    \centering
    \begin{subfigure}[t]{0.495\textwidth}
        \centering
        \includegraphics[width=\textwidth]{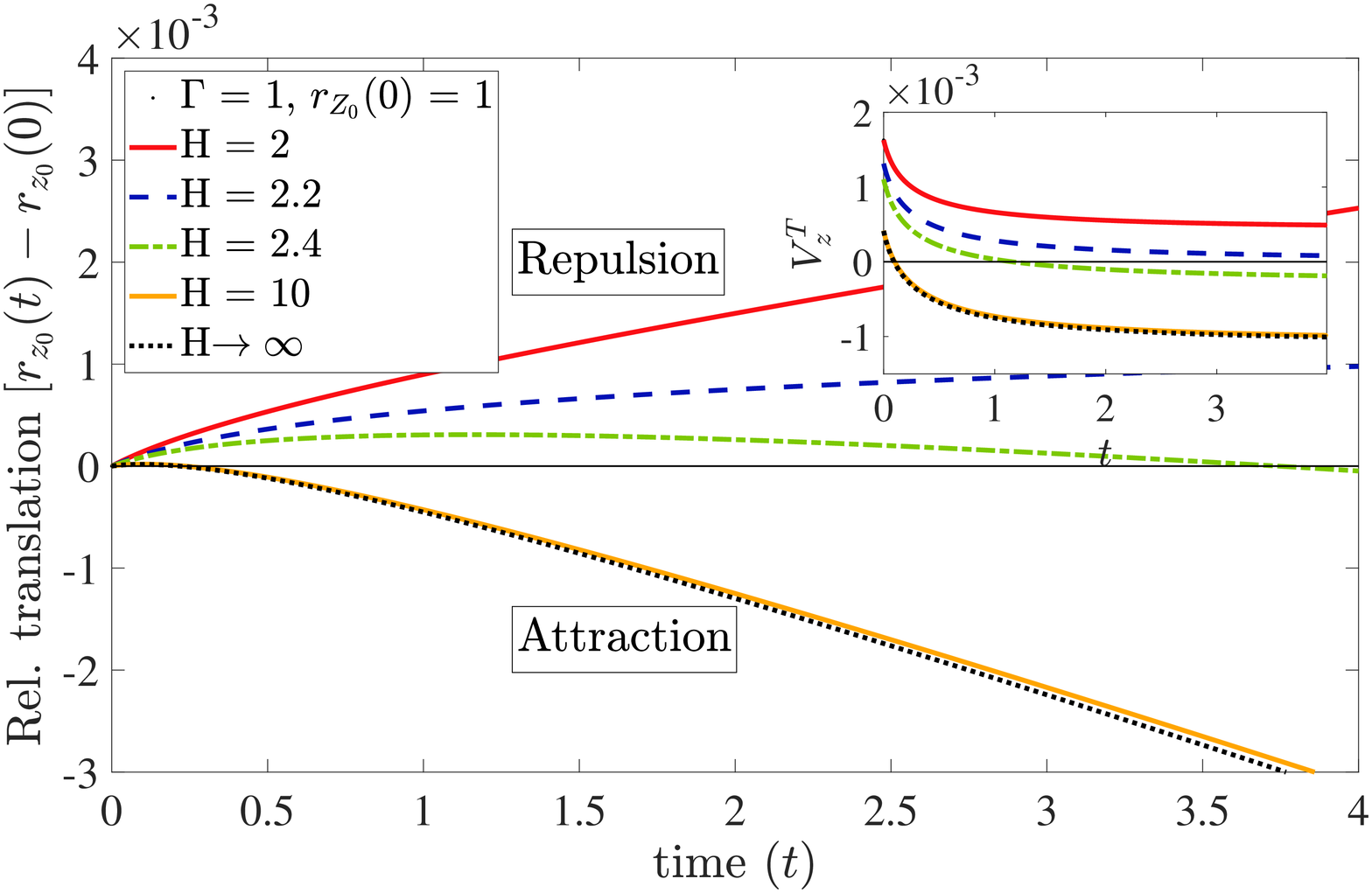}
        \caption{}
    \end{subfigure}%
    ~
    \begin{subfigure}[t]{0.495\textwidth}
        \centering
        \includegraphics[width=\textwidth]{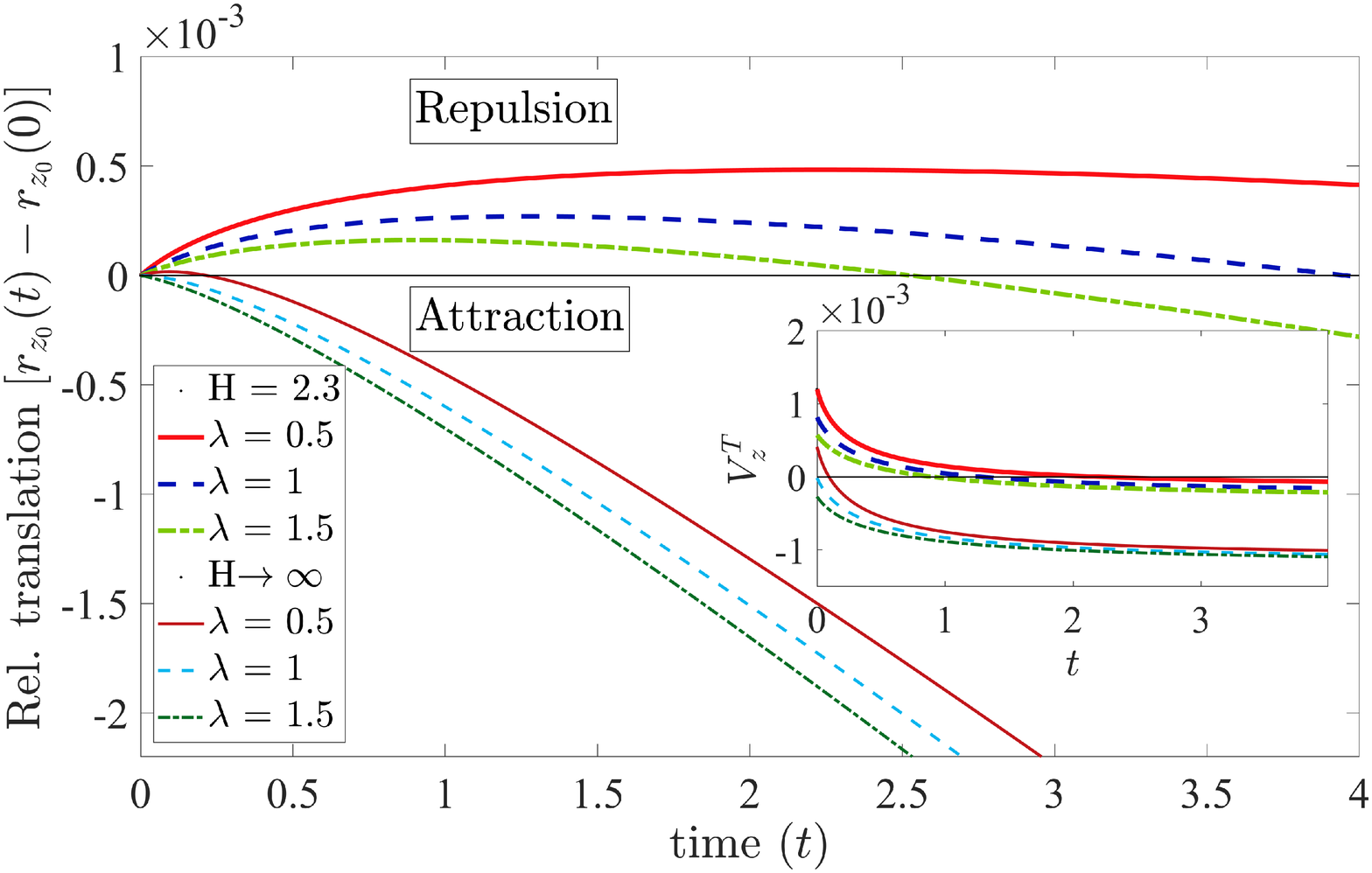}
        \caption{}
    \end{subfigure}   
    \caption{(a) The relative swimmer trajectory $r_{z_0}(t)-r_{z_0}(0)$ of a pusher plotted as a function of time for different distance $H$ between the interface and the rigid boundary: $H =$ 2, 2.2, 2.4, 10, when the viscosity ratio is $\lambda = 0.5$. (b) $r_{z_0}(t)-r_{z_0}(0)$ of a pusher plotted as a function of time for $\lambda =$ 0.5, 1 and 1.5, and fixing $H$ = 2.3. The insets contains the corresponding vertical component of the translation velocity $V_z^T\equiv \mathrm{d} r_{z_0}/\mathrm{d}t$. The \enquote{repulsion} and \enquote{attraction} are to be interpreted relative to the interface. In both figures, the initial swimmer distance to the interface is $\left\vert r_{z_0}\right\vert = 1$ and the ratio of surface tension to bending stress is $\Gamma =$ 1. The attraction and repulsion are specified relative to the deformable interface.}
\label{fig:rel_swim_traj_1_rigid}
\end{figure}

In Fig. \ref{fig:rel_swim_traj_1_rigid}a we plot the relative swimmer trajectory $r_{z_0}(t)-r_{z_0}(0)$ as a function of the distance between the boundaries $H$,
for $r_{z_0}(0) = 1$ and $\lambda = 0.5$.  In the limit $H\gg 1$, we recover the trajectory corresponding to that of a swimmer in the vicinity of a single interface as discussed in Sec. \ref{subsec_swimmers_parallel_results}. Upon increased confinement by reducing $H$ and increasing confinement, the residence time of the swimmer in the repulsive state increases. When the swimmer begins at the midpoint, it always moves towards the rigid boundary. This is also shown in the inset of Fig. \ref{fig:rel_swim_traj_1_rigid}a, where for $H=2$, $V_z^T$ does not undergo a zero-crossing in finite time. Physically, a rigid boundary provides an infinite resistance to both bending and shear, and thus facilitates a stronger attraction relative to an interface that only supports finite bending \cite{abdallah2019}. In other words, the flow field induced by the rigid surface dominates that of the deformable interface.

In Fig. \ref{fig:rel_swim_traj_1_rigid}b we plot the relative swimmer trajectory $r_{z_0}(t)-r_{z_0}(0)$ for three different viscosity ratios $\lambda$ = 0.5, 1, 1.5, with the distance between the interface and the rigid boundary fixed to $H = 2.3$. Interestingly, unlike a swimmer in the vicinity of a standalone interface, the swimmer experiences a transient attraction and repulsion for all values of $\lambda$. In Sec. \ref{subsec_swimmers_parallel_results} we explained that for $\lambda<1$ the swimmer translation had both repulsive and attractive component due to the difference in the instantaneous contribution, dominant at short times, and the interface deformation dependent contributions of $V_z^T$, dominant at long times (see Eq. \eqref{eq:VzT_parallel}). Here, however, the rigid boundary creates repulsion, through an $H$-dependent positive shift of the instantaneous component of $V_z^T$, the largest contribution being proportional to $(1+\lambda)^2$, and thus positive for any $\lambda$. As the confinement increases, the contribution of this term increases, and eventually for $H\gtrsim 2 r_{z_0}$, the swimmer only moves towards the rigid boundary. Although not shown, the dependence on the ratio of surface tension to bending stress, $\Gamma$, remains the same as that for a swimmer near a single deformable interface, and seen in Eqs. \eqref{eq:Interface_1rigid} and \eqref{eq:VzT_parallel_1rigid}.

\section{\label{sec_conclusion} Summary and Conclusion}

We have investigated the coupled hydrodynamics between finite sized orientable swimmers and a deformable interface. By treating the swimmers as slender bodies, we have gone beyond a far-field picture, solving the hydrodynamics appropriate at $O(1)$ swimmer lengths from the interface. Moreover, we predict the hydrodynamics for a robust range of the ratio of the viscous stress to bending stress, $\eta_1 V_s L^2/\kappa_\beta$, and the capillary number, $\eta_1 V_s/\gamma$. Our analysis reveals that the swimmer orientation plays a crucial role in the response of both the swimmer and deformable interface. Because parallel and perpendicular orientations exhibit differing overall trajectories, when swimmers take an arbitrary orientations we find a rich dynamical behavior. Importantly, given that an arbitrarily oriented pusher (puller) preferentially rotates to a perpendicular (parallel) orientation, the migration pattern for both swimmer types remains the same. However, given the difference in the migration speed in the two orientations, a pusher and puller that start from the same location with any orientation will spatially segregate.

When confined to swimming between a rigid and a deformable boundary, pushers closer to the latter interface experience repulsion that is extended to time scales longer than the elastic response time. Such confinement, between rigid and soft boundaries, provide new controls on swimmer migration.  For example, the different rotational preferences near deformable versus rigid boundaries provides one such confinement-dependent control of swimmers.

In general, the dynamics of swimmer translation is extremely sensitive to the distance from the interface. Even an $O(1)$ increase in this distance results in a large slowdown in the migration towards or away from the interface. The interfacial properties play a crucial role in migration, second only to the distance from the interface. Therefore, it is possible that two scenarios in which swimmers have different distances from the interface display similar migration patterns. Owing to the strong sensitivity of swimmer migration to distance from the interface, even a small change in the latter would require an $O(1)$ change in the interface properties to obtain a nearly matching migration pattern. This is particularly relevant from a practical standpoint, because interfacial properties are not necessarily spatially homogeneous. In consequence, swimmers in a suspension at different distances from the interface can, in principle, redistribute in a spatially similar manner.

We find that the viscosity ratio plays a controlling role in whether a swimmer is attracted or repelled from the interface. Indeed, viscosity variations may be dictate the migration of planktonic biota in marine ecology, by modifying both swimming and the nutrient dispersion \cite{Stocker_2012, Simspon_2015, Simpson_2021}. On the other hand, whereas interfacial deformation is clearly controlled by the relative importance of bending elasticity and surface tension, it also depends on the swimmer orientation and distance from the interface.  However, independent of the swimmer orientation, the interfacial deformation in the interfacial plane extends well beyond distances of order the swimmer size $L$.  Moreover,  the far-field scaling of the deformation in the interfacial plane remains $O\left(r_{\Vert}\right)^{-3}$ for swimmers oriented parallel or perpendicular to the interface. Even in the dilute limit, such far-field scaling can have important consequences in analyzing the deformation due to a suspension of microswimmers near deformable interfaces. This is because inclusion of the collective contribution from the swimmers upon the interface involves an integral over the domain volume, which decays as $O\left(r_{\Vert}\right)^{-3}$, implying the potential for long-ranged radial contributions.

It is important to note that time scales short relative to the elastic response time, $t_{c 1}\left(=\eta_{1} L^{3} / \kappa_{\beta}\right)$, are still usually very long compared to the time scales of swimming. For typical soft interfaces, the bending modulus $\kappa_{\beta} \approx 10^{-19}$ $J$ \cite{Libchaber_1997, freund2014numerical}, so if one considers a viscosity approximately that of water, $\eta_{1} \approx 10^{-3}$ $N s/m^2$, and swimmers of length $L \approx 10$ $\mu m$ (cell body+flagellar bundle combined), then $t_{c 1} \approx 10$ $s$. In general, microscopic swimmers also respond to external cues, such as an externally imposed flow \cite{Marcos4780, rusconi2014bacterial, costanzo2014motility, stocker_2015_1, yeomans2016, Wioland_2016, Lauga_annRev_2016, vennamneni2020shear} or some form of chemical actuation \cite{adler_1972, pedley_hill_kessler_1988, vincent_hill_1996, Cisneros2010, PAINTER2011363, subramanian_2011, bharat_Saintillan_2012, desai_ardikani_2019}, that can influence their migration on such long time scales. Typically, even in the absence of such an external driving force, the reorientations take seconds, as most microscopic swimmers have an inherent mechanism allowing them to change their orientation \cite{berg2008coli, koch2011collective, Lauga_annRev_2016, saintillan_rev_2019}. We note that when $\lambda<1$, a swimmer aligned with the interface might only sense either a repulsion (pusher) or an attraction (puller), before reorienting itself. Therefore, the nature of the short time dynamics shown in Fig. \ref{fig:swim_traj} implies the possibility of pushers migrating away from the interface on average. A similar argument can be made for other swimmer orientations.

Although we have considered the motion of a single swimmer near a deformable interface and confined between a deformable interface and a rigid boundary, one can readily extend this study to a dilute suspension of non-interacting swimmers. This requires a more careful consideration of the single-swimmer statistics developed here, as the expressions for the instantaneous field variables depend on the swimmer configuration. Moreover, given the complex dynamics involved, deriving a position-orientation space swimmer distribution function is challenging. Finally, a host of new dynamical processes will be revealed when extending this approach beyond the linear regime.

\begin{acknowledgments}
The authors gratefully acknowledge support from the Swedish Research Council, under grant no. 638-2013-9243. Nordita is partially supported by Nordforsk.
\end{acknowledgments}

\appendix
\section{\label{sec_appendix_A}Two-dimensional Fourier transform summary - Swimmers parallel to interface}
Here we outline the two-dimensional Fourier transform technique used in the main text to solve the governing equations and boundary conditions for a slender swimmer translating parallel to a deformable interface. For a more detailed description of the method, the reader is directed to \citet{bickel2007} and \citet{abdallah2016}.

\subsection{\label{subsec_appendix_A_governing_eqns}Formulating the  governing equations in Fourier space}
For a swimmer translating parallel to the interface, the Stokes equations and the continuity equation in fluid region 1 given in Eqs. \eqref{eq:stokes_continuity_nd_a} and \eqref{eq:stokes_continuity_nd_b} are:
\begin{subequations}
\label{eq:stokes_fluid_1_nd_FT}
	\begin{eqnarray}
		-2\pi \iu k_x \hat{P}_1 + \frac{\partial^2 \hat{v}_{1x}}{\partial r_z^2} - 4\pi^2 k^2 \hat{v}_{1x} &=& \frac{2 D\, p_x}{\pi \iu \bm{k}\cdot\bm{p}\ln\kappa}\delta(r_z)\sin^2\left(\frac{\pi}{2}\bm{k}\cdot\bm{p}\right), \label{eq:stokes_fluid_1_nd_FT_a}\\
		-2\pi \iu k_y \hat{P}_1 + \frac{\partial^2 \hat{v}_{1y}}{\partial r_z^2} - 4\pi^2 k^2 \hat{v}_{1y} &=& \frac{2 D\, p_y}{\pi \iu \bm{k}\cdot\bm{p}\ln\kappa}\delta(r_z)\sin^2\left(\frac{\pi}{2}\bm{k}\cdot\bm{p}\right), \label{eq:stokes_fluid_1_nd_FT_b}\\
		- \frac{\partial \hat{P}_1}{\partial r_z} + \frac{\partial^2 \hat{v}_{1z}}{\partial r_z^2} - 4\pi^2 k^2 \hat{v}_{1z} &=& 0 ,\;\; \text{and} \label{eq:stokes_fluid_1_nd_FT_c}\\		
		2\pi \iu (k_x \hat{v}_{1x} + k_y \hat{v}_{1y}) + \frac{\partial \hat{v}_{1z}}{\partial r_z} &=& 0.		\label{eq:stokes_fluid_1_nd_FT_d}
	\end{eqnarray}
\end{subequations}
In fluid region 2, the right-hand side of the corresponding Stokes equations for $\hat{\bm{v}}_{2}$, $P_2$ vanishes. It is convenient to solve this system of equations in Fourier space in a coordinate system aligned with the wavevector $\bm{k}$ \cite{bickel2007, abdallah2016}. We define the longitudinal and transverse coordinate systems $\hat{\bm{l}}$ and $\hat{\bm{t}}$, as
\begin{subequations}
\label{eq:lhat_that}
	\begin{eqnarray}
		\hat{\bm{l}} &=& \frac{k_x}{k}\hat{\bm{x}} + \frac{k_y}{k}\hat{\bm{y}} ,\;\; \text{and} \\
		\hat{\bm{t}} &=& \frac{k_y}{k}\hat{\bm{x}} - \frac{k_x}{k}\hat{\bm{y}},
	\end{eqnarray}
\end{subequations}
respectively. This choice of coordinates has a few advantages. First, it is evident from Eq. \eqref{eq:stokes_fluid_1_nd_FT_d} that the continuity equation provides a direct relation between the longitudinal and the $z$-component of the fluid velocity. Second, as will be described below, the pressure gradient term drops out of the $\hat{t}$-component of the Fourier transformed Stokes equation. Lastly, one can obtain a single $4^{th}$-order differential equation for the $z$-component of the fluid velocity, again without the pressure gradient term, thereby decoupling the flow-field field variables. In the ($\hat{\bm{l}}, \hat{\bm{t}}, \bm{z}$) coordinates Eqs. \eqref{eq:stokes_fluid_1_nd_FT_a}-\eqref{eq:stokes_fluid_1_nd_FT_d} become
\begin{subequations}
\label{eq:stokes_fluid_1_nd_FT_lt}
	\begin{eqnarray}
		-2\pi \iu k \hat{P}_1 + \frac{\partial^2 \hat{v}_{1l}}{\partial r_z^2} - 4\pi^2 k^2 \hat{v}_{1l} &=& \frac{2 D}{\pi \iu k\ln\kappa}\delta(r_z)\sin^2\left(\frac{\pi}{2}k p_l \right), \label{eq:stokes_fluid_1_nd_FT_lt_a} \\
		\frac{\partial^2 \hat{v}_{1t}}{\partial r_z^2} - 4\pi^2 k^2 \hat{v}_{1t} &=& \frac{2 D\, p_t}{\pi \iu k p_l\ln\kappa}\delta(r_z)\sin^2\left(\frac{\pi}{2}k p_l\right), \label{eq:stokes_fluid_1_nd_FT_lt_b} \\
		- \frac{\partial \hat{P}_1}{\partial r_z} + \frac{\partial^2 \hat{v}_{1z}}{\partial r_z^2} - 4\pi^2 k^2 \hat{v}_{1z} &=& 0 ,\;\; \text{and} \label{eq:stokes_fluid_1_nd_FT_lt_c}\\		
		2\pi \iu k \hat{v}_{1l} + \frac{\partial \hat{v}_{1z}}{\partial r_z} &=& 0,		\label{eq:stokes_fluid_1_nd_FT_lt_d}
	\end{eqnarray}
\end{subequations}
where we note that the $z$-component is the same as Eq. \eqref{eq:stokes_fluid_1_nd_FT_c}. In Eqs. \eqref{eq:stokes_fluid_1_nd_FT_lt_a} and \eqref{eq:stokes_fluid_1_nd_FT_lt_b}, $p_l\equiv \bm{p}\cdot \hat{\bm{l}}$ is the component of the orientation vector parallel to the wavevector and $p_t \equiv \bm{p}\cdot\hat{\bm{t}}$ is that transverse to the wavevector. Using Eqs. \eqref{eq:stokes_fluid_1_nd_FT_lt_a} and \eqref{eq:stokes_fluid_1_nd_FT_lt_c} to eliminate the pressure term, and Eq. \eqref{eq:stokes_fluid_1_nd_FT_lt_d} to represent $\hat{v}_{1l}$ in terms of $\hat{v}_{1z}$, one arrives at the following differential equation to solve for $\hat{v}_{1z}$:
\begin{equation}
		\frac{\partial^4 \hat{v}_{1z}}{\partial r_z^4} -8\pi^2 k^2 \frac{\partial^2 \hat{v}_{1z}}{\partial r_z^2} + 16\pi^4 k^4 \hat{v}_{1z} = -\frac{4\, D}{\ln\kappa}\delta^\prime(r_z)\sin^2\left(\frac{\pi}{2}k p_l\right),	
\label{eq:stokes_fluid_1_nd_FT_lt_2a}
\end{equation}
where $\delta^\prime(r_z)$ is the derivative of $\delta(r_z)$. Similar to the above derivation, in fluid region 2 we have the following set of equations:
\begin{subequations}
\label{eq:stokes_fluid_1_nd_FT_lt_F2}
	\begin{eqnarray}
		\frac{\partial^4 \hat{v}_{2z}}{\partial r_z^4} -8\pi^2 k^2 \frac{\partial^2 \hat{v}_{2z}}{\partial r_z^2} + 16\pi^4 k^4 \hat{v}_{2z} &=& 0, \label{eq:stokes_fluid_1_nd_FT_lt_F2_a}\\
		\frac{\partial^2 \hat{v}_{2t}}{\partial r_z^2} - 4\pi^2 k^2 \hat{v}_{2t} &=& 0 ,\;\;\text{and} \label{eq:stokes_fluid_1_nd_FT_lt_F2_b}\\
		2\pi \iu k \hat{v}_{2l} + \frac{\partial \hat{v}_{2z}}{\partial r_z} &=& 0.		\label{eq:stokes_fluid_1_nd_FT_lt_F2_c}
	\end{eqnarray}
\end{subequations}
Once $\hat{v}_{\alpha z}$ is determined, we can use the continuity equations \eqref{eq:stokes_fluid_1_nd_FT_lt_d} and  \eqref{eq:stokes_fluid_1_nd_FT_lt_F2_c} to determine the longitudinal component $\hat{v}_{\alpha l}$, where $\alpha\in [1, 2]$ refers to the two fluid regions.

\subsection{\label{subsec_appendix_A_bc}Formulating boundary conditions in Fourier space}
The equations for $\hat{v}_{\alpha z}$ and $\hat{v}_{\alpha t}$ are solved subject to the appropriate boundary conditions; the impenetrability and the no-slip conditions at the interface, and the continuity of tangential stress and the normal-stress jump across the interface. The Fourier transformed velocity boundary conditions are:
\begin{subequations}
\label{eq:bc_lt_1}
	\begin{eqnarray}
		\left.\hat{v}_{1 z}\right\vert_{r_{z_0}^+} &=& \left.\hat{v}_{2 z}\right\vert_{r_{z_0}^-}, \label{eq:bc_lt_1_a}\\
		\left.\hat{v}_{1 t}\right\vert_{r_{z_0}^+} &=& \left.\hat{v}_{2 t}\right\vert_{r_{z_0}^-} \;\; \text{and} \label{eq:bc_lt_1_b}\\
		\left.\frac{\partial \hat{v}_{1 z}}{\partial r_z}\right\vert_{r_{z_0}^+} &=& \left.\frac{\partial \hat{v}_{2 z}}{\partial r_z}\right\vert_{r_{z_0}^-},		\label{eq:bc_lt_1_c}
	\end{eqnarray}
\end{subequations}
where we have used the continuity equations \eqref{eq:stokes_fluid_1_nd_FT_lt_d} and \eqref{eq:stokes_fluid_1_nd_FT_lt_F2_c} to arrive at Eq. \eqref{eq:bc_lt_1_c} as an alternative to the no-slip boundary condition.

The continuity of tangential stress given by Eqs. \eqref{eq:bc_nd_c} and \eqref{eq:bc_nd_d} in the main text can be simplified in the $\hat{\bm{l}}$-$\hat{\bm{t}}$ coordinates to yield the following pair of equations for $\hat{v}_{\alpha z}$ and $\hat{v}_{\alpha t}$ at the interface
\begin{subequations}
\label{eq:bc_lt_2}
	\begin{eqnarray}
		\left.\left(\frac{\partial^2 \hat{v}_{1z}}{\partial r_z^2} + 4\pi^2 k^2 \hat{v}_{1z} \right)\right\vert_{r_{z_0}^+} &=& \lambda \left.\left(\frac{\partial^2 \hat{v}_{2z}}{\partial r_z^2} + 4\pi^2 k^2 \hat{v}_{2z} \right)\right\vert_{r_{z_0}^-} \;\;\text{and} \label{eq:bc_lt_2_a}\\
		\left.\frac{\partial \hat{v}_{1t}}{\partial r_z}\right\vert_{r_{z_0}^+} &=& \lambda\left.\frac{\partial \hat{v}_{2t}}{\partial r_z}\right\vert_{r_{z_0}^-}.		\label{eq:bc_lt_2_b}
	\end{eqnarray}
\end{subequations}
The Fourier transformed normal stress boundary condition at the interface give by Eq. \eqref{eq:bc_nd_e} is
\begin{equation}
\label{eq:bc_ft_normal}
		-\left(\left.\hat{P}_1\right\vert_{r_{z_0}^+} - \left.\hat{P}_2\right\vert_{r_{z_0}^-}\right) + 2\left(\left.\frac{\partial \hat{v}_{1z}}{\partial r_z}\right\vert_{r_{z_0}^+}-\lambda \left.\frac{\partial v_{2z}}{\partial r_z}\right\vert_{r_{z_0}^-}\right) = 4\pi^2 k^2\frac{\gamma}{\eta_1 V_s}\hat{u}_z + 16\pi^4 k^4\frac{\kappa_\beta L^2}{\eta_1 V_s}\hat{u}_z.
\end{equation}
Eq. \eqref{eq:bc_ft_normal} can be further simplified by using Eq. \eqref{eq:stokes_fluid_1_nd_FT_c} (and its equivalent equation for $\hat{\bm{v}}_2$ and $P_2$) along with the continuity equations \eqref{eq:stokes_fluid_1_nd_FT_lt_d} and \eqref{eq:stokes_fluid_1_nd_FT_lt_F2_c} to find:
\begin{eqnarray}
\label{eq:bc_ft_normal2}
		\left(\left.\frac{\partial^3 \hat{v}_{1z}}{\partial r_z^3}\right\vert_{r_{z_0}^+}-\lambda\left. \frac{\partial^3 v_{2z}}{\partial r_z^3}\right\vert_{r_{z_0}^-}\right) &+& 12\pi^2 k^2 \left( \left.\frac{\partial \hat{v}_{1z}}{\partial r_z}\right\vert_{r_{z_0}^+}-\lambda \left.\frac{\partial v_{2z}}{\partial r_z}\right\vert_{r_{z_0}^-} \right) = 16\pi^4 k^4 \left(\frac{\kappa_\beta L^2}{\eta_1 V_s}\right) \left(4\pi^2 k^2 + \Gamma \right) \hat{u}_z,
\end{eqnarray}
for the boundary condition of the normal component of the stress without the pressure term; note that $\Gamma\equiv \gamma L^2/\kappa_\beta$ in Eq. \eqref{eq:bc_ft_normal2}.

\subsection{\label{subsec_appendix_A_solution}Solving for the velocity field in Fourier space}
Here, we first solve for the transverse components of the Fourier transformed velocity fields and then the normal and the longitudinal components. A general solution for equations of the form of Eqs. \eqref{eq:stokes_fluid_1_nd_FT_lt_b} and  \eqref{eq:stokes_fluid_1_nd_FT_lt_F2_b} is: $\hat{v}_{\alpha t} = A \exp(\pm 2\pi k r_z)$ \cite[e.g.,][]{bickel2007, bender2013advanced}. In the specific problem considered here, the disturbance flow field must decay in the far field, and hence, the transverse velocity components are
\begin{subequations}
\label{eq:vt}
\begin{eqnarray}
\hat{v}_{1t} &=& 
    \begin{cases}
      A_1 \exp(-2\pi k r_z) ; r_z>0 \\
      A_2 \exp(-2\pi k r_z) + A_3 \exp(2\pi k r_z) ; r_{z_0}<r_z<0 ,\;\; \text{and} \\
    \end{cases} \\    
\hat{v}_{2t} &=& A_4 \exp(2\pi k r_z) ; r_z<r_{z_0}  .  
\end{eqnarray}
\end{subequations}
Now, the right-hand side of Eq. \eqref{eq:stokes_fluid_1_nd_FT_lt_b} is proportional to $\delta(r_z)$, and hence, we seek the Greens function of the differential equation. From Appendix \ref{subsec_appendix_A_bc}, we have two boundary conditions for $\hat{v}_{\alpha t}$, namely, the no-slip condition given by Eq . \eqref{eq:bc_lt_1_b} and the continuity of tangential stress given by Eq. \eqref{eq:bc_lt_2_b}, and four unknown constants $A_1$-$A_4$ to determine. Therefore, we seek two conditions to uniquely determine the constants. To extract additional boundary conditions, we use the continuity properties of the Greens function; (a) about the location of forcing ($r_z=0$), and (b) the first derivative with respect to $r_z$ exhibiting a finite-jump discontinuity about $r_z=0$ \cite{bickel2007, bender2013advanced}. These additional conditions are:
\begin{subequations}
\label{eq:bc_t_3}
	\begin{eqnarray}
		\left.\hat{v}_{1t}\right\vert_{r_z=0^+} &=& \left.\hat{v}_{1t}\right\vert_{r_z=0^-} ,\;\; \text{and} \label{eq:bc_t_3_a}\\
		\left.\frac{\partial \hat{v}_{1t}}{\partial r_z}\right\vert_{r_z=0^+} -  \left.\frac{\partial \hat{v}_{1t}}{\partial r_z}\right\vert_{r_z=0^-} &=& \frac{2 D\, p_t}{\pi \iu k p_l\ln\kappa} \sin^2\left(\frac{\pi}{2}k p_l\right).		\label{eq:bc_t_3_b}
	\end{eqnarray}
\end{subequations}
Using Eqs. \eqref{eq:bc_lt_1}b, \eqref{eq:bc_lt_2_b} and \eqref{eq:bc_t_3}, the four constants are uniquely determined, and we obtain the following expressions for the transverse velocity components in the two fluid region:
\begin{subequations}
\label{eq:vhat_t}
	\begin{eqnarray}
		\hat{v}_{1t} &=& \frac{\iu p_t}{2\pi^2 k^2 p_l \ln\kappa}\sin^2\left(\frac{\pi}{2} k p_l\right) \left[\exp(-2\pi k \vert r_z \vert) + \left(\frac{1-\lambda}{1+\lambda}\right) \exp(4\pi k r_{z_0}) \exp(-2\pi k r_z) \right]\nonumber\\&&; r_z>r_{z_0} ,\;\;\text{and} \\
		\hat{v}_{2t} &=& \frac{\iu p_t}{\pi^2 k^2 p_l (1+\lambda) \ln\kappa}\sin^2\left(\frac{\pi}{2} k p_l\right) \exp(2\pi k r_z); r_z<r_{z_0}.		
	\end{eqnarray}
\end{subequations}

Next, we consider the normal velocity components in Fourier space $\hat{v}_{\alpha z}$. A general solution for the $4^{th}$-order differential equations Eqs. \eqref{eq:stokes_fluid_1_nd_FT_lt_2a} and \eqref{eq:stokes_fluid_1_nd_FT_lt_F2_a} can be written as: $\hat{v}_{\alpha z} = (B + C r_z)\exp(\pm 2\pi k r_z)$ \cite{bickel2007, abdallah2016, bender2013advanced}. Again, noting that the disturbance flow field must decay in the far-field, we write the solutions as:
\begin{subequations}
\label{eq:vz}
\begin{eqnarray}
\hat{v}_{1z} &=& 
    \begin{cases}
      (B_1 + B_2 r_z) \exp(-2\pi k r_z) ; r_z>0 \\
      (B_3 + B_4 r_z) \exp(2\pi k r_z) + (B_5 + B_6 r_z) \exp(-2\pi k r_z) ; r_{z_0}<r_z<0 ,\;\;\text{and} \\
    \end{cases} \\    
\hat{v}_{2z} &=& (B_7 + B_8 r_z) \exp(2\pi k r_z) ; r_z<r_{z_0}  .  
\end{eqnarray}
\end{subequations}
In Appendix \ref{subsec_appendix_A_bc}, we have four boundary conditions for the normal velocity components, namely, the impenetrability of the velocity across the interface Eq. \eqref{eq:bc_lt_1_a}, the no-slip velocity at the interface Eq. \eqref{eq:bc_lt_1_c}, the continuity of tangential stress Eq. \eqref{eq:bc_lt_2_a} and the normal stress jump Eq. \eqref{eq:bc_ft_normal2}. Therefore, we require four relations to uniquely determine the eight unknown constants $B_1-B_8$, and we follow the protocol outlined above for $\hat{v}_{\alpha t}$, using the Greens function properties. In this case, however, the forcing on the right-hand side of Eq. \eqref{eq:stokes_fluid_1_nd_FT_lt_2a} is proportional to $\delta^\prime (r_z)$. Therefore, the additional boundary conditions are the continuity of $\hat{v}_{1 z}$, its first and third derivative with respect $r_z$ at $r_z=0$, whereas the second derivative undergoes a jump discontinuity \cite{bickel2007, abdallah2016, bender2013advanced}. Note that if the third derivative exhibited a finite jump discontinuity, then a $\delta(r_z)$ would appear on the right-hand side of Eq. \eqref{eq:stokes_fluid_1_nd_FT_lt_2a}. The four additional boundary conditions are
\begin{subequations}
\label{eq:bc_z_3}
	\begin{eqnarray}
		\left.\hat{v}_{1z}\right\vert_{r_z=0^+} &=& \left.\hat{v}_{1z}\right\vert_{r_z=0^-}, \label{eq:bc_z_3_a} \\
		\left.\frac{\partial \hat{v}_{1z}}{\partial r_z}\right\vert_{r_z=0^+} &=&  \left.\frac{\partial \hat{v}_{1z}}{\partial r_z}\right\vert_{r_z=0^-}, \label{eq:bc_z_3_b} \\
		\left.\frac{\partial^2 \hat{v}_{1z}}{\partial r_z^2}\right\vert_{r_z=0^+} -  \left.\frac{\partial^2 \hat{v}_{1z}}{\partial r_z^2}\right\vert_{r_z=0^-} &=& -\frac{4\, D}{\ln\kappa}\sin^2\left(\frac{\pi}{2}k p_l\right) ,\;\; \text{and} \label{eq:bc_z_3_c}\\
		\left.\frac{\partial^3 \hat{v}_{1z}}{\partial r_z^3}\right\vert_{r_z=0^+} &=&  \left.\frac{\partial^3 \hat{v}_{1z}}{\partial r_z^3}\right\vert_{r_z=0^-}, \label{eq:bc_z_3_d}		
	\end{eqnarray}
\end{subequations}
allowing $\hat{v}_{\alpha z}$ to be uniquely determined as:
\begin{subequations}
\label{eq:vhat_z}
	\begin{eqnarray}
		\hat{v}_{1z} &=& \frac{D}{2\pi k \ln\kappa}\sin^2\left(\frac{\pi}{2} k p_l\right) \nonumber\\ && \left[ r_z\exp(-2\pi k \vert r_z \vert) + \left(\frac{1-\lambda}{1+\lambda}\right) (r_z + 4\pi k r_{z_0}(r_z-r_{z_0})) \exp(4\pi k r_{z_0}) \exp(-2\pi k r_z) \right]\nonumber\\ &&- \left(\frac{\kappa_\beta}{\eta_1 V_s L^2}\right)\frac{\pi k \hat{u}_z}{(1+\lambda)}\left[1 + 2\pi k (r_z- r_{z_0})\right] \left(4\pi^2 k ^2 + \Gamma\right) \exp(-2\pi k (r_z-r_{z_0}))\nonumber\\ &&; r_z>r_{z_0} ,\;\; \text{and} \\
		\hat{v}_{2z} &=& \frac{D}{\pi k (1+\lambda) \ln\kappa}\sin^2\left(\frac{\pi}{2} k p_l\right) r_z \exp(2\pi k r_z)\nonumber\\ && - \left(\frac{\kappa_\beta}{\eta_1 V_s L^2}\right)\frac{\pi k \hat{u}_z}{(1+\lambda)}\left[1 + 2\pi k (r_{z_0}-r_z)\right] \left(4\pi^2 k ^2 + \Gamma\right) \exp(-2\pi k (r_{z_0}-r_z))\nonumber\\ && ; r_z<r_{z_0}.		
	\end{eqnarray}
\end{subequations}
We now use the continuity equations \eqref{eq:stokes_fluid_1_nd_FT_lt_d} and \eqref{eq:stokes_fluid_1_nd_FT_lt_F2_c} to determine the longitudinal velocity components $\hat{v}_{\alpha l}$. Note that the longitudinal component of the fluid velocity field is independent of the interface deformation at $r_z=r_{z_0}$.

\section{\label{sec_appendix_B}Swimmers orthogonal to interface - Two-dimensional Fourier transform summary}
Here, for swimmers oriented orthogonal to the interface we summarize the expressions for the disturbance flow field in Fourier space. In this case, $p_x=p_y=0$, whereas $p_z =+1 (-1)$, for swimmers oriented away from (towards) the interface. While the continuity equations given by Eqs. \eqref{eq:stokes_fluid_1_nd_FT_lt_d} and \eqref{eq:stokes_fluid_1_nd_FT_lt_F2_c} remain unchanged, the Stokes equations become:
\begin{subequations}
\label{eq:stokes_fluid_1_nd_FT_ortho}
	\begin{eqnarray}
		-2\pi \iu k_x \hat{P}_1 + \frac{\partial^2 \hat{v}_{1x}}{\partial r_z^2} - 4\pi^2 k^2 \hat{v}_{1x} &=& 0, \\
		-2\pi \iu k_y \hat{P}_1 + \frac{\partial^2 \hat{v}_{1y}}{\partial r_z^2} - 4\pi^2 k^2 \hat{v}_{1y} &=& 0 ,\;\; \text{and} \\
		- \frac{\partial \hat{P}_1}{\partial r_z} + \frac{\partial^2 \hat{v}_{1z}}{\partial r_z^2} - 4\pi^2 k^2 \hat{v}_{1z} &=& \frac{D}{\ln\kappa}\text{sgn}(r_z)H\left[\frac{1}{2}- \frac{r_z}{p_z}\right] H\left[\frac{1}{2}+ \frac{r_z}{p_z}\right],	
	\end{eqnarray}
\end{subequations}
where $\text{sgn}(x)$ is the sign-function and $H[x]$ is the Heaviside step function \cite{abramowitz1970tables}. The corresponding equations for $\hat{\bm{v}}_2$ and $P_2$ have a right-hand side equal to zero. Following the steps described in Appendix \ref{subsec_appendix_A_governing_eqns}, in the ($\hat{\bm{l}}, \hat{\bm{t}}, \bm{z}$) coordinates, the Stokes equations become:
\begin{subequations}
\label{eq:stokes_fluid_1_nd_FT_lt_F2_ortho}
	\begin{eqnarray}
		\frac{\partial^4 \hat{v}_{1z}}{\partial r_z^4} -8\pi^2 k^2 \frac{\partial^2 \hat{v}_{1z}}{\partial r_z^2} + 16\pi^4 k^4 \hat{v}_{1z} &=& -\frac{D}{\ln\kappa}4\pi^2 k^2\text{sgn}(r_z)H\left[\frac{1}{2}-\frac{r_z}{p_z}\right]H\left[\frac{1}{2}+\frac{r_z}{p_z}\right], \label{eq:stokes_fluid_1_nd_FT_lt_F2_ortho_a} \\
		\frac{\partial^4 \hat{v}_{2z}}{\partial r_z^4} -8\pi^2 k^2 \frac{\partial^2 \hat{v}_{2z}}{\partial r_z^2} + 16\pi^4 k^4 \hat{v}_{2z} &=& 0 , \;\;\text{and} \label{eq:stokes_fluid_1_nd_FT_lt_F2_ortho_b} \\		
		\frac{\partial^2 \hat{v}_{1t}}{\partial r_z^2} - 4\pi^2 k^2 \hat{v}_{1t} &=& 0,		\label{eq:stokes_fluid_1_nd_FT_lt_F2_ortho_c} 
	\end{eqnarray}
\end{subequations}
with the corresponding continuity equations being given by Eqs. \eqref{eq:stokes_fluid_1_nd_FT_lt_d} and \eqref{eq:stokes_fluid_1_nd_FT_lt_F2_c}. We note that unlike swimmers with a non-zero $p_x$ and/or $p_y$, in this case the $\hat{v}_{\alpha t}$ satisfy a homogeneous second order differential equation. Therefore, the no-slip and the tangential stress boundary conditions are sufficient to determine the transverse component of the velocity field, and give  $\hat{v}_{\alpha t} = 0$.

To determine the normal component, we first determine the Green's function of the differential operator in fluid region 1, which is
\begin{equation}
\frac{\partial^4 \hat{v}_{1z}}{\partial r_z^4} -8\pi^2 k^2 \frac{\partial^2 \hat{v}_{1z}}{\partial r_z^2} + 16\pi^4 k^4 \hat{v}_{1z} = F_1\delta(r_z-r_z^\prime),
 \label{eq:stokes_fluid_1_ortho_Greens}
\end{equation}
for an arbitrary forcing $F_1$, and then express the solution of Eq. \eqref{eq:stokes_fluid_1_nd_FT_lt_F2_ortho_a} as a convolution integral with the Green's function \cite{bender2013advanced}. In other words, we solve for Eqs. \eqref{eq:stokes_fluid_1_nd_FT_lt_F2_ortho_b} and \eqref{eq:stokes_fluid_1_ortho_Greens}. The solution form for $\hat{v}_{\alpha z}$ remains the same as described in Appendix \ref{subsec_appendix_A_solution}. In this case, the additional boundary conditions are \cite{bickel2007, bender2013advanced}:
\begin{subequations}
\label{eq:bc_z_3_ortho}
	\begin{eqnarray}
		\left.\hat{v}_{1z}\right\vert_{r_z=r_z^{\prime +}} &=& \left.\hat{v}_{1z}\right\vert_{r_z=r_z^{\prime -}}, \label{eq:bc_z_3_ortho_a}\\
		\left.\frac{\partial \hat{v}_{1z}}{\partial r_z}\right\vert_{r_z=r_z^{\prime +}} &=&  \left.\frac{\partial \hat{v}_{1z}}{\partial r_z}\right\vert_{r_z=r_z^{\prime -}}, \label{eq:bc_z_3_ortho_b}\\
		\left.\frac{\partial^2 \hat{v}_{1z}}{\partial r_z^2}\right\vert_{r_z=r_z^{\prime +}} &=&  \left.\frac{\partial^2 \hat{v}_{1z}}{\partial r_z^2}\right\vert_{r_z=r_z^{\prime -}} ,\;\; \text{and} \label{eq:bc_z_3_ortho_c}\\
		\left.\frac{\partial^3 \hat{v}_{1z}}{\partial r_z^3}\right\vert_{r_z=r_z^{\prime +}} -  \left.\frac{\partial^3 \hat{v}_{1z}}{\partial r_z^3}\right\vert_{r_z=r_z^{\prime -}} &=& F_1.		\label{eq:bc_z_3_ortho_d}
	\end{eqnarray}
\end{subequations}
We note that the additional boundary conditions obtained here remain true for any orientation of the swimmer with $p_z\neq 0$, as the $\bm{\delta}$-function in Eq. (\ref{eq:stokes_continuity}) integrates out. Therefore, swimmers oriented parallel to the interface are a special case. Using Eqs. \eqref{eq:bc_lt_1_a}, \eqref{eq:bc_lt_1_c}, \eqref{eq:bc_lt_2_a}, \eqref{eq:bc_ft_normal2} and \eqref{eq:bc_z_3_ortho} yields:
\begin{subequations}
\label{eq:vhat_z_g_ortho}
	\begin{eqnarray}
		\hat{v}_{1z-G}(r_z\vert r_z^\prime) &=& \frac{F_1}{32\pi^3 k^3}\left(1 + 2\pi k \left\vert r_z -r_z^\prime\right\vert \right)\exp(-2\pi k \left\vert r_z-r_z^\prime\right\vert) \nonumber\\ &&+ \frac{F_1}{32\pi^3 k^3} \left(\frac{1-\lambda}{1+\lambda}\right)\left[1+2\pi k r_z^\prime + 2\pi k r_z -4\pi k r_{z_0}+8\pi^2 k^2(r_z-r_{z_0})(r_z^\prime - r_{z_0})\right]\nonumber\\ && \qquad\qquad \exp(4\pi k r_{z_0})\exp(-2\pi k (r_z+r_z^\prime)) \nonumber\\ &&- \left(\frac{\kappa_\beta}{\eta_1 V_s L^2}\right)\frac{\pi k}{(1+\lambda)}\left[1 + 2\pi k (r_z- r_{z_0})\right] \left(4\pi^2 k ^2 + \Gamma\right) \hat{u}_z \exp(-2\pi k (r_z-r_{z_0}))\nonumber\\ &&; r_z>r_{z_0} ,\;\;\text{and} \\
		\hat{v}_{2z-G}(r_z\vert r_z^\prime) &=& \frac{F_1}{(1+\lambda)}\frac{1}{16\pi^2 k^3}(1+2\pi k (r_z^\prime - r_z))\exp(-2\pi k (r_z^\prime -r_z))\nonumber\\ && - \left(\frac{\kappa_\beta}{\eta_1 V_s L^2}\right)\frac{\pi k}{(1+\lambda)}\left[1 + 2\pi k (r_{z_0}- r_z)\right] \left(4\pi^2 k ^2 + \Gamma\right) \hat{u}_z \exp(2\pi k (r_z-r_{z_0}))\nonumber\\ && ; r_z<r_{z_0},		
	\end{eqnarray}
\end{subequations}
where the subscript $G$ refers to the Green's function. We can now obtain $\hat{v}_{\alpha z}$ by convolving $\hat{v}_{\alpha z-G}$ with the original forcing, that is, replacing $F_1$ by the right-hand side of Eq. \eqref{eq:stokes_fluid_1_nd_FT_lt_F2_ortho_a} and integrating over $r_z^\prime$. Thus, we solve
\begin{equation}
\hat{v}_{\alpha z} = -\frac{4\pi^2 k^2 D}{\ln\kappa}\int \mathrm{d}r_z^\prime \hat{v}_{\alpha z-G}(r_z\vert r_z^\prime) \text{sgn}(r_z^\prime)H\left[\frac{1}{2}- \frac{r_z^\prime}{p_z}\right] H\left[\frac{1}{2}+ \frac{r_z^\prime}{p_z}\right],
\label{eq:z_ortho_convolution}
\end{equation}
where only the terms involving $F_1$ in Eq. \eqref{eq:vhat_z_g_ortho} are to be integrated over. This yields
\begin{subequations}
\label{eq:vhat_z_ortho}
	\begin{eqnarray}
		\hat{v}_{1z} &=&  -\left(\frac{\kappa_\beta}{\eta_1 V_s L^2}\right)\frac{\pi k}{(1+\lambda)}\left[1 + 2\pi k (r_z- r_{z_0})\right] \left(4\pi^2 k ^2 + \Gamma\right) \hat{u}_z \exp(-2\pi k (r_z-r_{z_0})) \nonumber\\ && +  \frac{D}{8\pi^2 k^2\ln\kappa}\left(\frac{1-\lambda}{1+\lambda}\right)\exp(4\pi k r_{z_0})\exp(-2\pi k r_z)\nonumber\\ &&\qquad\qquad\bigg[2(\cosh(\pi k)-1)(1+3\pi k r_z - 4\pi k r_{z_0} \nonumber\\ &&\qquad\qquad - 4\pi^2 k^2 r_{z_0}(r_z-r_{z_0})) - \pi k \sinh(\pi k)(1+ 4\pi k (r_z-r_{z_0})) \bigg] \nonumber\\ && - \frac{D}{8\pi^2 k^2\ln\kappa} \begin{cases}
      \exp(2\pi k r_z)\Big[2(\pi k r_z -1)(\cosh(\pi k)-1) + \pi k \sinh(\pi k) \Big]\\ \qquad\qquad; r_z<-\left\vert \frac{ p_z}{2}\right\vert \\ \\
      \left(1 + \frac{\pi k}{2}(1+2 r_z)\right)\exp(-\pi k(1+2 r_z)) \\- \left(1 + \frac{\pi k}{2}(1-2 r_z)\right)\exp(-\pi k(1-2 r_z)) + 2\text{sgn}(r_z) \\ -\left(2\text{sgn}(r_z) +2\pi k r_z \right)\exp\left(-2\pi k\left\vert r_z \right\vert\right); r_z\in\left[-\left\vert \frac{ p_z}{2}\right\vert, \left\vert \frac{ p_z}{2}\right\vert\right] \\ \\
      \exp(-2\pi k r_z)\Big[2\pi k r_z (\cosh(\pi k)-1) - \pi k \sinh(\pi k) \Big]; r_z>\left\vert\frac{ p_z}{2}\right\vert ,\;\;\text{and} \\
    \end{cases} \\ 
		\hat{v}_{2 z} &=& - \left(\frac{\kappa_\beta}{\eta_1 V_s L^2}\right)\frac{\pi k}{(1+\lambda)}\left[1 + 2\pi k (r_{z_0}- r_z)\right] \left(4\pi^2 k ^2 + \Gamma\right) \hat{u}_z \exp(2\pi k (r_z-r_{z_0})) \nonumber\\ && + \frac{D}{4\pi^2 k^2  \ln\kappa}\frac{\exp(2\pi k r_z)}{(1+\lambda)} \left[2(1-\pi k r_z)\left(\cosh(\pi k)-1\right) - \pi k \sinh(\pi k)\right] ; r_z<r_{z_0}.		
	\end{eqnarray}
\end{subequations}
Again, using $\hat{v}_{\alpha z}$ from Eq. \eqref{eq:vhat_z_ortho} in the continuity equations \eqref{eq:stokes_fluid_1_nd_FT_lt_d} and \eqref{eq:stokes_fluid_1_nd_FT_lt_F2_c} readily yield the longitudinal velocity components $\hat{v}_{\alpha l}$.

\section{\label{sec_appendix_C}Swimmers orthogonal to interface - Validation of the kinematic boundary approximation}
Here, for swimmers oriented orthogonal to the interface we compare the kinematic boundary condition from Sec. \ref{subsec_nondim} given by Eq. \eqref{eq:Interface_3}, which contains the non-linear terms, with the approximate condition Eq. \eqref{eq:Interface_4} that we have used in the main manuscript. The decoupling of the hydrodynamics for this swimmer configuration enables us to solve both of the equations numerically in a convenient manner. The expanded form Eq. \eqref{eq:Interface_3} in Cartesian coordinate system is:
\begin{equation}
\frac{\partial u_z}{\partial t} + \left(\frac{\eta_1 L^2 V_s}{\kappa_\beta}\right)\bigg[v_{x} \frac{\partial u_z}{\partial r_{x}} + v_y \frac{\partial u_z}{\partial r_y} - u_z\frac{\partial v_z}{\partial r_z}\bigg] =\left(\frac{\eta_1 L^2 V_s}{\kappa_\beta}\right) \Big[\left. v_z\right\vert_{r_{z_0}} + p_z \Big].
\label{eq:Interface_App_1}
\end{equation}
Given that $\hat{v}_{\alpha t}$ is zero for this swimmer configuration ($\alpha\in [1, 2]$ for the two fluid regions), the Fourier transformed planar velocity components are proportional only to $\hat{v}_{\alpha l}$, and are $\hat{v}_{\alpha x} = k_x \hat{v}_{\alpha l}/k$ and $\hat{v}_{\alpha y} = k_y \hat{v}_{\alpha l}/k$. We can then readily carry out the inverse Fourier transform of the planar and normal components of the disturbance velocity and the gradient of the normal velocity components to get the following expressions for $v_x$, $v_y$, $v_z$ and $\partial v_z/\partial r_z$ at $r_{z_0}$:
\begin{eqnarray}
\label{eq:Interface_App_1_components}
    \left. v_x\right\vert_{r_{z_0}} &=& - \frac{D r_x}{2\pi (1+\lambda)\ln\kappa} \left[\frac{1}{\left(4 r_{\Vert}^2 + \left(1 -2 r_{z_0}^2 \right)^2\right)^{\frac{1}{2}}} + \frac{1}{\left(4 r_{\Vert}^2 + \left(1 +2 r_{z_0}^2 \right)^2\right)^{\frac{1}{2}}} - \frac{1}{\left(r_\Vert^2 + r_{z_0}^2\right)^{\frac{1}{2}}}\right],\nonumber\\&& \\
    \left. v_y\right\vert_{r_{z_0}} &=& - \frac{D r_y}{2\pi (1+\lambda)\ln\kappa} \left[\frac{1}{\left(4 r_{\Vert}^2 + \left(1 -2 r_{z_0}^2 \right)^2\right)^{\frac{1}{2}}} + \frac{1}{\left(4 r_{\Vert}^2 + \left(1 +2 r_{z_0}^2 \right)^2\right)^{\frac{1}{2}}} - \frac{1}{\left(r_\Vert^2 + r_{z_0}^2\right)^{\frac{1}{2}}}\right],\nonumber\\&& \\    
	\left. v_z\right\vert_{r_{z_0}} &=&   \frac{D}{2\pi (1+\lambda)\ln\kappa} \int_0^\infty \mathrm{d}k \frac{1}{k} J_0\left (2 k r_{\Vert}\right) \exp\left(2 k r_{z_0}\right) \Big[2\left(1 - k r_{z_0}\right)\left(\cosh (k)-1\right) - k \sinh(k)\Big] ,\;\;\text{and}\\
	\left. \frac{\partial v_z}{\partial r_z}\right\vert_{r_{z_0}} &=&   \frac{D}{2\pi (1+\lambda)\ln\kappa} \left[\frac{2\left(2 r_\Vert^2 + \left(1 +2 r_{z_0}^2 \right)^2\right)}{\left[4 r_{\Vert}^2 + \left(1 +2 r_{z_0}^2 \right)^2\right]^{\frac{3}{2}}} + \frac{2\left(2 r_\Vert^2 + \left(1 -2 r_{z_0}^2 \right)^2\right)}{\left [4 r_{\Vert}^2 + \left(1 -2 r_{z_0}^2 \right)^2\right]^{\frac{3}{2}}} - \frac{r_{\Vert}^2 + 2 r_{z_0}^2}{\left[r_{\Vert}^2 + r_{z_0}^2\right]^{\frac{3}{2}}} \right].\nonumber\\&&	
\end{eqnarray}

The radial symmetry of $v_x$ and $v_y$ allows us to solve Eq. \eqref{eq:Interface_App_1} in cylindrical polar coordinates ($\hat{\bm{r}}_{\Vert}$, $\hat{\bm{\theta}}$, $\hat{\bm{z}}$). Because $v_\theta = 0$, there is a dimensional reduction in the polar coordinate system; $r_{\Vert}^2 = r_x^2 + r_y^2$ and $\theta = \arctan(r_y/r_x)$. Thus, we solve:
\begin{equation}
\frac{\partial u_z}{\partial t} + \left(\frac{\eta_1 L^2 V_s}{\kappa_\beta}\right)\bigg[v_{r_{\Vert}} \frac{\partial u_z}{\partial r_{\Vert}} - u_z\frac{\partial v_z}{\partial z}\bigg] =\left(\frac{\eta_1 L^2 V_s}{\kappa_\beta}\right) \Big[\left. v_z\right\vert_{r_{z_0}} + p_z \Big].
\label{eq:Interface_App_2}
\end{equation}
where $v_{r_\Vert} = r_x v_x/r + r_y v_y/r$. Exploiting radial symmetry, we calculate $\partial u_z/\partial r_\Vert$ as an inverse Hankel transform of $\hat{u}_z$ with the Bessel function $J_1(2\pi k r_\Vert)$. Here, we refer to the interface deformation obtained from solving the approximate kinematic condition Eq. \eqref{eq:Interface_9} as $u_{z-approx}$.

\begin{figure}
\includegraphics[width=0.6\textwidth]{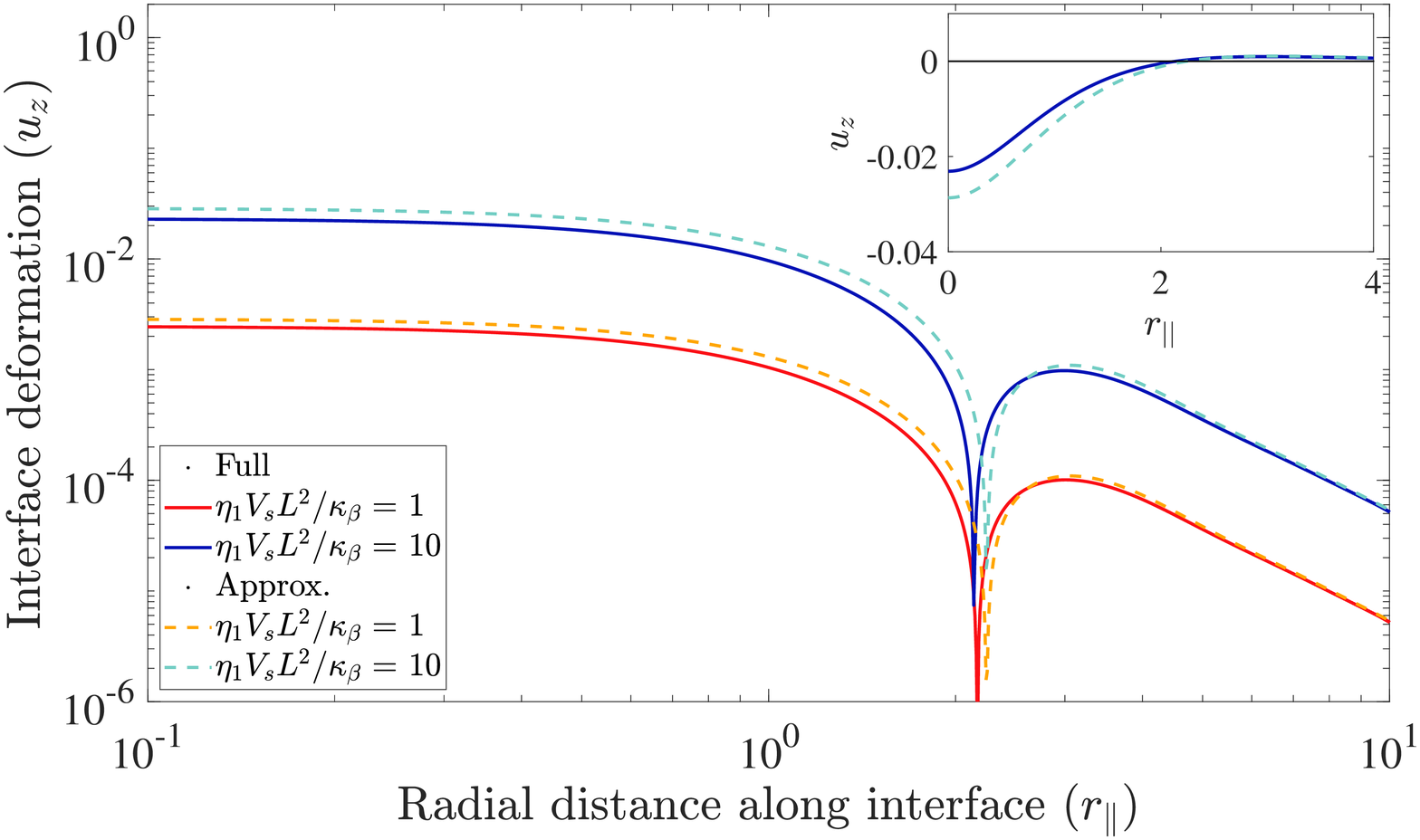}
\caption{The interface deformation $u_z$ due to a pusher plotted as a function of the distance parallel to the interface $r_{\Vert}$ on a log-log scale for the ratio of viscous stress to bending stress: $\eta_1 V_s L^2/\kappa_\beta = $ 1 and 10, at $t=1$ when $\lambda = 0.5$ and $r_{z_0}(0) = 1$. Solid lines represent the full solution obtained from Eq. \eqref{eq:Interface_App_2}, and the dashed lines the approximate solution from Eq. \eqref{eq:Interface_9}. The inset shows a linear plot of $u_z$ for $\eta V_s L^2/\kappa_\beta = 10$.}
\label{fig:appendix_bc_validation}
\end{figure}

In the discussion preceeding Eq. \eqref{eq:Interface_4}, we argued that the approximation to the kinematic boundary condition will hold when the ratio of the viscous stress to the bending stress $\left(\eta_1 V_s L^2/\kappa_\beta\right)$  was $O(1)$ or smaller. As this ratio becomes larger, one would expect the largeness in $\left(\eta_1 V_s L^2/\kappa_\beta\right)$ to dominate over the weak perturbation of the inverse logarithm of the slender swimmer aspect ratio $\kappa$, and hence necessitate the inclusion of the neglected nonlinear terms. In Fig. \ref{fig:appendix_bc_validation} we compare $u_z$ obtained from Eq. \eqref{eq:Interface_App_2} with $u_{z-approx}$ (both solved in along with Eq. \eqref{eq:VzT_perp} for $V_z^T$), for $\eta_1 V_s L^2/\kappa_\beta = 1, 10$ to assess the relative importance of the nonlinear terms on the left-hand side of Eq. \eqref{eq:Interface_App_2}. We focus only on equations specific to shakers, and thus $p_z=0$ on the right-hand side of both Eqs. \eqref{eq:Interface_9} and \eqref{eq:Interface_App_2}. It is clear that even for an initial distance of the swimmer from the interface of $r_{z_0}(0) = 1$, $u_{z-approx}$ provides a very good approximation to the full solution obtained from Eq. \eqref{eq:Interface_App_2}. In the inset, a linear plot is shown for $\eta_1 V_s L^2/\kappa_\beta = 10$, showing that the deviation between the two is small, with a maximum in the region of $r_{\Vert} = 0$ having a relative error of about $0.24$. Although not shown here, $u_{z-approx}$ approaches the $u_z$ obtained form Eq. \eqref{eq:Interface_App_2} as $r_{z_0}$ increases. This validates the boundary approximation for a slender swimmer even at $O(1)$ distances from the interface and for values of $\eta_1 V_s L^2/\kappa_\beta\sim O(10)$. 

\section{\label{sec_appendix_B2}Swimmers arbitrarily oriented to the interface - Two-dimensional Fourier transform summary}
To solve for a swimmer arbitrarily oriented to the interface, we follow the same approach to that of swimmers oriented orthogonal to the interface. Here, the equations for $\hat{v}_z$ and $\hat{v}_t$ are similar to that for swimmers orthogonal to the interface, as given by Eq. \eqref{eq:stokes_fluid_1_nd_FT_lt_F2_ortho} in Appendix \ref{sec_appendix_B}. For an arbitrarily oriented slender swimmer, we have the following set of equations to solve for in the two fluid regions:
\begin{subequations}
\label{eq:stokes_fluid_1_nd_FT_lt_F2_arbit}
	\begin{eqnarray}
		\frac{\partial^4 \hat{v}_{1z}}{\partial r_z^4} -8\pi^2 k^2 \frac{\partial^2 \hat{v}_{1z}}{\partial r_z^2} + 16\pi^4 k^4 \hat{v}_{1z} &=& -2\pi  \iu k \frac{D}{\ln\kappa}\frac{1}{p_z^2}\exp\left(-2\pi \iu k r_z \frac{p_l}{p_z}\right)\bigg\{ 2 p_l p_z  \; \delta(r_z) H\left[\frac{1}{2}-\frac{r_z}{p_z}\right]H\left[\frac{1}{2}+\frac{r_z}{p_z}\right] \nonumber\\ && \qquad + p_l\; \text{sgn}(r_z)  H\left[\frac{1}{2}-\frac{r_z}{p_z}\right]  \delta\left[\frac{1}{2}+\frac{r_z}{p_z}\right] - p_l \; \text{sgn}(r_z)  \delta\left[\frac{1}{2}-\frac{r_z}{p_z}\right]  H\left[\frac{1}{2}+\frac{r_z}{p_z}\right] \nonumber\\ && \qquad - 2\pi \iu k (p_z^2 + p_l^2) \;  \text{sgn}(r_z)H\left[\frac{1}{2}-\frac{r_z}{p_z}\right]  H\left[\frac{1}{2}+\frac{r_z}{p_z}\right] \bigg\} , \label{eq:stokes_fluid_1_nd_FT_lt_F2_arbit_a} \\
		\frac{\partial^4 \hat{v}_{2z}}{\partial r_z^4} -8\pi^2 k^2 \frac{\partial^2 \hat{v}_{2z}}{\partial r_z^2} + 16\pi^4 k^4 \hat{v}_{2z} &=& 0 , \;\;\text{and} \label{eq:stokes_fluid_1_nd_FT_lt_F2_arbit_b} \\		
		\frac{\partial^2 \hat{v}_{1t}}{\partial r_z^2} - 4\pi^2 k^2 \hat{v}_{1t} &=& \frac{D}{\ln\kappa}\frac{p_t}{p_z}\exp\left(-2\pi \iu k r_z \frac{p_l}{p_z}\right) \text{sgn}(r_z)H\left[\frac{1}{2}-\frac{r_z}{p_z}\right]H\left[\frac{1}{2}+\frac{r_z}{p_z}\right].		\label{eq:stokes_fluid_1_nd_FT_lt_F2_arbit_c} 
	\end{eqnarray}
\end{subequations}
After some algebra, we obtain the following expressions for the transverse component of the disturbance flow field:
\begin{subequations}
\label{eq:vhat_t_arbit}
	\begin{eqnarray}
		\hat{v}_{1t} &=&  - \frac{1}{2\pi^2 k^2}\frac{D}{\ln\kappa}\frac{p_t \;(\iu p_l - p_z)}{p_l^2 + p_z^2} \left(\frac{1-\lambda}{1+\lambda}\right)\exp(4\pi k r_{z_0})\exp(-2\pi k r_z) \sinh^2\left(\frac{\pi}{2} k(\iu p_l+ p_z)\right)     \nonumber\\ && - \frac{1}{4\pi^2 k^2}\frac{D}{\ln\kappa}\frac{p_t}{p_l^2 + p_z^2} \begin{cases}
      2(\iu p_l-p_z)\exp(2\pi k r_z) \sinh^2\left(\frac{\pi}{2} k(\iu p_l+ p_z)\right); r_z<-\left\vert \frac{ p_z}{2}\right\vert \\ \\
      \frac{1}{2}\Big[ (\iu p_l-p_z)\exp(2\pi k r_z)\exp(-\pi k \;\text{sgn}(p_z)(\iu p_l+p_z)) \\ \quad +  (\iu p_l+p_z)\exp(-2\pi k r_z)\exp(\pi k \;\text{sgn}(p_z)(\iu p_l-p_z)) \\ \quad- 2(\iu p_l +\text{sgn}(r_z)p_z)\exp(-2\pi k \vert r_z\vert) + 2\;\text{sgn}(r_z)p_z\exp\left(-2\pi \iu k r_z \frac{p_l}{p_z}\right)\Big]\\ ; r_z\in\left[-\left\vert \frac{ p_z}{2}\right\vert, \left\vert \frac{ p_z}{2}\right\vert\right] \\ \\
      2 (\iu p_l+p_z) \exp(-2\pi k r_z) \sinh^2\left(\frac{\pi}{2} k(\iu p_l - p_z)\right); r_z>\left\vert\frac{ p_z}{2}\right\vert ,\;\;\text{and} \\
    \end{cases} \\ 
		\hat{v}_{2 t} &=& - \frac{1}{\pi^2 k^2}\frac{D}{(1+\lambda)\ln\kappa}\frac{p_t \;(\iu p_l - p_z)}{p_l^2 + p_z^2}\exp(2\pi k r_z) \sinh^2\left(\frac{\pi}{2} k(\iu p_l+ p_z)\right) ; r_z<r_{z_0}.		
	\end{eqnarray}
\end{subequations}
For a swimmer oriented orthogonal to the interface, $p_t = 0$, implying that $\hat{v}_{\alpha t} = 0$, as pointed out in Appendix \ref{sec_appendix_B}.

The Green's function for the normal components $\hat{v}_{\alpha z}$ remain the same as given by Eq. \eqref{eq:vhat_z_g_ortho}, where $\alpha \in[1, 2]$ for the two fluid regions. Convolving the Green's function with the right-hand side of Eq. \eqref{eq:stokes_fluid_1_nd_FT_lt_F2_arbit_a} gives the normal component of the velocity. The longitudinal components $\hat{v}_{\alpha l}$ can then be obtained from the use of the continuity equations \eqref{eq:stokes_fluid_1_nd_FT_lt_d} and \eqref{eq:stokes_fluid_1_nd_FT_lt_F2_c}. We omit writing the rather cumbersome expressions for the normal and longitudinal components of the fluid velocity that emerge in this swimmer configuration.

\section{\label{sec_appendix_D}Swimmers confined between a deformable interface and a rigid boundary - Two-dimensional Fourier transform summary}
Here, we briefly describe the formulation for a swimmer confined between a rigid boundary and an initially undeformed interface, specifically when its oriented parallel to the boundaries. We follow the development in Appendix \ref{sec_appendix_A}, and outline only the additional boundary conditions that emerge owing to the additional rigid boundary. For this swimmer orientation only the $z$-component of the velocity is relevant, given that $\dot{\bm{p}} = 0$ as mentioned in Sec. \ref{sec_swimmers_confined_results}. We therefore focus solely on the additional boundary conditions for $\hat{v}_{\alpha z}$, where $\alpha \in [1, 2]$ for the two fluid regions. For a swimmer of arbitrary orientation, however, the rotation rate would be non-trivial and it would be essential to derive the transverse and longitudinal velocity components, $\hat{v}_{\alpha t}$ and $\hat{v}_{\alpha l}$.

The equations for $\hat{v}_{\alpha z}$ are solved subject to the appropriate boundary conditions; the impenetrability and the no-slip conditions at the deformable interface and the rigid boundary, the continuity of tangential stress and the normal-stress jump across the deformable interface. The boundary conditions at the deformable interface are as in Appendix \ref{subsec_appendix_A_bc}, namely Eqs. \eqref{eq:bc_lt_1_a}, \eqref{eq:bc_lt_1_c}, \eqref{eq:bc_lt_2_a} and \eqref{eq:bc_ft_normal2}. The additional Fourier transformed velocity boundary conditions owing to the rigid boundary are:
\begin{subequations}
\label{eq:bc_l_1}
	\begin{eqnarray}
				\left.\hat{v}_{1 z}\right\vert_{H^-} &=& 0, \label{eq:bc_l_1_b}\\
				\left.\frac{\partial \hat{v}_{1 z}}{\partial r_z}\right\vert_{H^-} &=& 0 \label{eq:bc_l_1_c}.
	\end{eqnarray}
\end{subequations}

A general solution for the $4^{th}$-order differential equations Eqs. \eqref{eq:stokes_fluid_1_nd_FT_lt_2a} and \eqref{eq:stokes_fluid_1_nd_FT_lt_F2_a}, noting that the disturbance flow field must decay in the far-field, is
\begin{subequations}
\label{eq:vz_confined}
\begin{eqnarray}
\hat{v}_{1z} &=& 
    \begin{cases}
      (B_1 + B_2 r_z) \exp(-2\pi k r_z) +  (B_3 + B_4 r_z) \exp(-2\pi k r_z); H>r_z>r_{z_0} \\
      (B_5 + B_6 r_z) \exp(2\pi k r_z) + (B_7 + B_8 r_z) \exp(-2\pi k r_z) ; 0<r_z<r_{z_0} ,\;\;\text{and} \\
    \end{cases} \\    
\hat{v}_{2z} &=& (B_9 + B_{10} r_z) \exp(2\pi k r_z) ; r_z<0  .  
\end{eqnarray}
\end{subequations}
We use the six boundary conditions mentioned above, in addition to the four conditions emerging from the properties of the Green's function, which remain the same as in Eq. \eqref{eq:bc_z_3}, to obtain the constants $B_1-B_{10}$. We omit for brevity the cumbersome expressions for $\hat{v}_{\alpha z}$.

\bibliography{refer}

\end{document}